\documentclass[12pt,a4paper]{article}
\usepackage[utf8]{inputenc}
\usepackage[T1]{fontenc}
\usepackage{amsmath}
\usepackage{amsthm}
\numberwithin{equation}{section}
\usepackage{amsfonts}
\usepackage{mathtools}
\usepackage{mathrsfs}
\usepackage{mathpazo}
\usepackage{multirow}
\usepackage{amssymb}
\usepackage{graphicx}
\usepackage{ifpdf}
\usepackage{braket}
\usepackage{cancel}
\usepackage[dvipsnames]{xcolor}
\usepackage{multicol}

\usepackage{cite}
\usepackage[bookmarks=true,colorlinks=true,linkcolor=black,citecolor=orange,urlcolor=orange,bookmarksnumbered]{hyperref}

\usepackage[left=2cm, right=2cm, top=2cm, bottom=3cm]{geometry}

\usepackage[normalem]{ulem}

\usepackage{accents}

\newcommand{\di}[2]{\ensuremath{\arraycolsep=1pt\begin{array}{c} \scriptstyle #1 \\[-0.2em] \scriptstyle #2\end{array}}}

\begin{document}

\renewcommand*{\thefootnote}{\fnsymbol{footnote}}
\begin{center}

{\LARGE \bf Gauged Extended Field Theory \\
and Generalised Cartan Geometry}

\vspace{0.5truecm}

Falk Hassler, David Osten and Alex Swash

\vspace{0.5truecm}

{Institute for Theoretical Physics (IFT), University of Wroc\l aw \\
pl. Maxa Borna 9, 50-204 Wroc\l aw, Poland}

\vspace{0.2truecm}

{{\tt \{falk.hassler,david.osten,alex.swash\}@uwr.edu.pl}}

\vspace{0.5truecm}

\end{center}

\begin{abstract}
Cartan geometry provides a unifying algebraic construction of curvature and torsion, based on an underlying model Lie algebra -- a viewpoint that can be extended naturally to the higher algebraic structures underlying supergravity. We present a Cartan–geometric framework for generalised  geometries governed by a differential graded Lie algebra, extending previous results. The extended tangent bundle admits the action of both a global duality group $\mathcal{G}$ and a local gauge group $H$. This algebraic structure is implemented via a brane current algebra -- the phase space Poisson structure of $p$-branes. Within this Cartan-inspired framework, we define a hierarchy of generalised connections and compute their linearised torsion and curvature tensors, including the higher curvatures required by the tensor hierarchy. This provides a systematic construction of curvature and torsion tensors in generic generalised geometries.
\end{abstract}

\vspace*{-0.5cm}

\renewcommand*{\thefootnote}{\arabic{footnote}}
\setcounter{footnote}{0}

\newpage
\tableofcontents

\section{Introduction} \label{chap:Intro}

Geometry has always been central to our understanding of the physical world—from the curvature of spacetime in Einstein’s general relativity to the intricate gauge symmetries of particle physics. A clear manifestation of this connection appears in (non-linear) $\sigma$-models, which are theories whose fields are maps $X:\  \Sigma \rightarrow M$ from a $p$-dimensional world-volume $\Sigma$ to a $d$-dimensional target space $M$, typically equipped with a metric $g$. In their various incarnations, $\sigma$-models capture a wide range of phenomena, from fundamental interactions in particle physics to effective descriptions in condensed matter systems.

In the Hamiltonian formulation of $\sigma$-models, the phase space often carries a global symmetry group $\mathcal{G}$, known as the duality group. For example, the bosonic string $\sigma$-model enjoys $\mathcal{G} = \mathrm{O}(d,d)$ T-duality symmetry \cite{Duff:1989tf,Tseytlin:1990nb,Tseytlin:1990va,Siegel:1993xq,Siegel:1993th,Siegel:1993bj,Hull:2004in,Hull:2006va,Hitchin:2004ut,Gualtieri:2003dx,Grana:2008yw,Hull:2009mi,Zwiebach:2011rg,Geissbuhler:2013uka,Aldazabal:2013sca,Berman:2013eva,Blair:2013noa,Hohm:2013bwa,Plauschinn:2018wbo}. Also for several higher dimensional $\sigma$-models, reorganising the degrees of freedom so that a $\mathcal{G}$-covariance, typically $\mathcal{G} = E_{d(d)}$, is manifest -- either at the level of the Lagrangian \cite{Sakatani:2016sko,Blair:2017hhy,Sakatani:2017vbd,Arvanitakis:2018hfn,Blair:2019tww,Sakatani:2020umt} or in the Hamiltonian framework \cite{Duff:1990hn,Alekseev:2004np,Berman:2010is,Hatsuda:2012uk,Hatsuda:2012vm,Hatsuda:2013dya,Hatsuda:2020buq,Duff:2015jka,Sakatani:2020iad,Sakatani:2020wah,Strickland-Constable:2021afa,Osten:2021fil,Arvanitakis:2021wkt,Hatsuda:2022zpi,Osten:2023iwc,Osten:2024mjt} --- has proven to be a powerful organisational principle. This reformulation not only streamlines the analysis of symmetries and dynamics, but also reveals deep connections to extensions of classical geometry, such as (exceptional) generalised geometry \cite{Hull:2007zu,Pacheco:2008ps,Berman:2010is,Berman:2011jh,Coimbra:2011ky,Berman:2012vc,Berman:2012uy,Coimbra:2012af,Hohm:2013pua,Lee:2014mla,Hohm:2014qga}, see \cite{Berman:2020tqn,Sterckx:2024vju,Samtleben:2025fta} for recent reviews.

In many situations, the $\sigma$-model also possesses a local gauge symmetry $H$, leading to gauged $\sigma$-models. The presence of this gauge symmetry reduces the physical phase space: only $H$-equivariant configurations are physical. This reduction naturally fits into the language of constrained Hamiltonian systems and symplectic reduction \cite{Marsden:1974dsb}, where the physical phase space is obtained as a quotient by the $H$-action. In this work, we seek a unifying geometric framework that makes \textit{both} $\mathcal{G}$-covariance and $H$-gauge symmetry manifest.

A natural source of inspiration for such a framework is Cartan geometry. Originally developed to generalise Riemannian geometry, Cartan geometry describes a manifold $M$ by modelling its tangent spaces as homogeneous spaces $G/H$, where $G$ is a Lie group and $H$ a closed subgroup \cite{Sharpe:1997,cap2009parabolic}. This perspective encompasses a wide range of geometric structures, including Riemannian, conformal, and projective geometries, and is inherently tied to the theory of Lie algebroids \cite{Mackenzie_2005,Blaom:2006,crampin2016cartan,Attard:2019pvw}.

Generalised geometry provides another, orthogonal, extension of the classical picture: instead of working solely with the tangent bundle $TM$, one considers an extended tangent bundle, such as $TM \oplus T^*M$ or more generally $TM \oplus \bigwedge^p T^*M \oplus \dots$, to incorporate gauge transformations of form fields into the geometric framework. This can be formulated in terms of $\mathcal{G}$-algebroids—such as Courant or Leibniz algebroids \cite{Liu:1995lsa,baraglia2012leibniz,Bugden:2021wxg}—which generalise the Lie algebroid structure underlying Cartan geometry.

Our main motivation is to extend Cartan geometry beyond Lie algebroids, replacing them by more general algebroid structures appropriate to generalised or exceptional geometry. Such an extension not only implements the symmetries and dualities of string and M-theory into the Cartan geometric language, but also provides a unified description of the metric, $B$-field, and other higher-form fields. This broader framework -- bridging Cartan geometry and generalised geometry -- promises new insights into the geometric foundations of duality-symmetric theories and opens new directions for both physical applications and pure mathematical developments. In the following, we will review the state of the art in the relevant subfields.

\paragraph{Curvature in Generalised Geometries.} 
In classical differential geometry, the curvature of an affine connection plays a central role. It allows one to define the Ricci scalar and tensor, which appear in the Einstein--Hilbert action and its equations of motion. Requiring the connection to be torsion-free and metric-compatible uniquely determines it as the Levi--Civita connection. In generalised geometry, however, this uniqueness is lost: the natural analogues of these two conditions fail to fix the generalised connection uniquely, leading instead to a family of \emph{generalised Levi--Civita connections} \cite{Siegel:1993th,Coimbra:2011nw,Coimbra:2011ky,Hohm:2012mf,Garcia-Fernandez:2016ofz,Cortes:2025lns}. This forces us to extend the standard notions of torsion and curvature. A na\"ive substitution of the Lie bracket by the Dorfman or Courant bracket in the usual formulas produces non-tensorial objects \cite{Gualtieri:2007bq,Hohm:2011si}. For the torsion, this issue can be resolved by introducing the \emph{Gualtieri torsion}, which contains an additional term precisely cancelling the non-tensorial contribution \cite{Gualtieri:2007bq}. 

Curvature in generalised geometry can likewise be redefined so that it is genuinely tensorial, though it still depends on the choice of generalised connection \cite{Siegel:1993th,Hohm:2011si,Hohm:2012mf,Jurco:2016emw}. This construction crucially relies on the existence of the $\mathrm{O}(d,d)$ metric on the generalised tangent bundle and therefore does not directly extend to arbitrary generalised geometries. Moreover, in this setting the curvature can no longer be interpreted as the commutator of covariant derivatives. An alternative approach is to restrict attention to certain subbundles of the generalised tangent bundle on which the non-covariance vanishes \cite{Garcia-Fernandez:2013gja,Coimbra:2011nw}. Remarkably, one can still define a \emph{generalised Ricci scalar} and a \emph{generalised Ricci tensor} which are covariant and independent of the particular choice for the connection \cite{Coimbra:2011nw,Coimbra:2011ky,Cavalcanti:2024uky}. This enables the construction of an action principle analogous to the Einstein--Hilbert action, reproducing the bosonic NS--NS sector of 10d type II or heterotic supergravity (for $\mathrm{O}(d,d)$ and $\mathrm{O}(d,d+n)$) and the internal $d$-dimensional sector of 10d type II supergravity and 11d supergravity (for $E_{d(d)}$ with $d \le 7$) \cite{Coimbra:2011nw,Coimbra:2011ky,Garcia-Fernandez:2013gja,Cederwall:2013naa}. For incorporating $\alpha'$ corrections, however, an analogue of the Riemann tensor in generalised geometry is required \cite{Hohm:2012mf}. 

An elegant, alternative construction of the generalised curvature tensor was given by Pol\'a\v{c}ek and Siegel \cite{Polacek:2013nla}, who computed it by commuting covariant derivatives on an \emph{extended space} including coordinates for the local symmetry. As we will explain below, this mirrors the structure of \emph{Cartan geometry}, where the Cartan connection unifies the vielbein and spin connection, whose curvature contains both torsion and Riemann components \cite{Sharpe:1997,cap2009parabolic,Hassler:2024hgq}. To generalise this construction beyond $\mathrm{O}(d,d)$, one requires a \emph{hierarchy} of connections, naturally organised in the framework of tensor hierarchies \cite{Hassler:2023axp}. Related perspectives on torsion and curvature also arise from the graded geometry approach \cite{Aschieri:2019qku,Cueca_2020,Batakidis:2020ylk,Chatzistavrakidis:2023otk,Chatzistavrakidis:2024utp}. 

\paragraph{Cartan Geometry.} Cartan geometry provides a unifying framework encompassing Riemannian, projective, conformal, and more general \emph{Klein geometries} \cite{Sharpe:1997,cap2009parabolic}. It replaces the flat tangent space of Riemannian geometry with a homogeneous space $G/H$, where $G$ is a Lie group and $H$ is a closed subgroup. Non-homogeneous spaces are then viewed as \emph{infinitesimally} Kleinian: Locally they are modelled on $G/H$, but their curvature breaks the global homogeneity. For instance, in general relativity the model is Minkowski space $\mathrm{ISO}(1,d-1)/\mathrm{SO}(1,d-1) \cong \mathbb{R}^{1,d-1}$, and the Cartan connection unifies the translational vielbein and rotational spin connection into a single geometric object. 

Formally, let $M$ be a $d$-dimensional manifold, and $\mathfrak{g}$, $\mathfrak{h}$ the Lie algebras of $G$ and $H$ with $\dim(\mathfrak{g}/\mathfrak{h}) = d$. A \emph{Cartan geometry} modelled on $G/H$ is a principal $H$-bundle $\pi: P \to M$ with a $\mathfrak{g}$-valued one-form $\theta$ on $P$ (the \emph{Cartan connection}) satisfying the following properties:
\begin{enumerate}
    \item \textbf{Absolute parallelism:} $\theta |_p : T_p P \to \mathfrak{g}$ is a linear isomorphism for every $p\in P$;
    \item \textbf{$H$-equivariance:} $R_h^* \theta = \mathrm{Ad}_{h^{-1}}\theta$ for all $h\in H$;
    \item \textbf{Reproduction of fundamental fields:} $\theta(X_\xi) = \xi$ for $\xi \in \mathfrak{h}$, where $X_\xi$ is the corresponding fundamental vector field.
\end{enumerate}
The Cartan curvature is the $\mathfrak{g}$-valued two-form
\begin{equation}
\Theta = - \mathrm{d}\theta + \tfrac12 [\theta,\theta] .
\end{equation}
For \emph{reductive} Cartan geometries -- those with a decomposition $\mathfrak{g} = \mathfrak{h} \oplus \mathfrak{g}/\mathfrak{h}$ invariant under $\mathrm{Ad}(H)$ -- the connection decomposes as $\theta = \omega + e$, where $\omega$ is an Ehresmann connection (e.g. spin connection) and $e$ a coframe. The Cartan curvature similarly splits into an $\mathfrak{h}$-valued curvature $R^\omega$ and a $\mathfrak{g}/\mathfrak{h}$-valued torsion $T^\omega$ for the connection $\omega$. 

\paragraph{Generalised Cartan Geometry.} Motivated by generalised geometry, the notion of Cartan geometry can itself be \emph{generalised} \cite{Hassler:2024hgq}, building on earlier work by Pol\'a\v{c}ek and Siegel \cite{Polacek:2013nla}. There, techniques inspired by Cartan geometry were used to systematically construct covariant torsion and curvature tensors for generalised geometry and double field theory -- previously obtained only via ad hoc methods. 

In this setting, absolute parallelism is replaced by a linear isomorphism
\begin{equation}
\theta |_p : T_p P \oplus T_p^* P \longrightarrow \mathfrak{d},
\end{equation}
where $\mathfrak{d}$ is a $2(d+\dim \mathfrak{h})$-dimensional Lie algebra with a non-degenerate split symmetric ad-invariant bilinear form $\eta$, and $\mathfrak{h} \subset \mathfrak{d}$ is isotropic. The connection preserves $\eta$, and $\mathfrak{d}$ is assumed to admit an $\mathrm{Ad}(H)$-invariant decomposition
\[
\mathfrak{d} \cong \mathfrak{h} \oplus \widetilde{\mathfrak{d}} \oplus \mathfrak{h}^*,
\]
with $\widetilde{\mathfrak{d}}$ modelling the generalised tangent space $T_x M \oplus T_x^* M$. The generalised Cartan connection then takes a block form involving $(\Omega, E)$---the analogues of $(\omega, e)$---together with an additional field $\rho \in \wedge^2 \mathfrak{h}$, which acts as a compensating gauge field for the dual $\mathfrak{h}^*$. 

Its curvature decomposes into generalised torsion $T^\Omega$, curvature $R^\Omega$ of the generalised spin connection $\Omega$, and a higher curvature $R^\rho$ for $\rho$. The structure mirrors that of ordinary Cartan geometry, but with an extra layer: just as the torsion constructed from $e$ alone is made covariant by introducing $\omega$, the curvature of $\Omega$ becomes covariant only upon introducing $\rho$, whose own curvature is then covariant on its own, without the need for additional connections. This reflects the inherent \emph{tensor hierarchy} structure of extended geometries \cite{Hassler:2023axp}.

\hspace{2pt}

Based on these ideas, we introduce a \emph{minimal} Cartan-geometric framework adapted to extended geometries, generalising the standard theory by replacing the underlying Lie algebroid of $TP$ with more general algebroids over the principal bundle $P$ governed by a differential graded Lie algebra. This yields a covariant and systematic formulation of generalised connections, torsion, and curvature---naturally fitting into a tensor hierarchy---relevant to both double and exceptional field theories.

Section~\ref{chap:ExtExFT} defines a class of generalised geometries based on $H\times \mathcal{G}$, with $H$ the local frame group and $\mathcal{G}$ a global duality group. Section~\ref{chap:CurrentAlgebra} realises these geometries in the phase space of branes, showing how their current algebras encode the extended tangent bundle. Section~\ref{chap:Curvature} presents our main geometric results: explicit linearised curvatures for the full generalised Cartan connection, including higher gauge fields like $\rho$, and their organisation into a tensor hierarchy. Section~\ref{chap:Outlook} concludes with applications and open directions.

\section{\texorpdfstring{$H \times \mathcal{G}$}{H x G} Generalised Geometry} \label{chap:ExtExFT}

In this section, we introduce an extended notion of generalised geometry, which geometrises the action of a gauge group $H$ in addition to a duality group $\mathcal{G}$. To this end, we define an extended notion (in the Cartan geometry sense) of the generalised tangent bundle. Moreover, we show which and how notions  of generalised geometry, such as the tensor hierarchy and generalised Lie derivative, can be defined on this bundle. In section \ref{chap:CurrentAlgebra}, we demonstrate how this geometry is naturally realised on the phase space of $p$-branes.

\subsection{Input from \texorpdfstring{$\mathcal{G}$}{G}-generalised Geometry}
We assume a generalised geometry associated with a duality group $\mathcal{G}$, the typical examples with applications in supergravity and string theory being $\mathcal{G} = $ O$(d,d)$ or $E_{d(d)}$. Let us introduce the central objects and conventions used in this article. A key role is played by the tensor hierarchy (algebra) associated with $\mathcal{G}$. This is understood here as the graded vector space
    \begin{equation}
        \mathcal{T} = \bigoplus_{p \geq 1} \mathcal{R}_p, 
    \end{equation}
where $\mathcal{R}_p$ denote representations of $\mathcal{G}$ or their associated bundles. Elements of $\mathcal{R}_p$ have grading $p-1$, and we will use $K_p,L_p,M_p, \dots$ to denote their indices.
$\mathcal{T}$ can be understood as an infinity-enhanced Leibniz algebra, or alternatively (after degree shift) a differential graded Lie algebra\footnote{It becomes a differential graded Lie algebra only when including $\mathcal{R}_0$. Without it, we need to have a Leibniz product on $\mathcal{R}_1$ as an additional algebraic input.}\cite{Palmkvist:2013vya,Greitz:2013pua, Cederwall:2018aab,Cederwall:2019qnw, Cederwall:2019bai,Bonezzi:2019ygf,Bonezzi:2019bek, Lavau:2017tvi, Lavau:2019oja, Lavau:2020pwa}. The two ingredients for this are the $\bullet$-product (corresponding to the algebra bracket) $\bullet: \ \mathcal{R}_p \otimes \mathcal{R}_q \rightarrow \mathcal{R}_{p+q}$, and the differential $\partial: \ \mathcal{R}_{p} \rightarrow \mathcal{R}_{p-1}$,
    \begin{align}
       (V \bullet W)^{L_{p+q}} &= {\eta^{L_{p+q}}}_{M_p N_q} V^{M_p} W^{N_q} \label{eq:EtaDef} \\
    (\partial V)^{L_{p-1}} &= {D_{K_{p}}}^{L_{p-1} M_1} \partial_{M_1} V^{K_{p}}
    \end{align}
for $V \in \mathcal{R}_p$, $W\in \mathcal{R}_q$. The partial derivative $\partial_{M_1}$ is understood with respect to generalised coordinates $X^{M_1}$. These algebraic objects are characterised by structure constants, called $\eta$- and $D$-symbols in the following. Explicit expressions for $\mathcal{G} = E_{d(d)}$ can be found in \cite{Sakatani:2017xcn}. The $\eta$-symbols are graded symmetric\footnote{For the application to exceptional generalised geometry in arbitrary dimensions or arbitrary high representations in the tensor hierarchy, this condition has to be relaxed as some of the $\eta$-symbols do not have fixed symmetry, starting from $\mathcal{R}_{9-d}$ for $E_{d(d)}$ generalised geometry.} in the sense
\begin{equation}
    {\eta^{L_{p+q}}}_{M_p N_q} = (-1)^{(p-1)(q-1)} {\eta^{L_{p+q}}}_{N_q M_p}. \label{eq:EtaGradedSymmetry}
\end{equation} 

Let us note that the $D$-symbols are often (but not always) the dual objects to the $\eta$-symbols \cite{Osten:2024mjt}. The following conditions on the $\eta$- and $D$-symbols are needed here:
    \begin{align}
{D_{K_{2}}}^{L_1 M_1} \partial_{L_1} \otimes \partial_{M_1} &= 0 \label{eq:SectionCondition}\\
{D_{K_{p+1}}}^{L_p N_1} {D_{M_{p+2}}}^{K_{p+1} P_1} \partial_{(N_1} \otimes \partial_{P_1)} &= 0, \label{eq:IdentityDD}
        \end{align}
in addition to relations that can be understood from graded Jacobi (Leibniz) identities
\begin{align}
    0 &= {\eta^{N_{p+q}}}_{K_p L_q} {\eta^{P_{p+q+r}}}_{N_{p+q} M_r} + (-1)^{(p-1)} {\eta^{N_{q+r}}}_{L_q M_r} {\eta^{P_{p+q+r}}}_{K_p N_{q+r}} \nonumber \\
    &\qquad + (-1)^{(q-1)(r-1)} {\eta^{N_{p+r}}}_{K_p M_r} {\eta^{P_{p+q+r}}}_{N_{p+r} L_q} \, ,
\label{eq:IdentityEtaEta}
\end{align}   
when understanding the tensor hierarchy as a differential graded Lie (Leibniz) algebra \cite{Bonezzi:2019ygf,Bonezzi:2019bek,Lavau:2017tvi, Lavau:2020pwa}. Furthermore, compatibility between the differential $\partial$ and the $\bullet$-product implies the following equations relating the $\eta$- and $D$-symbols:
      \begin{align}
           &{} \left( {\eta^{M_{p+q}}}_{L_q K_p} {D_{P_{p+1}}}^{K_p N_1} + {D_{K_{p+q+1}}}^{M_{p+q} N_1} {\eta^{K_{p+q+1}}}_{L_q P_{p+1}}\right) \partial_{N_1} = \left\lbrace \begin{array}{cc}
            \delta^{M_{p+1}}_{P_{p+1}} \partial_{L_1} & \text{ for } q=1 \\
            0  & \text{ for } q>1 \label{eq:IdentityEtaD}
        \end{array} \right. .
    \end{align}
Moreover, let us comment on the relation between $D$- and $\eta$-symbols. In many relevant cases, one can identify the $D$-symbols with the (canonical) duals of the $\eta$-symbols. For instance, for $\mathcal{G} = E_{d(d)}$ this holds for the representations $\mathcal{R}_p$ up to $p<9-d$. Beyond this point, the relevant representations are typically reducible, e.g. $\mathcal{R}_{9-d} = \mathbb{1} \oplus \mathbf{adj}$, in which case the $D$-symbol is a linear combination of the relevant dual $\eta$-symbols.
    
The first condition \eqref{eq:SectionCondition} is the \textit{section condition} that restricts physical functions to only depend on a subset $x^m$ of the coordinates $X^{M_1}$
\begin{equation}
    \partial_{M_1} = (\partial_m , 0 , \dots),
\end{equation}
while the second condition \eqref{eq:IdentityDD} is a consequence of nilpotency of the differential $\partial^2 = 0$. 
As usual after solving the section condition, the representations $\mathcal{R}_p$ correspond to bundles over some underlying manifold $M$ with coordinates $x^m$. Their concrete form will not be relevant in the following. The $\mathcal{R}_1$-representation corresponds to the so-called \textit{generalised tangent bundle} $\mathcal{R}_1 = TM \oplus \dots$, where "$\dots$" corresponds to various $p$-forms and mixed-symmetry tensors, depending on the concrete example. On this generalised tangent bundle, one can define a \textit{generalised Lie derivative}:
\begin{equation}
    \mathcal{L}_{V} W^{N_1} = V^{M_1} \partial_{M_1} W^{N_1} - W^{M_1} \partial_{M_1} V^{N_1} + Y^{M_1 N_1}{}_{K_1 L_1} \partial_{M_1} V^{K_1} W^{L_1} \label{eq:GeneralisedLieDerivative}
\end{equation}
for $V,W \in \mathcal{R}_1$. The $Y$-tensor\footnote{As explained in \cite{Osten:2024mjt}, the $Y$-tensors of $E_{7(7)}$ and $E_{8(8)}$ can also be written in this form.} is defined in terms of the $\eta$- and $D$-symbols as:
\begin{equation}
    Y^{M_1 N_1}{}_{K_1 L_1} = D_{P_2}{}^{M_1 N_1} \eta^{P_2}{}_{K_1 L_1}.
\end{equation} 
The algebra of \textit{generalised diffeomorphisms}, generated by the generalised Lie derivative \eqref{eq:GeneralisedLieDerivative} closes into the so-called $C$-bracket \cite{Berman:2012vc}
\begin{equation}
    [\mathcal{L}_{V_1},\mathcal{L}_{V_2}] W^{N_1} = \mathcal{L}_{[V_1 , V_2]_C} W^{N_1}, \qquad [V_1 , V_2]_C = \frac{1}{2} (\mathcal{L}_{V_1} V_2 - \mathcal{L}_{V_2} V_1).
 \label{eq:CBracket}
\end{equation}
Similarly, there is an action of generalised vectors $\Lambda \in \mathcal{R}_1$ on sections $\Phi$ of an $\mathcal{R}_p$-bundle for $p > 1$, defined in terms of the tensor hierarchy differential $\partial$ and $\bullet$-product \cite{Cederwall:2013naa,Hohm:2015xna, Wang:2015hca}:
\begin{equation}
    \mathcal{L}_{\Lambda}\Phi = \Lambda \bullet (\partial \Phi) + \partial (\Lambda \bullet \Phi).
\end{equation}
Again this closes into the $C$-bracket, $\left[\mathcal{L}_{\Lambda_1} ,\mathcal{L}_{\Lambda_2} \right] \Phi = \mathcal{L}_{[\Lambda_1 , \Lambda_2 ]_C} \Phi$.

\subsection{Extended Generalised Geometry}\label{sec:extGG}
Generalised geometry geometrises the action of the duality group $\mathcal{G}$ in the sense that all quantities appearing are phrased as $\mathcal{G}$-representations. The goal now is to geometrise an \textit{additional} gauge symmetry $\mathfrak{h} = \text{Lie} (H)$. When comparing to general relativity, $\mathcal{G}$ would correspond to GL$(d)$, and $H$ could be taken as the Lorentz group O$(1, d-1)$. Alternatively, $\mathfrak{h}$ could correspond to an arbitrary algebra of generalised Killing vectors.

\paragraph{Action of an additional symmetry group $H$.} The Lie group $H$ with Lie algebra $\mathfrak{h}$ (with structure constants $f_{\alpha \beta}\mathstrut^\gamma$) describes an additional gauge symmetry that is associated with an action of $\mathfrak{h}$ on the representations $\mathcal{R}_p$
\begin{equation}
    f_{\alpha \beta}\mathstrut^\gamma f_{\gamma M_p}\mathstrut^{N_p} = - 2 f_{[\alpha | M_p}\mathstrut^{K_p} f_{| \beta] K_p}\mathstrut^{N_p}.
    \label{eq:StructureConstantsRepresentation}
\end{equation}
Furthermore, we assume that $\mathfrak{h}$ leaves the $\eta$ and $D$-symbols invariant:
\begin{align}
    f_{\alpha M_p}\mathstrut^{K_p} \eta^{L_{p+q}}\mathstrut_{K_p N_q} + f_{\alpha N_q}\mathstrut^{K_q} \eta^{L_{p+q}}\mathstrut_{M_p K_q} &= f_{\alpha K_{p+q}}\mathstrut^{L_{p+q}} \eta^{K_{p+q}}\mathstrut_{M_p N_q},
\label{eq:Identityfeta} \\
    f_{\alpha K_p}\mathstrut^{M_p} D_{L_{p+1}}\mathstrut^{K_p N_1} + f_{\alpha K_1}\mathstrut^{N_1} D_{L_{p+1}}\mathstrut^{M_p K_1} &= f_{\alpha L_{p+1}}\mathstrut^{K_{p+1}} D_{K_{p+1}}\mathstrut^{M_p N_1}.
\label{eq:IdentityfD}
\end{align}
A typical example of $\mathfrak{h}$ will be the maximal compact subgroup of $\mathcal{G}$.

\paragraph{Representations.} As a first step, we define an extended $\mathfrak{R}_1$-representation as the vector space
\begin{equation}
    \mathfrak{R}_1 = \mathfrak{h} \oplus \mathcal{R}_1 \oplus (\mathcal{R}_2 \otimes \mathfrak{h}^*) \oplus (\mathcal{R}_3 \otimes \mathfrak{h}^* \wedge \mathfrak{h}^*) \oplus \dots = \mathfrak{h} \oplus \bigoplus_{q \geq 0} \Big( \mathcal{R}_{1+q} \otimes \bigwedge\nolimits^{q} \mathfrak{h}^* \Big) \label{eq:ExtendedR1}
\end{equation}
that geometrises both the action of $\mathcal{G}$ via the appearance of $\mathcal{G}$-representations $\mathcal{R}_p$ and the action of $H$ with $\mathfrak{h}$ and $\mathfrak{h}^*$ transforming in the adjoint and coadjoint representation, and $\mathcal{R}_p$ transforming as above \eqref{eq:StructureConstantsRepresentation}. For an object in this representation, we introduce the notation
\begin{align}
    \mathcal{M}_1 &= \left( \mu, M_1, \di{\mu}{M_2}, \di{\mu_1 \mu_2}{M_3}, \dots , \di{\mu_1 \dots \mu_{q-1}}{M_q} ,\dots \right)
\end{align}
for its index. Also for $p>1$, this can be extended to a hierarchy of representations $\mathfrak{R}_p$
\begin{equation}
    \mathfrak{R}_p = \mathcal{R}_p \oplus (\mathcal{R}_{p+1} \otimes \mathfrak{h}^*) \oplus (\mathcal{R}_{p+2} \otimes \mathfrak{h}^* \wedge \mathfrak{h}^* ) \oplus \dots = \bigoplus_{q \geq 0} \Big( \mathcal{R}_{p+q} \otimes \bigwedge\nolimits^{q} \mathfrak{h}^* \Big) \label{eq:ExtendedRp}
\end{equation}
with indices
\begin{align}
    \mathcal{M}_p &= \left( M_p , \di{\mu}{M_{p+1}}, \dots , \di{\mu_1 \dots \mu_{q}}{M_{p+q}}, \dots \right).
\end{align}
Note that despite being seemingly infinite-dimensional, these representations are finite-dimensional (given that all representations $\mathcal{R}_p$ in the original tensor hierarchy are finite-dimensional). They are restricted by the dimensionality $n$ of the gauge algebra $\mathfrak{h}$, which cuts-off all form indices beyond $\mu_1 \dots \mu_n$. From the point of view of supergravity, the ordinary tensor hierarchy of $E_{d(d)}$ typically ends at $\mathcal{R}_{9-d}$ \cite{Samtleben:2008pe}, whereas, from the point of view of branes, the ordinary tensor hierarchy $E_{d(d)}$ is restricted by the $P$-brane dimension to $\mathcal{R}_{P+1}$.

For $\mathcal{G} = E_{d(d)}$, the approach here is closely related to the 'megaspace' exceptional field theory in \cite{Hassler:2023axp}. There, an extended geometry is proposed that naturally corresponds to generalised geometry associated to $\mathcal{G} = E_{d+n(d+n)}$ where $n = \text{dim}(\mathfrak{h})$. The $\mathfrak{R}_1$-representation \eqref{eq:ExtendedR1} proposed here, could be understood as a subset of the $\mathcal{R}_1$ of $E_{d+n(d+n)}$:
\begin{equation}
    \mathfrak{R}_1 (E_{d(d)},\mathfrak{h}) \subset \mathcal{R}_1( E_{d+n(d+n)} ) \, ,
\end{equation}
in both the finite- and infinite-dimensional cases. $\mathfrak{R}_1 (E_{d(d)},\mathfrak{h})$ appears as leading contribution in the level decomposition of $\mathcal{R}_1( E_{d+n(d+n)} )$. In particular, $\mathfrak{R}_1$ ignores the mixed-symmetry tensors that appear in the GL$(d)$-decomposition of $E_{d+n(d+n)}$ for $\mathcal{R}_p$ with $p> 9 - (d+n)$. 

The approach in \cite{Hassler:2023axp} has the effect that it geometrises not only duality transformations by elements in $\mathcal{G}=E_{d(d)}$ and the gauge group $H$, but full duality transformations by $E_{d+n(d+n)}$. This larger group allows to capture generalised dualities \cite{YuhoFalkNew}, but the trade-off is that for generic choices of $d < 9$ and $n$ one ends up with irregular, infinite-dimensional Lie algebras\footnote{Though see \cite{Bossard:2017aae,Bossard:2018utw,Bossard:2021jix,Cederwall:2021ymp,Cederwall:2025muh,Bossard:2019ksx,Bossard:2021ebg} for recent progress in extended geometries based on infinite-dimensional (affine/Kac-Moody) Lie algebras.} $\mathfrak{e}_{d(d)}^{++\dots+}$. To circumvent this problem, we propose a \textit{minimal} extended symmetry algebra here that is determined $\textit{solely}$ by the global $\mathcal{G}$- and local $H$-covariance.

\paragraph{Extending $\mathfrak{R}_p$ to $p=0$.} In section~\ref{chap:Curvature} we will introduce the Cartan connection. To deal with it in an effective way, it is practical to define $\mathfrak{R}_0$ as
\begin{equation}
    \mathfrak{R}_0 = \bigoplus_{q \geq 1} \Big( \mathcal{R}^*_{q} \otimes \bigwedge\nolimits^{q} \mathfrak{h} \Big) \oplus (\mathfrak{h} \oplus \mathcal{R}_0) \oplus \bigoplus_{q \geq 1} \Big( \mathcal{R}_{q} \otimes \bigwedge\nolimits^{q} \mathfrak{h}^* \Big)\,.
\end{equation}
This is motivated by the existence and similar structure of an $\mathcal{R}_0$-representation of the duality group $\mathcal{G}$, which is simply the Lie algebra $\mathfrak{g}$ of $\mathcal{G}$. Here, we introduce it by hand. It is clearly a representation of $H \times \mathcal{G}$. The representation $\mathcal{R}_0$ is the Lie algebra $\mathfrak{g}$ of the duality group $\mathcal{G}$. A special role will be played by a parabolic subalgebra 
\begin{equation}\label{eq:R0tilde}
    \widetilde{\mathfrak{R}}_0 = \bigoplus_{q \geq 1} \Big( \mathcal{R}_{q} \otimes \bigwedge\nolimits^{q} \mathfrak{h}^* \Big),
\end{equation}
in which the physical fields, in our cases the generalised spin connection and higher versions, are contained. Only for them, we would like to compute torsion, curvatures and the corresponding Bianchi identities. Therefore, it will be practical to limit the discussion to $\widetilde{\mathfrak{R}}_0$. We can define the following Lie algebra structure of $\widetilde{\mathfrak{R}}_0$,
\begin{equation}
    [R^{A_p}_{\alpha_1 \dots \alpha_p} , R^{B_q}_{\beta_1 \dots \beta_q}] = \frac{1}{(p+q)!} \eta_{C_{p+q}}{}^{A_p B_q} \delta_{\alpha_1 \dots \alpha_p \beta_1 \dots \beta_q}^{\gamma_1 \dots \gamma_{p+q}} R^{C_{p+q}}_{\gamma_1 \dots \gamma_{p+q}} \label{eq:Generators},
\end{equation}
in terms of the canonical duals of the $\eta$-symbols \eqref{eq:EtaDef} subject to the normalisation 
\begin{equation}
    \eta_{C_{p+q}}{}^{A_p B_q}\eta^{D_{p+q}}{}_{A_p B_q} = \delta_{C_{p+q}}^{D_{p+q}},
\end{equation}
with the convention that the generalised Kronecker delta takes values $\pm 1$ for its non-zero components. 
Here $R_{\alpha_1 \dots \alpha_p}^{A_p}$ denote the generators of $\mathcal{R}_p \otimes \bigwedge^p \mathfrak{h}^*$. The commutator in \eqref{eq:Generators} should be viewed as an ad hoc definition of a Lie algebra structure; nevertheless, consistency is ensured by the graded Jacobi identity satisfied by the dual $\eta$-symbols.

\paragraph{$\eta$- and $D$-symbols.} In principle, these are simply inherited from the $\eta$- and $D$-symbols of $\mathcal{G}$-generalised geometry. The extended $\eta$-symbol $\eta^{\mathcal{L}_2}\mathstrut_{\mathcal{M}_1 \mathcal{N}_1}: \mathfrak{R}_1 \otimes \mathfrak{R}_1 \rightarrow \mathfrak{R}_2$ has the following non-vanishing components for $p,q,r \geq 1$,
{\small\begin{align}
    \eta^{  \di{\lambda_1 \dots \lambda_{p-2}}{L_p}} {}_{\mu, \di{\nu_1 \dots \nu_{q-1}}{N_q}} =  \eta^{  \di{\lambda_1 \dots \lambda_{p-2}}{L_p}} {}_{\di{\nu_1 \dots \nu_{q-1}}{N_q} , \mu} &=  \left\lbrace \begin{array}{cc}
       -(-1)^{q} \delta^{\nu_1 \dots \nu_{q-1}}_{\mu \lambda_1 \dots \lambda_{q-2}} \delta^{L_q}_{N_q}
       & ,\text{ for } p=q \geq 2 \\
       0 & \text{else}
    \end{array} \right. \label{eq:eta_mu_decomp} \\
    \eta^{  \di{\lambda_1 \dots \lambda_{r-2}}{L_{r}}} \mathstrut_{\di{\mu_1 \dots \mu_{p-1}}{M_p} , \di{\nu_1 \dots \nu_{q-1}}{N_q}} &= \left\lbrace \begin{array}{cc}
       -(-1)^{pq} \eta^{L_{p+q}}{}_{M_p  N_q} \delta^{\mu_1 \dots \mu_{p-1} \nu_1 \dots \nu_{q-1}}_{\lambda_1 \dots \lambda_{p+q-2}}  & ,\text{ for } r=p+q \\
    0 & \text{else} 
    \end{array} \right. \nonumber
\end{align}}
In particular, we note that ${\eta^{\mathcal{L}_2}}_{\mu \nu} =0$ and ${\eta^{\mathcal{L}_2}}_{\mu N_1} = {\eta^{\mathcal{L}_2}}_{N_1 \mu} = 0$. We adopt conventions in which contractions of form indices include the standard combinatorial factors. For example,
\begin{equation}
    X^{\mathcal{L}_2} Y_{\mathcal{L}_2} = X^{L_2} Y_{L_2} + X^{\di{\lambda}{L_3}} Y_{\di{\lambda}{L_3}} + \ldots + \frac{1}{(p-2)!} X^{\di{\lambda_1 \dots \lambda_{p-2}}{L_{p}}} Y_{\di{\lambda_1 \dots \lambda_{p-2}}{L_{p}}} + \dots \nonumber
\end{equation}
The $H \times \mathcal{G}$ $D$-symbols are defined similarly in terms of the $\mathcal{G}$ $D$-symbols:
{\small \begin{align*}
    D_{  \di{\lambda_1 \dots \lambda_{p-2}}{L_p}} {}^{\mu, \di{\nu_1 \dots \nu_{q-1}}{N_q}} =  D_{  \di{\lambda_1 \dots \lambda_{p-2}}{L_p}} {}^{\di{\nu_1 \dots \nu_{q-1}}{N_q} , \, \mu} &=  \left\lbrace \begin{array}{cc}
       - (-1)^{q} \delta_{\nu_1 \dots \nu_{q-1}}^{\mu \lambda_1 \dots \lambda_{q-2}} \delta^{N_q}_{L_q}
       & ,\text{ for } p=q \geq 2 \\
       0 & \text{otherwise}
    \end{array} \right. \label{eq:D_mu_decomp} \\
    D_{  \di{\lambda_1 \dots \lambda_{r-2}}{L_{r}}} \mathstrut^{\di{\mu_1 \dots \mu_{p-1}}{M_p} , \di{\nu_1 \dots \nu_{q-1}}{N_q}} &= \left\lbrace \begin{array}{cc}
       -(-1)^{pq} D_{L_{p+q}}{}^{M_p  N_q} \delta_{\mu_1 \dots \mu_{p-1} \nu_1 \dots \nu_{q-1}}^{\lambda_1 \dots \lambda_{p+q-2}}  & ,\text{ for } r=p+q \\
    0 & \text{otherwise}
    \end{array} \right.
\end{align*}}
for $p,q \geq 1$ and $r \geq 2$. Here, we have formally introduced a $D$-symbol $D_{L_{p+q}}{}^{M_p  N_q}$ for $q>1$. Such objects do not usually appear in the exceptional field theory literature. However, they are not needed in practice: in all relevant expressions, the $D$-symbols are contracted with derivatives. Hence, these hypothetical components will not appear explicitly, due to the section condition which is explained below.

The extended $\eta$- and $D$-symbols are used to define the extension of the $Y$-tensor:
\begin{equation}
    Y^{\mathcal{K}_1 \mathcal{L}_1}{}_{\mathcal{M}_1 \mathcal{N}_1} = {D_{\mathcal{P}_2}}^{\mathcal{K}_1 \mathcal{L}_1} {\eta^{\mathcal{P}_2}}_{\mathcal{M}_1 \mathcal{N}_1} .\label{eq:ExtendedYtensor}
\end{equation}
In a similar vein, one can introduce $\eta$- and $D$-symbols that map between different representations $\mathfrak{R}_p$.

\paragraph{Section condition.} As usual, we associate generalised coordinates $X^{\mathcal{M}_1}$ to the representation $\mathfrak{R}_1$. A section is defined by the constraint on the coordinate dependence of functions, namely
\begin{equation}
    {D_{\mathcal{L}_2}}^{\mathcal{M}_1 \mathcal{N}_1} \partial_{\mathcal{M}_1} \otimes \partial_{\mathcal{N}_1} = 0.
\end{equation}
The form of the $\eta$-symbols \eqref{eq:eta_mu_decomp} makes it clear that the only solution to the section condition at this point is one with coordinates $y^\mu$ associated to $\mathfrak{h}$ and coordinates $X^M$ associated to the usual $\mathcal{R}_1$-representation, implying
\begin{equation}
    \partial_{\mathcal{M}_1} = (\partial_\mu , \partial_{M_1} , 0 ,. \dots)
\end{equation}
where $\partial_{M_1}$ is subject to the usual $\mathcal{G}$-section condition
\begin{equation}
    {D_{L_2}}^{M_1 N_1} \partial_{M_1} \otimes \partial_{N_1} = 0.
\end{equation}
Hence, an appropriate setup for a generalised notion of Cartan geometry uses the coordinates
\begin{equation}
    X^{\mathcal{M}} = (y^\mu , X^{M_1} , 0 , \dots)
\end{equation}
on section, where $y^\mu$ corresponds to the coordinates of the gauge group $H$ and $X^{M_1}$ to the generalised coordinates of $\mathcal{G}$-generalised geometry. In the context of generalised Cartan geometry, the former become the fibre coordinates of the $H$-principal bundle $P \rightarrow M$, and the latter would be extended coordinates associated with the generalised tangent bundle of $M$.

\subsection{Generalised Lie derivative}\label{sec:genLieDeriv}
The natural generalised Lie derivative, or Dorfman bracket, of sections $\mathcal{V},\mathcal{W}$ of the bundle associated to $\mathfrak{R}_1$ would be\footnote{Depending on the context we will also write $\mathcal{M} \equiv \mathcal{M}_1$, when there is no reason for confusion.}
\begin{equation}\label{eq:genLieFullGen}
   \mathcal{L}_\mathcal{V} \mathcal{W}^\mathcal{M} = \mathcal{V}^\mathcal{N} \partial_\mathcal{N} \mathcal{W}^\mathcal{M} - \mathcal{W}^\mathcal{N} \partial_\mathcal{N} \mathcal{V}^\mathcal{M} + {Y^{\mathcal{MN}}}_{\mathcal{KL}} \partial_\mathcal{N} \mathcal{V}^\mathcal{K} \mathcal{W}^\mathcal{L}.
\end{equation}
Closure of this generalised Lie derivative requires the following identities for the $Y$-tensor \cite{Berman:2012vc}:
\begin{align}
    \label{eq:YtensorIdentities-1}
    0 &= \partial_\mathcal{P} \partial_\mathcal{N} U^\mathcal{R} V^\mathcal{S} W^\mathcal{Q} \left( {Y^{\mathcal{MP}}}_{\mathcal{KQ}} {Y^{\mathcal{KN}}}_{\mathcal{RS}} - {Y^{\mathcal{MN}}}_{\mathcal{RS}} \delta^\mathcal{P}_\mathcal{Q} \right) \\
    0 &= \left( \partial_\mathcal{P}U^\mathcal{R} \partial_\mathcal{N} V^\mathcal{S} - \partial_\mathcal{N}U^\mathcal{S} \partial_\mathcal{P} V^\mathcal{R} \right) W^\mathcal{Q} \nonumber \\
    \label{eq:YtensorIdentities-2}
    &{} \quad \times \left( 2 {Y^{\mathcal{MN}}}_{\mathcal{SK}} {Y^{\mathcal{KP}}}_{\mathcal{RQ}} + {Y^{\mathcal{MN}}}_{\mathcal{KQ}} {Y^{\mathcal{KP}}}_{\mathcal{RS}} + 2 {Y^{\mathcal{MN}}}_{\mathcal{RQ}} \delta_\mathcal{S}^\mathcal{P} + {Y^{\mathcal{MN}}}_{\mathcal{SR}} \delta_\mathcal{Q}^\mathcal{P} \right)
\end{align}
While the first identity holds because the $Y$-tensor comes from a consistent tensor hierarchy algebra, the second one does not hold in general because terms proportional to $y$-derivatives could cause problems. A simple proof of \eqref{eq:YtensorIdentities-1} for $\mathcal{G}$-generalised geometry goes as follows: We know from \eqref{eq:IdentityEtaD} that 
\begin{equation}
    \left( 
    \eta^{M_2}\mathstrut_{L_1 K_1} D_{P_2}\mathstrut^{K_1 N_1} + D_{K_3}\mathstrut^{M_2 N_1} \eta^{K_3}\mathstrut_{P_2 L_1}
    \right) \partial_{N_1} = \delta^{M_2}_{P_2} \delta^{N_1}_{L_1} \partial_{N_1}
\end{equation}
holds. By tensoring this relation with $D_{M_2}\mathstrut^{R_1 S_1} \partial_{S_1}$ and symmetrising the derivatives, we see that the second term vanishes due to \eqref{eq:IdentityDD} and this eventually leaves us with
\begin{equation}
    \label{eq:IdentityYD}
    Y^{R_1 S_1}\mathstrut_{L_1 K_1} D_{P_2}\mathstrut^{K_1 N_1} \partial_{(N_1} \otimes \partial_{S_1)} = D_{P_2}\mathstrut^{R_1 S_1} \delta^{N_1}_{L_1} \partial_{(N_1} \otimes \partial_{S_1)}.
\end{equation}
Further contracting this with $\eta^{P_2}\mathstrut_{M_1 Q_1}$ gives us our desired result \eqref{eq:YtensorIdentities-1}.

To deal with \eqref{eq:YtensorIdentities-2}, we will introduce two additional assumptions on the generalised Lie derivative, that will lead to a consistent gauge structure. By committing to a larger duality group (in particular one which would rotate to a different choice of $\mathfrak{h}$) they could be lifted but this is not the aim of the present, \textit{minimal}, setup. It should be understood in contrast to the \textit{maximal extensions} of \cite{Hassler:2023axp} where the generalised Lie derivative is defined for the full $\mathfrak{R}_1$ representation. For $\mathcal{G} = E_{d(d)}$ the latter leads to $\mathfrak{R}_1$ being a representation of $E_{d+n(d+n)}$ for $n=$ dim$(\mathfrak{h})$. The fact that for $\mathcal{G}=$ O$(d,d)$ our definition of the $\mathfrak{R}_1$-structure coincides with O$(d+n,d+n)$, as shown in \cite{Polacek:2013nla,Polacek:2017hnq,Butter:2022iza,Hassler:2022egz,Hassler:2024hgq}, is a peculiarity of this specific duality group.

\paragraph{Assumptions:}
\begin{enumerate}
    \item Parameters of the generalised diffeomorphism are required to be of the form
        \begin{align}\label{eq:restrictedV}
            \mathcal{V} = v + V_1 \in \mathfrak{h} \oplus \mathcal{R}_1
        \end{align}
        because they describe the action of the original duality group $\mathcal{G}$ and the action of a gauge algebra $\mathfrak{h}$.
    \item We assume a special ansatz for the dependence on $y$. It only enters via a twist, depending on the representation of $\mathfrak{h}$ through
    \begin{equation}\label{eq:twist}
        T^{\mu \dots}_{M_p \dots}(y,X) = {e_\alpha}^\mu(y) \dots  {e_{M_p}}^{A_p}(y) T^{\alpha \dots}_{A_p \dots}(X),
    \end{equation}
    where ${e_\mu}^\alpha(y)$ is associated to the right-invariant Maurer-Cartan form on $H$. Moreover, ${e_{M_p}}^{A_p}(y)$ mediates the action of $H$ on the representations $\mathcal{R}_p$. 
\end{enumerate}
As hinted already by \eqref{eq:Identityfeta} and \eqref{eq:IdentityfD}, the $H$-action has to preserve the symbols $\eta$ and $D$ of $\mathcal{G}$, restricting the respective Lie algebra $\mathfrak{h}$ to be a subalgebra of $\mathfrak{g}$.  The role of this twist becomes clearer in the current algebra picture where it is explicitly derived in section~\ref{sec:addHsym}.

Acting with a $\mathcal{V}$ from \eqref{eq:restrictedV} on a general
\begin{align*}
    \mathcal{W} &= w + W_1 + W_2 + \ldots \sim (w^\mu , W_1^{M_1} , (W_2)^{M_2}_\mu , \dots)  \in \mathfrak{R}_1\,,
\end{align*}
the generalised Lie derivative in \eqref{eq:genLieFullGen} simplifies to
\begin{align}
   \mathcal{L}_{v+V_1} \mathcal{W} &= \left( [v,w]_{\mathfrak{h}}^\mu + \mathcal{L}_{V_1} w^\mu - \mathcal{L}_{W_1} v^\mu \right) \partial_\mu \nonumber \\
    &{} \quad + \left( \mathcal{L}_{V_1} W_1^{M_1} + \partial_{N_1} v^\nu (W_2)^{L_2}_\nu {D_{L_2}}^{M_1 N_1} \right) \partial_{M_1}. \nonumber \\
    &{} \quad + \left( \mathcal{L}_{V_1} (W_2)^{M_2}_{\mu} +  2 \partial_{N_1} v^\nu (W_3)^{L_3}_{\mu\nu} {D_{L_3}}^{M_2 N_1}
     - v^\kappa {f_{\kappa \mu}}^\nu (W_2)^{M_2}_\nu \right)
    \partial_{\di{\mu}{M_2}} \label{eq:hGgeneralisedDiff} \\
    &{} \quad + \dots \nonumber \\
    &{} \quad + \frac{1}{(p-1)!} \left( \mathcal{L}_{V_1} (W_p)^{M_p}_{\mu_1 \dots \mu_{p-1}} + p \partial_{N_1} v^\nu (W_{p+1})^{L_{p+1}}_{\mu_1 \dots \mu_{p-1} \nu} {D_{L_{p+1}}}^{M_p N_1} \right. 
    \nonumber \\ 
     &{} \qquad \left. - (p-1) v^\kappa {f_{\kappa [ \mu_{p-1}}}^\nu (W_p)^{M_p}_{\mu_1 \dots \mu_{p-2} ] \nu} \right) \partial_{\di{\mu_1 \dots \mu_{p-1}}{M_p}} \nonumber \\
    &{} \quad + \dots, \nonumber
\end{align}
with $[v,w]_\mathfrak{h}^\mu = {f_{\kappa \lambda}}^\mu v^\kappa w^\lambda $. Remarkably, the algebra of such extended generalised diffeomorphisms \eqref{eq:restrictedV} closes on arbitrary sections $\mathcal{W}$ of the $\mathfrak{R}_1$-bundle, 
\begin{equation}
    \label{eq:ClosureGenDiffeos}
    \mathcal{L}_{u+U_1} \mathcal{L}_{v+V_1} \mathcal{W} - \mathcal{L}_{v+V_1} \mathcal{L}_{u+U_1} \mathcal{W} = \mathcal{L}_{[u+U_1,v+V_1]_C} \mathcal{W}
\end{equation}
with the extended Courant bracket
\begin{equation}
    [u+U_1,v+V_1]_C = \left([u,v]_\mathfrak{h} + \mathcal{L}_{U_1} v - \mathcal{L}_{V_1} u \right) + [U_1 , V_1]_C \, ,
\end{equation}
where the last term is the ordinary $C$-bracket associated to the duality group $\mathcal{G}$.  The proof of this closure crucially depends on the identities \eqref{eq:IdentityDD} and \eqref{eq:IdentityEtaD} and is presented in appendix~\ref{sec:ClosureExtGenLd}.

\section{Derivation from Brane Current Algebra}  \label{chap:CurrentAlgebra}

In the previous section, we introduced a new generalised geometry that makes a duality group $\mathcal{G}$ and a gauge symmetry algebra $\mathfrak{h}$ manifest. Here, we will show that this setting has a very natural origin in the realisation of the phase space of branes and its parametrisation in terms of $\mathcal{G}$-covariant currents.

\subsection{Review: brane currents in \texorpdfstring{$\mathcal{G}$}{G}-generalised geometry}

For the realisation of $\mathcal{G}$-generalised geometry on the phase space of $\frac{1}{2}$-BPS $P$-branes, the following dictionary has been established in \cite{Osten:2024mjt}:
\begin{itemize}
    \item We consider the Hamiltonian formulation of a $P$-brane in $d$-dimensional space.\footnote{The extension to $(11=d+n)$- or $(10=d+n-1)$-dimensional supergravity backgrounds is straightforward and has been discussed in \cite{Osten:2024mjt}.}
    
    The phase space variables can be put into the form of a hierarchy of currents, associated to the tensor hierarchy of $\mathcal{G}$ (usually this will be $E_{d(d)}$) for $P>1$, namely
    \begin{equation}
        \text{ $(P-p+1)$-forms } t_{M_p} \in \mathcal{R}_p\,. 
    \end{equation}
    Typically, these $t_{M_p}$ are constructed from the canonical momentum density of the $P$-brane, the embedding coordinate fields and world-volume gauge fields. Examples of explicit realisations of the brane currents for the Hamiltonian formulation of $D$-branes, $M$-branes or Kaluza-Klein monopole world-volume theories have been presented in \cite{Osten:2021fil, Osten:2024mjt}.
    
    For example, for M2-branes, the typical currents are
    \begin{equation}
        t_{M_1} = (p_m , \mathrm{d}x^m \wedge \mathrm{d}x^{m^\prime}, 0 , \dots) , \quad t_{M_2} = (\mathrm{d}x^m , 0 , \dots) , \quad t_{M_3} = ( 1,0, \dots).
    \end{equation}
    They arise after employing the M-theory decomposition of $\mathcal{R}_p$-indices into GL$(d)$-indices $m,m^\prime$.
    \item The Poisson brackets of these $P$-form currents, from now on referred to as \textit{current algebra}, take the form
    \begin{equation}
    \{ t_{A_p}(\sigma) , t_{B_q}(\sigma^\prime) \} = - \eta^{C_{p+q}}{}_{A_p B_q} t_{C_{p+q}} (\sigma') \wedge \mathrm{d}' \delta (\sigma - \sigma') . \label{eq:CurrentAlgebraSmall}
    \end{equation} 
    Additionally, there can be also spatial world-volume boundary contributions $\int \mathrm{d} (\dots)$. They have been discussed in detail in \cite{Osten:2019ayq,Osten:2021fil} for the string and the membrane. Here, we will always neglect them.
    
    There are two underlying gradings for a current $t_{M_p}$ -- one from its nature as a spatial $(P-p+1)$-form (hence depending on $P$) and as an element of $\mathcal{R}_p$ in the tensor hierarchy/infinity-enhanced Leibniz algebroid. With respect to the former, the current algebra bracket of $(P-p+1)$-form currents is in fact graded skew-symmetric (again up to world-volume boundary terms) with degree $-P$
    \begin{equation}
        \{ t_{A_p} (\sigma) , t_{B_q} (\sigma^\prime) \} = - (-1)^{(p-1)(q-1)} \{ t_{B_q} (\sigma^\prime) , t_{A_p} (\sigma) \} .
    \end{equation}
    This grading is consistent with the graded symmetry of the $\eta$-symbols \eqref{eq:EtaGradedSymmetry}.
    \item A correspondence between generalised geometry and this current algebra structure comes from a section $\Phi$ of the generalised tangent bundle or a general $\mathcal{R}_p$-bundle, which can be associated to the object
    \begin{align}
        \Phi = \int \Phi^{A_p}(X(\sigma)) t_{A_p}(\sigma)
    \end{align}
    where the integral is to be understood over the spatial part of the world-volume. In general, the Poisson bracket of two such sections, for \emph{generic currents} satisfying \eqref{eq:CurrentAlgebraSmall}, does \textit{not} automatically reproduce the generalised Lie derivative $\mathcal{L}$
    \begin{align}
        \{ \Phi , \Lambda \} = \mathcal{L}_{\Lambda} \Phi
    \end{align}
    for a section $\Lambda = \int \Lambda^{A_1} (X(\sigma)) t_{A_1} (\sigma) \in \mathcal{R}_1$. In order to obtain this correspondence between Poisson structure and generalised Lie derivative (up to total derivatives), the hierarchy of currents additionally has to satisfy the so-called \emph{brane charge constraints}
    \begin{equation}
        t_{M_p} \wedge \mathrm{d} X^{N_1}\partial_{N_1} = {D_{M_p}}^{K_{p-1} N_1} t_{K_{p-1}} \partial_{N_1}, \label{eq:BraneChargeSmall}
    \end{equation}
    together with the assumption $t_{M_1} = (p_m , \dots)$, where $p_m$ is the canonical momentum of the brane. Remarkably, solutions to the constraints \eqref{eq:BraneChargeSmall} are in one-to-one correspondence with a Hamiltonian formulation of $\frac{1}{2}$-BPS branes.
    \item Hamiltonian and spatial diffeomorphism constraints are characterised by a generalised metric $\mathcal{H}^{A_1 B_1}$, in which also the coupling to a background metric and $p$-form gauge fields is encoded, 
    \begin{align}
        H = \mathcal{H}^{A_1 B_1} t_{A_1} \wedge \star t_{B_1} \approx 0 , 
    \end{align}
    and the $D$-symbols
    \begin{align}
        \qquad D_{C_2}{}^{A_1 B_1} t_{A_1} \wedge \star t_{B_1} \approx 0.
    \end{align}
\end{itemize}
In principle, this is just a reformulation of $\mathcal{G}$-generalised geometry, but in the following we use it as a computational device. Moreover, if we neglect world-volume boundary contributions, the algebraic structure becomes simpler -- it gives rise to a Lie algebroid. In particular, the Jacobi identity of the current algebra \eqref{eq:CurrentAlgebraSmall} corresponds to the graded Jacobi identity of the $\eta$-symbols \eqref{eq:IdentityEtaEta}.

\subsection{Adding \texorpdfstring{$\mathfrak{h}$}{h}-gauge symmetry}\label{sec:addHsym}
Following the idea of \cite{Polacek:2013nla}, pursued further in \cite{Hassler:2024hgq, Hatsuda:2023dwx}, we supplement the above brane current algebra with the local action of a gauge group $H\subset \mathcal{G}$. It is generated on the brane phase space by the currents $s_\alpha$ with the Poisson brackets
\begin{align}
    \{ s_\alpha (\sigma), s_\beta (\sigma') \} &= f_{\alpha \beta}\mathstrut^\gamma s_\gamma (\sigma) \delta (\sigma - \sigma'), \label{eq:CurrentAlgebraSeed}\\
    \{ s_\alpha (\sigma), t_{M_p} (\sigma') \} &= f_{\alpha M_p}\mathstrut^{N_p} t_{N_p} (\sigma) \delta (\sigma - \sigma'), \nonumber 
\end{align}
which are governed by the structure constants $f_{\alpha \beta}\mathstrut^\gamma$ and the $\mathfrak{h}$-action $f_{\alpha M_p}\mathstrut^{N_p}$ from section~\ref{sec:extGG}. However, the extended algebra of the currents $(s_\alpha, t_{A_1}, t_{A_2}, \dots)$ does not close into a Poisson algebra any more. Fortunately, the authors of \cite{Polacek:2013nla} noticed that the Jacobi identity can be restored after introducing dual gauge symmetry generators $\Sigma^\alpha$ into the current algebra in a specific way. Here, we do the same and introduce dual generators such that the extended current algebra becomes an (infinite-dimensional) Lie algebra,
{\small \begin{align}
     \{s_\alpha ( \sigma ), s_\beta ( \sigma' )\} &= f_{\alpha \beta}\mathstrut^\gamma s_\gamma (\sigma) \delta (\sigma - \sigma') \nonumber \\
     \{ s_\alpha ( \sigma ), t_{M_p} ( \sigma' ) \} &= f_{\alpha M_p}\mathstrut^{N_p} t_{N_p} (\sigma) \delta (\sigma - \sigma') \nonumber
     \\
     \{ s_\alpha ( \sigma ), \Sigma^\beta(\sigma^\prime) \} &= \delta_\alpha^\beta \mathrm{d}\delta(\sigma-\sigma^\prime) +  f_{\gamma \alpha}\mathstrut^{\beta} \Sigma^\gamma (\sigma) \delta (\sigma - \sigma') \label{eq:CurrentAlgebraFlat}
     \\
     \{ \Sigma^\beta ( \sigma ), t_{M_p} ( \sigma' ) \} &= 0 = \{\Sigma^\alpha(\sigma) , \Sigma^\beta(\sigma^\prime) \} \nonumber \\
     \{t_{M_p} (\sigma), t_{N_q}  (\sigma')\} &= - \eta^{L_{p+q}} \mathstrut_{M_p N_q} \, t_{L_{p+q}} (\sigma^\prime) \wedge \mathrm{d}' \delta (\sigma - \sigma') + f_{\alpha M_p}\mathstrut^{K_p} \eta^{L_{p+q}}\mathstrut_{K_p N_q} t_{L_{p+q}} \wedge \Sigma^{\alpha} (\sigma) \delta (\sigma - \sigma'). \nonumber
 \end{align}}
Written in this way, the last equation is not manifestly skew-symmetric. But it turns out (up to neglected boundary terms) that it is, after assuming
 \begin{equation}
    \mathrm{d}t_{M_p} = - {f_{\alpha M_p}}^{N_p} \Sigma^\alpha \wedge t_{N_p}.
 \label{eq:Identitydt}
\end{equation}
Due to the Maurer-Cartan equation $\mathrm{d} \Sigma^\gamma = \frac{1}{2} f_{\alpha \beta}\mathstrut^\gamma \Sigma^\alpha \wedge \Sigma^\beta$, similar identities hold for
 \begin{equation}
    (-1)^{P-p+1}\mathrm{d}t^\alpha_{M_p} = - {f_{\beta M_p}}^{N_p} t_{N_p}^{\beta \alpha} +  \frac{1}{2} f_{\beta \gamma}\mathstrut^\alpha t_{M_p}^{\beta\gamma},
 \label{eq:Identitydtalpha}
 \end{equation}
where the prefactor $(-1)^{P-p+1}$ is due to the form degree of $t_{M_p}$.

Combined with this relation, the current algebra \eqref{eq:CurrentAlgebraFlat} should be seen as a \textit{minimal extension} of \eqref{eq:CurrentAlgebraSeed} such that it is a Poisson algebra, which satisfies the Jacobi identity as is shown in appendix \ref{App:Jacobi}. It can be easily understood as a twisted version of a trivial extension of the current algebra \eqref{eq:CurrentAlgebraSmall} associated to $\mathcal{G}$-generalised geometry, as shown in appendix \ref{App:Derivation}.

From the currents in \eqref{eq:CurrentAlgebraFlat}, one can construct
\begin{equation}\label{eq:SigmaComposition}
    t^{\alpha_1 \dots \alpha_q}_{M_p} \coloneq t_{M_p} \wedge \Sigma^{\alpha_1} \wedge \ldots \wedge \Sigma^{\alpha_q},
\end{equation}
where $\Sigma^\alpha = {e_\mu}^\alpha(y) \mathrm{d}y^\mu$ is the right-invariant Maurer-Cartan form and the coordinates $y^\mu$ are canonically conjugate to $s_\alpha$, defined by
\begin{align}
    \{{e_\mu}^\alpha s_\alpha (\sigma) , y^\nu(\sigma^\prime) \} = -\delta^\nu_\mu \delta(\sigma - \sigma^\prime).
\end{align}
Moreover, we define the dual frame fields ${e_\alpha}^\mu (y)$ with $e_\alpha{}^\mu(y) e_\mu{}^\beta (y) =\delta_\alpha^\beta$. At this point, we have found a higher dimensional motivation for the second assumption \eqref{eq:twist} from section~\ref{sec:genLieDeriv}. The twist described there arises after combining \eqref{eq:Identitydt} with \eqref{eq:SigmaComposition} -- the first of these two equations gives rise to the $e_\alpha{}^\mu$ factors, while the second results in the $e_{M_p}{}^{A_p}$ twists.

From the $P$-brane current point of view, $\mathfrak{R}_1$ will consist of all the possible spatial $P$-forms
\begin{equation}
    t_{\mathcal{M}_1} = \left(s_\mu,t_{M_1},t_{M_2}^\mu,t_{M_3}^{\mu_1 \mu_2} , \dots \right).
\end{equation}
Similarly, $\mathfrak{R}_{k>1}$ will consist of all possible $(P-k+1)$-forms
\begin{equation}
    t_{\mathcal{M}_k} = \left(t_{M_k},t_{M_{k+1}}^\mu,t_{M_{k+2}}^{\mu_1 \mu_2} , \dots \right) .
\end{equation} The relevant bracket for these currents in the $\mathfrak{R}_1$-representation\footnote{For the generic case one obtains:
\begin{align*}
    &{} \quad \{ t_{M_p}^{\alpha_1 \dots \alpha_r}(\sigma) ,  t_{N_q}^{\beta_1 \dots \beta_s}(\sigma^\prime) \} \\
    &= (-1)^{(rq + s)} \left( - \eta^{L_{p+q}} \mathstrut_{M_p N_q} \, t_{L_{p+q}}^{\alpha_1 \dots \alpha_r \beta_1 \dots \beta_s} (\sigma^\prime) \wedge \mathrm{d}' \delta (\sigma - \sigma') + f_{\gamma M_p}\mathstrut^{K_p} \eta^{L_{p+q}} \mathstrut_{K_p N_q} t_{L_{p+q}}^{\alpha_1 \dots \alpha_r \beta_1 \dots \beta_s \gamma} (\sigma) \delta (\sigma - \sigma') \right. \\
    &\left. + (-1)^{(r+s)} \frac{r}{2} \eta^{L_{p+q}}\mathstrut_{M_p N_q} \, f_{\gamma \delta}\mathstrut^{[\alpha_1} t_{L_{p+q}}^{\alpha_2 \dots \alpha_{r}] \beta_1 \dots \beta_s \gamma \delta} (\sigma) \delta (\sigma - \sigma') \right).
\end{align*} 
} $t_{M_p}^{\alpha_1 \dots \alpha_{p-1}}$
can be obtained as
\begin{align}
    &{} \quad \{ t_{M_p}^{\alpha_1 \dots \alpha_{p-1}}(\sigma) ,  t_{N_q}^{\beta_1 \dots \beta_{q-1}}(\sigma^\prime) \} \nonumber \\
    &= (-1)^{pq} \left( \eta^{L_{p+q}} \mathstrut_{M_p N_q} \, t_{L_{p+q}}^{\alpha_1 \dots \alpha_{p-1} \beta_1 \dots \beta_{q-1}} (\sigma^\prime) \wedge \mathrm{d}' \delta (\sigma - \sigma') \right. \nonumber \\
    &{} \qquad \left. - f_{\gamma M_p}\mathstrut^{K_p} \eta^{L_{p+q}} \mathstrut_{K_p N_q} t_{L_{p+q}}^{\alpha_1 \dots \alpha_{p-1} \beta_1 \dots \beta_{q-1} \gamma} (\sigma) \delta (\sigma - \sigma') \right. \label{eq:CurrentAlgebraExtR1} \\
    &{} \qquad \left. - \, (-1)^{(p+q)} \eta^{L_{p+q}}\mathstrut_{M_p N_q} \, \frac{(p-1)}{2!} f_{\gamma \delta}\mathstrut^{[\alpha_1} t_{L_{p+q}}^{\alpha_2 \dots \alpha_{p-1}] \beta_1 \dots \beta_{q-1} \gamma \delta} (\sigma) \delta (\sigma - \sigma') \right), \nonumber
\end{align}
and
\begin{align}
      \{s_\alpha (\sigma), t^{\beta_1 \dots \beta_{p-1}}_{M_p}( \sigma' )\}  &= (p-1) \delta_\alpha^{[\beta_{p-1}} t_{M_p}^{\beta_1 \dots \beta_{p-2}]} (\sigma^\prime) \wedge \mathrm{d} \delta (\sigma - \sigma') \nonumber \\
      &{} \qquad + f_{\gamma \alpha}\mathstrut^{[\underline{\beta_1}} t^{\gamma \underline{\beta_2 \dots \beta_{p-1}}]}_{M_p} (\sigma) \delta (\sigma - \sigma') \label{eq:CurrentAlgebraExtH} \\
     &{} \qquad + f_{\alpha M_p }\mathstrut^{N_p} t^{\beta_1 \dots \beta_{p-1}}_{N_p} (\sigma) \, \delta (\sigma - \sigma') \nonumber.
\end{align}
In front of the $\mathrm{d}\delta$-terms, we recognise the 'extended' $\eta$-symbols defined in \eqref{eq:eta_mu_decomp}. Making use of them, we are eventually left with the compact current algebra
\begin{equation}\label{eq:CurrentAlgebraTwistedModelAlgebra}
\boxed{
     \{ t_{\mathcal{M}_1}(\sigma) , t_{\mathcal{N}_1}(\sigma^\prime) \} = - \eta^{\mathcal{L}_2}{}_{\mathcal{M}_1\mathcal{N}_1} t_{\mathcal{L}_2}(\sigma^\prime) \wedge \mathrm{d}^\prime \delta(\sigma - \sigma^\prime) + f_{\mathcal{M}_1\mathcal{N}_1}{}^{\mathcal{L}_1} t_{\mathcal{L}_1} (\sigma) \delta(\sigma- \sigma^\prime)
}\,.
\end{equation}
Similarly to above one can define representations of $\mathfrak{R}_1 [P]$ as brane world-volume \textit{general} sections of 
        \begin{equation*}
             \mathcal{V} = \int \left( v^\mu s_\mu (\sigma) + V_1^{M_1} t_{M_1} (\sigma) + V_2^{M_2}{}_\alpha t_{M_2}^\alpha(\sigma) + \dots \right).
        \end{equation*}
In direct analogy with the $\mathcal{G}$-generalised geometry case, one can reproduce the extended generalised Lie bracket between such objects as
\begin{equation}
    \{ \mathcal{W},\mathcal{V} \} = \mathcal{L}_\mathcal{V} \mathcal{W}
\end{equation}
if an \textit{extended brane charge condition} holds:
\begin{equation}
    t_{\mathcal{M}_p} \wedge \mathrm{d}X^{\mathcal{N}_1} \partial_{\mathcal{N}_1}= D_{\mathcal{M}_p}{}^{\mathcal{L}_{p-1} \mathcal{N}_1} t_{\mathcal{L}_{p-1}} \partial_{\mathcal{N}_1} 
\end{equation}
for the currents $t_{\mathcal{M}_p}$. Decomposing it into $\mathfrak{h}$- and $\mathcal{R}_p$-indices, this simply reduces to the brane charge condition in $\mathcal{G}$-generalised geometry \eqref{eq:BraneChargeSmall} and the condition \eqref{eq:SigmaComposition}.

Let us note that the brane current algebra always closes, not only when  \eqref{eq:restrictedV} is satisfied. Only the following identity holds: 
\begin{equation}
    \{ \mathcal{U} , \{ \mathcal{V} , \mathcal{W} \} \} + c.p. = \int \left( [\mathcal{L}_{\mathcal{V}} , \mathcal{L}_{\mathcal{W}}]\mathcal{U} - \mathcal{L}_{[\mathcal{V},\mathcal{W}]_C} \mathcal{U} + \text{boundary terms} \right) = 0.
\end{equation}
Hence, in general closure of the extended generalised Lie bracket \eqref{eq:genLieFullGen} is only guaranteed \textit{up to boundary terms}.\footnote{A side result of this is that the extended generalised Lie bracket closes with requiring \eqref{eq:restrictedV}, up to total derivative terms.} As shown in section \ref{sec:extGG}, closure without such boundary terms will only be possible if \eqref{eq:restrictedV} satisfied.

\subsection{The zero-mode algebra} \label{chap:ZeroModeAlgebra}
The structure constants $f_{\mathcal{M}_1\mathcal{N}_1}{}^{\mathcal{L}_1}$ in the current algebra \eqref{eq:CurrentAlgebraTwistedModelAlgebra} correspond to a Leibniz algebra (in particular they are not manifestly skew-symmetric). In the context of generalised Cartan geometry, this algebra plays the role of the \textit{model algebra} -- as explained below in section \ref{chap:Curvature}. In particular, we are going to show that the structure constants in \eqref{eq:CurrentAlgebraTwistedModelAlgebra} naturally correspond to a Leibniz algebra associated with a differential graded Lie algebra structure. The latter comes from the canonical Poisson structure of spatial $p$-forms and the canonical momentum. Crucially, this is analogy is only valid \emph{up to world-volume boundary terms} (associated with the world-volume de Rham differential).

Starting from the currents $t_{A_p}^{\alpha_r} \equiv t_{A_p}\wedge \Sigma^{\alpha_1} \wedge \dots \wedge \Sigma^{\alpha_r}$, we define their zero-modes $T_{A_p}^{\alpha_r} \coloneq \int t_{A_p}^{\alpha_r} (\sigma) \in \mathfrak{R}_{p-r} $ for $r < p$, and a differential $Q: \mathfrak{R}_{i} \rightarrow \mathfrak{R}_{i-1}$ acting on these by
\begin{equation}
    Q T_{A_p}^{\alpha_r} := (-1)^{P+1}\int \mathrm{d} t_{A_p}^{\alpha_r} (\sigma) = (-1)^{1+p+r} f_{\gamma A_p}\mathstrut^{B_p} T_{B_p}^{\alpha_r \gamma} + (-1)^p \frac{r}{2} f_{\beta \gamma}\mathstrut^{[\alpha_1} T_{A_p}^{\alpha_2 \dots \alpha_r] \beta \gamma}. \label{eq:QDef}
\end{equation}
This operator squares to zero thanks to \eqref{eq:StructureConstantsRepresentation} and the Jacobi identity of $\mathfrak{h}$.
We also define a graded symmetric product $\bullet : \mathfrak{R}_i \otimes \mathfrak{R}_j \rightarrow \mathfrak{R}_{i+j}$ for the zero-modes, explicitly acting on these as
\begin{equation}
    T_{A_p}^{\alpha_r} \bullet T_{B_q}^{\beta_s} = (-1)^{1 + rq + s} \eta^{C_{p+q}}\mathstrut_{A_p B_q} T_{C_{p+q}}^{\alpha_r \beta_s}.
\end{equation}
From this formula we read off the grading $| T_{A_p}^{\alpha_r} | = p-r-1 $, which agrees with the fact that $T_{A_p}^{\alpha_r}$ is an object of the vector space $\mathfrak{R}_{p-r}$.

Using \eqref{eq:Identityfeta}, we check that the differential $Q$ is a derivation of the product $\bullet$,
\begin{equation}
    Q (T_{A_p}^{\alpha_r} \bullet T_{B_q}^{\beta_s}) = Q T_{A_p}^{\alpha_r} \bullet T_{B_q}^{\beta_s} + (-1)^{|T_{A_p}^{\alpha_r}|} T_{A_p}^{\alpha_r} \bullet Q T_{B_q}^{\beta_s}.
\label{eq:IdentityQCompatibility}
\end{equation}
From the definition of $\bullet$ and the properties of the $\eta$-symbols \eqref{eq:IdentityEtaEta}, we also have a (non standard) graded Jacobi identity of the form
\begin{equation}
\text{\small$(-1)^k (T_{A_p}^{\alpha_r} \bullet T_{B_q}^{\beta_s}) \bullet T_{C_l}^{\gamma_k} + (-1)^{| T_{A_p}^{\alpha_r} | + r} T_{A_p}^{\alpha_r} \bullet ( T_{B_q}^{\beta_s} \bullet T_{C_l}^{\gamma_k} ) + (-1)^{| T_{B_q}^{\beta_s} | | T_{C_l}^{\gamma_k} | + s} (T_{A_p}^{\alpha_r} \bullet T_{C_l}^{\gamma_k} ) \bullet T_{B_q}^{\beta_s} = 0.$}
\label{eq:IdentityGradedJacobiModified}
\end{equation}
We see that setting $r = s = k = 0$ recovers the usual graded Jacobi identity for the product $\bullet$ (upon suspension) \cite{Bonezzi:2019bek}.

The differential $Q : \mathfrak{R}_{i} \rightarrow \mathfrak{R}_{i-1}$ is defined for $i \geq 2$, but extending the graded vector space to include a further space $\mathfrak{R}_0$, we can also extend the differential to $Q : \mathfrak{R}_1 \rightarrow \mathfrak{R}_0$. As was discussed in \cite{Bonezzi:2019ygf}, this extension allows us to take the differential $Q$ and the product $\bullet$ as the fundamental structures, forming a differential graded Lie algebra. In particular, we can define the Leibniz product $\circ: \mathfrak{R}_1 \otimes \mathfrak{R}_1 \rightarrow \mathfrak{R}_1$ as 
\begin{align}
    T_{A_p}^{\alpha_{p-1}} \circ T_{B_q}^{\beta_{q-1}} \coloneq& (-1)^{p + q} T_{B_q}^{\beta_{q-1}} \bullet Q T_{A_p}^{\alpha_{p-1}} \equiv (-1)^{p + q} Q T_{A_p}^{\alpha_{p-1}} \bullet T_{B_q}^{\beta_{q-1}} \nonumber \\
    =& (-1)^{(p-1) (q-1)} f_{\gamma A_p}\mathstrut^{D_p} \eta^{C_{p+q}}\mathstrut_{D_p B_q} T^{\alpha_{p-1} \beta_{q-1} \gamma}_{C_{p+q}} + \nonumber \\
    & (-1)^{pq} \left( \frac{p-1}{2} \right) \eta^{C_{p+q}}\mathstrut_{A_p B_q} f_{\gamma \delta}\mathstrut^{[\alpha_1} T^{\alpha_2 \dots \alpha_{p-1}] \beta_1 \dots \beta_{q-1} \gamma \delta}_{C_{p+q}} \nonumber \\\label{eq:genPoincare}
    =& f_{\di{A_p}{\alpha_p-1} \di{B_q}{\beta_{q-1}}}{}^{\di{C_{p+q}}{\gamma_{p+q-1}}} T_{\di{C_{p+q}}{\gamma_{p+q-1}}},
\end{align} 
where $T_{A_p}^{\alpha_{p-1}}$ and $T_{B_q}^{\beta_{q-1}}$ are zero-modes in $\mathfrak{R}_1$. The graded Jacobi identity of $\bullet$ \eqref{eq:IdentityGradedJacobiModified} and the compatibility of $Q$ with $\bullet$ \eqref{eq:IdentityQCompatibility} ensure the Leibniz identity for the product $\circ$,
{\small \begin{align*}
    T_{A_p}^{\alpha_{p-1}} \circ ( T_{B_q}^{\beta_{q-1}} \circ T_{C_l}^{\gamma_{l-1}} ) &= (-1)^{q + l} T_{A_p}^{\alpha_{p-1}} \circ ( T_{C_l}^{\gamma_{l-1}} \bullet Q T_{B_q}^{\beta_{q-1}} ) = (-1)^{p-1} (T_{C_l}^{\gamma_{l-1}} \bullet Q T_{B_q}^{\beta_{q-1}}) \bullet Q T_{A_p}^{\alpha_{p-1}} \\
    &= (-1)^{l-1} T_{C_l}^{\gamma_{l-1}} \bullet ( Q T_{B_q}^{\beta_{q-1}} \bullet Q T_{A_p}^{\alpha_{p-1}} ) + (-1)^{q+1} (T_{C_l}^{\gamma_{l-1}} \bullet Q T_{A_p}^{\alpha_{p-1}} ) \bullet Q T_{B_q}^{\beta_{q-1}} \\ 
    &= (T_{A_p}^{\alpha_{p-1}} \circ T_{B_q}^{\beta_{q-1}} ) \circ T_{C_l}^{\gamma_{l-1}} + T_{B_q}^{\beta_{q-1}} \circ ( T_{A_p}^{\alpha_{p-1}} \circ T_{C_l}^{\gamma_{l-1}} ).
\end{align*}}
If we take the maximal compact subgroup of the duality group $\mathcal{G}$, the model algebras we construct here can be understood as a generalisation of the Euclidean or Poincar\'e group -- hence, in the context of O$(d,d)$ in \cite{Hassler:2024hgq} this was dubbed \textit{generalised Poincar\'e} algebra. To see why, let us consider the simplest possible duality group $\mathcal{G}=$ GL($d$). It is special because its tensor hierarchy only has a non-trivial $\mathcal{R}_1$ representation, the fundamental of GL($d$). After identifying $H =$ O($d$) with its maximal compact subgroup, the structure constants in \eqref{eq:genPoincare} are those of the Euclidean group's Lie algebra.

\section{Curvatures and Torsions in the Linearised Theory} \label{chap:Curvature}

\subsection{Extended Generalised Cartan Geometry}
Given a manifold $M$, the aim is to define curvature tensors that are both covariant under generalised diffeomorphisms, i.e. admit an action of a duality group $\mathcal{G}$, and under (local) $H$-gauge transformations. Let us define the setting step-by-step, almost in full analogy with \cite{Hassler:2024hgq} where the same steps are discussed for O$(d,d)$ generalised geometry. The final curvature tensors agree with \cite{Hassler:2023axp}, as far as they have been calculated there.
\begin{enumerate}
    \item The underlying algebraic structure is constituted by:
    \begin{itemize}
        \item A Lie group $H$, called \textit{gauge group}, with Lie algebra $\mathfrak{h}$. Given the underlying duality group $\mathcal{G}$, $H$ is typically the maximal compact subgroup of $\mathcal{G}$ -- i.e. for O$(d,d)$ it is the double Lorentz group O$(1,d-1) \times $ O$(d-1,1)$ , or for $E_{4(4)}=$ SL$(5)$ it would be SO$(5)$. We assume the differential graded Lie algebra structure for the tensor hierarchy of $\mathcal{G}$, as described in section \ref{chap:ExtExFT}. $H \subset \mathcal{G}$ ensures that the identities in \eqref{eq:Identityfeta} hold. In standard Cartan geometry, the tangent bundle $TP$ of the principal $H$-bundle $P \rightarrow M$ plays a central role. Its generalisation in this context will be a bundle over $M$, associated to the $\mathfrak{R}_1$-representation of $H\times \mathcal{G}$ generalised geometry introduced in section \ref{chap:ExtExFT},
        \begin{equation}
            \mathfrak{R}_1 [P] \sim TH \oplus \mathcal{R}_1[M] \oplus (\mathcal{R}_2[M] \otimes T^* H) \oplus \dots,
        \end{equation}
        where $\mathcal{R}_p[M]$ represents the usual $\mathcal{R}_p$-bundles of $\mathcal{G}$-generalised geometry. Locally, which is all that we are interested in, this agrees with the result \cite{Hassler:2024hgq} that for O$(d,d)$ the extended generalised tangent bundle is $\mathfrak{R}_1 [P]=TP \oplus T^*P$.

        Crucially, as shown in sections \ref{chap:ExtExFT} and \ref{chap:CurrentAlgebra}, it is possible to define a generalised Lie derivative on this bundle. For sections $\mathcal{V}^{\mathcal{M}_1} = (v^\mu,V_1^{M_1},0, \dots)$ of $\mathfrak{R}_1 [P]$ acting on general sections $\mathcal{W}$ of $\mathfrak{R}_p[P]$, the bracket \eqref{eq:hGgeneralisedDiff} closes. On the other hand, it turned out that on the brane world-volume \textit{general} sections of $\mathfrak{R}_1 [P]$ have a well-defined algebra defined by current algebra \eqref{eq:CurrentAlgebraExtR1} and \eqref{eq:CurrentAlgebraExtH} -- \textit{up to world-volume boundary terms}. It is the latter structure that we will employ below in order to obtain curvature tensors.
        
      \item The \textit{model algebra} $\mathfrak{L}$ is a differential graded Lie algebra \cite{Bonezzi:2019ygf,Bonezzi:2019bek} with an isotropic subalgebra $\mathfrak{h} \subset \mathfrak{l}_1$. In particular, we assume the existence of a graded vector space
      \begin{equation}
          \mathfrak{L} = \mathfrak{l}_0 + \mathfrak{l}_1 + \mathfrak{l}_2 + \ldots = \bigoplus_{p \geq 0} \mathfrak{l}_p
      \end{equation}
      with a graded symmetric product $\bullet: \mathfrak{l}_p \times \mathfrak{l}_q \rightarrow \mathfrak{l}_{p+q}$ and a differential $Q: \mathfrak{l}_p \rightarrow \mathfrak{l}_{p-1}$, subject to the compatibility conditions in \cite{Bonezzi:2019ygf,Bonezzi:2019bek}. From this differential graded Lie algebra, one can derive a Leibniz bracket on $\mathfrak{l}_1$ via the derived bracket
      \begin{equation}
          a \circ b = - Q a \bullet b, \qquad \text{for} \; a,b \in \mathfrak{l}_1.
      \end{equation}
     As an example, we had demonstrated above in section \ref{chap:ZeroModeAlgebra} that the $H\times \mathcal{G}$ tensor hierarchy introduces exactly such a differential graded Lie algebra as zero-mode part of the current algebra (up to world-volume boundary terms). 

     For the extended generalised Cartan geometry setting, we also assume $\mathfrak{l}_1$ to contain $\mathfrak{h}$ as a (non-maximal) \textit{isotropic Lie subalgebra}: 
        \begin{equation}
          [ \mathfrak{h} , \mathfrak{h} ] \subset \mathfrak{h}, \qquad \text{and} \qquad \eta(\mathfrak{h},\mathfrak{h}) = 0.
        \end{equation}
Similar to the $\mathfrak{R}_1$-bundle, $\mathfrak{l}_1$ can be decomposed as a direct sum of vector spaces (but \textit{not} of Lie/Leibniz algebras):
\begin{equation}
	\mathfrak{l}_1 = \mathfrak{h} \oplus \mathcal{R}_1 \oplus \left(\mathfrak{h}^* \otimes \mathcal{R}_2 \right) \oplus \dots \,.
\end{equation}
In particular, since $\dim (\mathfrak{l}_1)= \dim (\mathfrak{R}_1)$, we will label its elements with $\mathfrak{R}_1$-indices $\mathcal{K},\mathcal{L}$. These satisfy
\begin{equation}
    T_{\mathcal{K}} \circ T_{\mathcal{L}} = f_{\mathcal{K} \mathcal{L}}{}^{\mathcal{M}} T_{\mathcal{M}}.
\end{equation}
    In this article, we consider the above minimal setting of a model algebra $\mathfrak{l}_1$, which is defined alone by the action of $\mathfrak{h}$ on the tensor hierarchy $\{ \mathcal{R}_p \}$ of $\mathcal{G}$, and the tensor hierarchy structure itself. Nevertheless, a more general Leibniz algebra structure (with $\mathfrak{h}$ an isotropic Lie subalgebra) should be easy to incorporate in what is to follow.
    \end{itemize}
    
    \item A \textit{generalised Cartan connection} $\theta$ is a pointwise isomorphism between the two algebraic settings introduced above. It is defined by
    \begin{equation}
        \theta^{(1)}|_p : \ \mathfrak{R}_1[P]_p \rightarrow \mathfrak{l}_1 \label{eq:GeneralisedCartanConnectionDefinition},
    \end{equation}
    and preserves the differential graded Lie algebra structure which is governed by the $\eta$- and $D$-symbols in \eqref{eq:eta_mu_decomp}. This implies that this isomorphism -- the generalised Cartan connection -- is in fact a (pointwise) isomorphism of differential graded Lie algebras 
    \begin{equation}
        \vartheta |_p : \bigoplus_q \mathfrak{R}_q [P]_p \rightarrow \mathfrak{L}
    \end{equation}
    consisting of a collection of maps $\theta^{(q)} |_p : \ \mathfrak{R}_q[P]_p \rightarrow \mathfrak{l}_q$. Nevertheless, all additional connections $\theta^{(q)}$ besides $\theta^{(1)}$ will not introduce new degrees of freedom in the definition of the generalised Cartan connection, as it is shown below in section \ref{chap:Connection}. Hence, we will typically refer to $\theta \equiv \theta^{(1)}$ as the generalised Cartan connection, although one should understand it as acting on the full hierarchy. 
    
    The further standard assumptions on $\theta$ are the same as in standard Cartan geometry and generalised Cartan geometry \cite{Hassler:2024hgq}:
 \begin{itemize}
        \item Left-invariant vector fields of $TH$ are identified with the Lie subalgebra $\mathfrak{h} \subset \mathfrak{l}_1$.
        
        \item \textit{Equivariance}. The right-action $R_h$ of $h \in H$ has to match the adjoint action of $h^{-1}$, through
        \begin{equation}
            R^*_h \theta = \mathrm{Ad}_{h^{-1}} \theta\,. \label{eq:EquivarianceGeneralised}
        \end{equation}
        Again, we will interpret this relation as gauge transformations for the components of $\theta$. 
    \end{itemize}
    It will turn out that, under these assumptions, a natural choice is to take $\theta$ generated by the parabolic subalgebra $\widetilde{\mathfrak{R}}_0$ introduced in \eqref{eq:R0tilde}:
    \begin{equation}
        \theta = \exp\left( \Omega_{A_1}^\alpha R^{A_1}_\alpha + \frac{1}{2} \rho_{A_2}^{\alpha_1 \alpha_2} R_{\alpha_1 \alpha_2}^{A_2} + \dots \right).
    \end{equation}
    This will be explained in detail in the next section where we will also construct representations $(R_{\alpha_1 \dots \alpha_p}^{A_p})_{\mathcal{B}_1} {}^{\mathcal{M}_1}$ of the generators of $\widetilde{\mathfrak{R}}_0$.

    \item For a definition of the \textit{generalised Cartan curvature} $\Theta$, we employ the brane current algebra \eqref{eq:CurrentAlgebraExtR1} and \eqref{eq:CurrentAlgebraExtH}. Consider the generalised Cartan connection $\theta$ represented as a brane current
    \begin{equation}
        \theta^{(p)}_{\mathcal{A}_p}(\sigma) = \theta_{\mathcal{A}_p}^{(p),\mathcal{M}_p}(\sigma) t_{\mathcal{M}_p}(\sigma). 
    \end{equation}
    As explained in the next section, we will only be concerned with $p=1$, as the other components do not contain any new information compared to $\theta^{(1)}$. The generalised Cartan curvature will then be given by the non-trivial components $\Theta_{\mathcal{A}_1 \mathcal{B}_1}{}^{\mathcal{C}_1}$ of the current algebra\footnote{The fact that the $\theta^{(p)}$ leave the $\eta$-symbols invariant, guarantees this form of the current algebra.}
    \begin{equation}
       \boxed{ \{ \theta^{(1)}_{\mathcal{A}_1}(\sigma) , \theta^{(1)}_{\mathcal{B}_1}(\sigma^\prime) \} = - \eta^{\mathcal{C}_2}{}_{\mathcal{A}_1 \mathcal{B}_1} \theta^{(2)}_{\mathcal{C}_2}(\sigma^\prime) \wedge \mathrm{d}^\prime \delta(\sigma - \sigma^\prime) + \Theta_{\mathcal{A}_1 \mathcal{B}_1}{}^{\mathcal{C}_1} \theta^{(1)}_{\mathcal{C}_1}(\sigma) \delta(\sigma-\sigma^\prime)}\,.  \label{eq:GeneralisedCartanCurvatureDefinition}
    \end{equation}
    Due to the additional twist by the generalised Cartan connection, it contains
    \begin{itemize}
        \item the model algebra \eqref{eq:CurrentAlgebraTwistedModelAlgebra} including the action of $\mathfrak{h}$ on the representations $\mathcal{R}_p$ of $\mathcal{G}$, because it is already contained in in \eqref{eq:CurrentAlgebraTwistedModelAlgebra}, and, additionally,
        \item \textit{generalised torsions and curvatures}, which are covariant under generalised diffeomorphisms and $\mathfrak{h}$-gauge transformations.
    \end{itemize}
   This data can be extracted from a decomposition of the $\mathfrak{R}_1$-indices of $\Theta_{\mathcal{A}_1 \mathcal{B}_1}{}^{\mathcal{C}_1}$ into $H \times \mathcal{G}$-indices. All required steps are detailed below in section \ref{chap:CartanCurvature}.
    
    The definition \eqref{eq:GeneralisedCartanCurvatureDefinition} generalises the one originating from string currents in the O$(d,d)$ case in \cite{Hassler:2024hgq}. But it is also closely related to the one in \cite{Hassler:2023axp}, which is based on a generalised Lie derivative or Dorfman bracket $[ \cdot , \cdot ]_{\mathfrak{d},D}$ (without referring to the string current algebra), twisted by the generalised model algebra $\mathfrak{d}$: $\Theta_{\mathcal{AB}} = - [\theta_{\mathcal{A}} , \theta_{\mathcal{B}} ]_{D,\mathfrak{d}} \in \mathfrak{d}$.
    
    As the connections $\theta^{(p)}$ in general, do not only contain components in $\mathfrak{h}$ and $\mathfrak{R}_1$, a similar definition for the extended generalised Cartan curvature is not possible in the framework of $H\times \mathcal{G}$-generalised geometry, because the generalised Lie derivative \eqref{eq:hGgeneralisedDiff} only closes after imposing the additional constraint \eqref{eq:restrictedV}\sout{does not close}. Therefore, we employ the brane current algebra in order to define the generalised Cartan curvature in \eqref{eq:GeneralisedCartanCurvatureDefinition}, which is well-defined up to boundary terms. In contrast \cite{Hassler:2023axp} is able to make use of the generalised Lie derivative directly because they are using the maximal extension to $E_{d+n(d+n)}$ for which the generalised Lie derivative closes without additional constraints.
\end{enumerate}

\subsection{Generalised Cartan Connection} \label{chap:Connection}
As an ansatz for the parametrisation of the $\mathfrak{R}_1$-connection $\theta \equiv \theta^{(1)}|_p : \mathfrak{R}_1[P]_p \rightarrow \mathfrak{l}_1$ consistent with our assumptions, we propose
 \begin{equation} \label{eq:BigVielbein}
    \theta_{\mathcal{A}_1}\mathstrut^{\mathcal{M}_1} = \begin{pmatrix}
    \delta_\alpha^\mu  & 0 & 0 & 0 & \cdots & 0 \\
    \Omega_{A_1}^{\mu} & E_{A_1}\mathstrut^{M_1} & 0 & 0 & \cdots & 0 \\
    \Omega_{A_2}^{\alpha,\mu} & \delta_\nu^\alpha \Omega_{A_2}^{M_1,\nu} & \delta_\mu^\alpha E_{A_2}\mathstrut^{M_2} & 0 & \cdots & 0 \\
    \Omega_{A_3}^{\alpha_1 \alpha_2 , \mu} & \delta_{\nu_1 \nu_2 }^{\alpha_1 \alpha_2} \Omega_{A_3}^{M_1, \nu_1 \nu_2} & \delta_{\mu \nu}^{\alpha_1 \alpha_2} \Omega_{A_3}^{M_2,\nu} & \delta_{\mu_1 \mu_2}^{\alpha_1 \alpha_2} E_{A_3}\mathstrut^{M_3} & \cdots & 0 \\
    \vdots & \vdots & \vdots & \vdots & \ddots & \vdots \\
\end{pmatrix}.
 \end{equation}
This form of the generalised Cartan connection extends both the \emph{megavielbein} from \cite{Butter:2021dtu} and the generalised Cartan connection in O$(d,d)$ generalised geometry \cite{Hassler:2024hgq}\footnote{These special cases arise from truncating the differential graded Lie algebra: the generalised Cartan connection of \cite{Hassler:2024hgq} corresponds to the case where the tower terminates at $\mathcal{R}_2 = \mathbf{1}$, while the ordinary Cartan connection is recovered when the tower terminates already at $\mathcal{R}_1$.}. Its components organise themselves into a tower of \emph{higher connections} $\Omega$, beginning with the spin connection $\Omega_{A_1}\mathstrut^\mu$. In particular, for $q<p$, we have
\begin{equation}
       \theta^{(1)}_{\di{\alpha_1 \dots \alpha_{p-1} }{A_p}}\mathstrut^{\di{\mu_1 \dots \mu_{q-1} }{M_q}} = \delta^{\alpha_1 \dots \alpha_{p-1}}_{\mu_1 \dots \mu_{q-1} \nu_1 \dots \nu_{p-q}} \Omega_{A_p}^{M_q,\nu_1 \dots \nu_{p-q}},
\end{equation}
and a dependence on the generalised frames $E_{A_p}{}^{M_p}$ of the $\mathcal{R}_p$-bundles of $\mathcal{G}$-generalised geometry. Similar to the Cartan connection, the physically relevant input is given by $E_{A_1}{}^{M_1}$, while all the other generalised frames can be derived from this quantity. They represent a natural generalisation of the solder form familiar from ordinary Cartan geometry.
 
\paragraph{Linearised Constraints from compatibility with $\eta$.} Not all the connection components are independent. There are relations between the $\Omega$'s because the connections $\theta^{(p)}$ are required to preserve the graded structure, and in particular the $\eta$-symbols
  \begin{equation}
     \eta^{\mathcal{C}_2}\mathstrut_{\mathcal{A}_1 \mathcal{B}_1} = \theta^{(1)}_{\mathcal{A}_1}\mathstrut^{\mathcal{M}_1} \theta^{(1)}_{\mathcal{B}_1}\mathstrut^{\mathcal{N}_1}  \theta^{(2)}_{\mathcal{L}_2}\mathstrut^{\mathcal{C}_2} \eta^{\mathcal{L}_2}\mathstrut_{\mathcal{M}_1 \mathcal{N}_1}.
     \label{eq:OrthConstraint2}
 \end{equation}
As such, the generalised Cartan connections give rise to a Lie group. But working with this Lie group directly is cumbersome -- one could choose different parametrisations and a preferred one is not immediately obvious. Instead, we rather focus here on the underlying Lie algebra. It arises when we work at the \textit{linearised level} in $\Omega$. Moreover, for the sake of brevity of the final expressions, we work in the metric formalism and assume $E_{A_p}{}^{M_p} = \delta_{A_p}^{M_p}$ for the generalised frame fields of $\mathcal{G}$-generalised geometry. After performing the variation, we are left with \begin{equation}\label{eq:variation}
    0 = \delta\theta^{(1)}_{\mathcal{A}_1}{}^{\mathcal{M}_1} \eta^{\mathcal{C}_2}{}_{\mathcal{M}_1\mathcal{B}_1} + \delta\theta^{(1)}_{\mathcal{B}_1}{}^{\mathcal{M}_1} \eta^{\mathcal{C}_2}{}_{\mathcal{A}_1\mathcal{M}_1} + \delta\theta^{(2)}_{\mathcal{M}_2}{}^{\mathcal{C}_2} \eta^{\mathcal{M}_2}{}_{\mathcal{A}_1\mathcal{B}_1} 
\end{equation}
to solve. For the sake of brevity, one might ignore changes of $E_{A_p}{}^{M_p}$ to circumvent the Lie algebra of the duality group $\mathcal{G}$ and instead focus on the generators for the connections that arise as generalisations of the affine connection. These assumptions are sufficient to analyse the independent components of the generalised Cartan connection, and to gain insights into the construction and consistency of generalised torsions and curvatures. As a consequence of the compatibility condition \eqref{eq:variation}, we conclude:
 \begin{itemize}
     \item The components of the connections $\theta^{(p)}$ for $p>1$ are fixed by the components of $\theta \equiv \theta^{(1)}$. Assuming $\mathcal{A}_1 = \di{\alpha_1 \dots \alpha_{p-1}}{A_p}$ for $p>1$, $\mathcal{B}_1 = \beta$ in \eqref{eq:variation} can be used to iteratively determine $\theta^{(p)}$. For example, for $\theta^{(2)}$ one finds the identity for $q<p$
            \begin{equation}
                {\delta\theta}^{(2)}_{\di{\alpha_1 \dots \alpha_{p-2} }{A_p}}\mathstrut^{\di{\mu_1 \dots \mu_{q-2} }{M_q}} =  (-1)^{(p-q)} \Omega_{A_p}^{M_q,\nu_1 \dots \nu_{p-q}} \delta^{\alpha_1 \dots \alpha_{p-2}}_{\mu_1 \dots \mu_{q-2} \nu_1 \dots \nu_{p-q}}.
            \end{equation} 
     
     \item All independent components of the Cartan connection are contained in the first column of \eqref{eq:BigVielbein},  which is formed by $\delta\theta_{\mathcal{A}_1}^\mu$. There are further restrictions which eventually only keep 
     \begin{equation}\label{eq:genConnections}
         \Omega_{M_1}^\mu, \ \rho_{M_2}^{\mu_1 \mu_2},\ \rho_{M_3}^{\mu_1 \mu_2 \mu_3}, \dots
     \end{equation}
     with
     \begin{align}
        \rho_{A_p}^{\beta \alpha_1 \dots \alpha_{p-1}} := \Omega_{A_p}^{[\beta,\alpha_1 \dots \alpha_{p-1}]} = \Omega_{A_p}^{\beta,\alpha_1 \dots \alpha_{p-1}} 
     \end{align}
     as independent components . To understand the derivation of this tensor structure, put $\mathcal{A}_1 = \di{\alpha_1 \dots \alpha_{p-1}}{A_p}$ for $p>1$, $\mathcal{B}_1 = \di{\beta}{B_2}$ in \eqref{eq:OrthConstraint2}, and project onto the $\mathcal{C}_2 = \di{\gamma_1 \dots \gamma_{q-2}}{C_q}$-component for $q>1$, then
        \begin{itemize}
            \item for $(p,q) = (2,2)$:
                \begin{equation}
                    \Omega_{A_2}^{\beta,\alpha} \delta_{B_2}^{C_2} + \Omega_{B_2}^{\alpha,\beta} \delta_{A_2}^{C_2} = {\eta^{D_4}}_{A_2 B_2} \Omega_{D_4}^{C_2 \alpha \beta}.
                \end{equation}
                After contraction with of $B_2$ and $C_2$, for example, one notices that
                \begin{equation}
                    \Omega_{A_2}^{(\alpha,\beta)} = 0, \qquad \Omega_{A_2}^{[\alpha,\beta]} \sim {\eta^{D_4}}_{A_2 B_2} \Omega_{D_4}^{B_2 \alpha \beta}
                \end{equation}
                showing that $\Omega_{A_2}^{\alpha,\beta}$ only has a skew-symmetric contribution. Moreover, we see that parts of the $\Omega_{D_4}^{A_2}$-component are fixed by this choice.
            \item for $p>q=2$ we have that:
                \begin{equation}
                    \Omega_{A_p}^{\beta,\alpha_1 \dots \alpha_{p-1}} \delta_{B_2}^{C_2} = (-1)^p \eta^{D_{p+2}}\mathstrut_{A_p B_2} \Omega_{D_{p+2}}^{C_2 , \alpha_1 \dots \alpha_{p-1} \beta}
                \end{equation}
            showing that $\Omega_{A_p}^{\beta,\alpha_1 \dots \alpha_{p-1}} = \Omega_{A_p}^{[\beta\alpha_1 \dots \alpha_{p-1}]} $.
        \end{itemize}
     
    \item Components in the remaining columns are connected to fundamental ones \eqref{eq:genConnections} in the first column via identities like
        \begin{align}
            \Omega_{D_{p+1}}^{C_p,\nu} \eta^{D_{p+1}}\mathstrut_{A_p B_1} &= \Omega_{A_p}^{D_{p-1}, \nu} \eta^{C_p}\mathstrut_{D_{p-1} B_1} - (-1)^p \Omega^\nu_{B_1} \delta^{C_p}_{A_p} \label{eq:ConnectionIdentity1} \\
             \Omega_{D_{p+1}}^{C_r, \nu_1 \dots \nu_{p-r+1}} {\eta^{D_{p+1}}}_{A_p B_1} &= \Omega_{A_p}^{D_{r-1},\nu_1 \dots \nu_{p-r+1}} {\eta^{C_r}}_{D_{r-1} B_1}, \quad 2 \leq r < p . \nonumber
        \end{align}
        This essentially determines the first sub-diagonal component of the Cartan connection, expressed in terms of $\Omega_{A_1}^\nu$.\footnote{From eq. \eqref{eq:OrthConstraint2}, these can be obtained by the choice $\mathcal{A}_1 = \di{\alpha_1 \dots \alpha_{p-1}}{A_p}$ for $p>1$ and $\mathcal{B}_1 = B_1$. Similar identities follow for other choices of $\mathcal{B}_1$.}
    \item These results suggest that the components of the generalised Cartan connection arise from the vector space $\widetilde{\mathfrak{R}}^*_0$ dual to the one defined in \eqref{eq:R0tilde}. We do not see the fact that $\mathcal{R}_0 = \mathfrak{g}$, where $\mathfrak{g}$ is the Lie algebra of the duality group, because we keep the physical frame fixed. Hence, from here on, we take the Cartan connection to be generated by $\widetilde{\mathfrak{R}}^*_0$. This is in agreement with the known results for the duality group O($d$,$d$) \cite{Polacek:2013nla,Hassler:2024hgq} and the alternative approach for exceptional duality groups presented in \cite{Hassler:2023axp}. 
    \item Non-linearity in the connection will \textit{not} impact the counting of the degrees of freedom of the generalised Cartan connection $\theta$. The only difference to what we have seen here is that contractions and $\bullet$-products of the independent components will appear.
\end{itemize}

\paragraph{Solving the constraints.} Still, we should check if the constraints \eqref{eq:ConnectionIdentity1} indeed admit solutions for all columns beyond the first. Taking a look at the first relation there, we find the solution
\begin{equation}
    \Omega_{L_{p+1}}^{K_p , \mu} = - (-1)^p D_{L_{p+1}}{}^{K_p M_1} \Omega_{M_1}^{\mu} . \label{eq:ConnectionIdentitySol}
\end{equation}
assuming that
\begin{equation}
    \left( D_{E_{p+1}}{}^{C_p B_1} \eta^{E_{p+1}}\mathstrut_{A_p D_1} + {\eta^{C_p}}_{E_{p-1} D_1} {D_{A_p}}^{E_{p-1} B_1} \right) \Omega_{B_1}^\nu = \Omega^\nu_{D_1} \delta^{C_p}_{A_p} \label{eq:ConnectionIdentity2}
\end{equation}
holds. Remarkably this relation is nothing else than the $\eta$-$D$ identity \eqref{eq:IdentityEtaD} where the partial derivative is replaced by $\Omega^\nu_{A_1}$. It constrains the $\mathcal{R}_1$ part of the generalised connection, while the $\mathfrak{h}$ factor is a mere spectator. But due to the $\mathcal{G}$-covariance of \eqref{eq:IdentityEtaD}, either this full $\mathcal{R}_1$ irreducible representation is compatible with it or nothing at all. Hence, we conclude that \eqref{eq:ConnectionIdentity2} holds generally.

Assume for a moment that $\rho_{M_p}^{\mu_1\dots\mu_p}=0$ and only $\Omega^\mu_{M_1}$ contributes to $\delta\theta_{\mathcal{A}_1}{}^{\mathcal{M}_1}$ through
\begin{equation}
    \delta\theta_{\mathcal{A}_1}{}^{\mathcal{M}_1} = \Omega^\nu_{B_1} (R_\nu^{B_1})_{\mathcal{A}_1}{}^{\mathcal{M}_1}
\end{equation}
with the generators $R_\nu^{B_1}$. As matrices, the latter only have entries on the first sub-diagonal, namely
\begin{align}
    (R_\nu^{B_1})_{A_1}{}^\mu &= \delta_\nu^\mu \delta_{A_1}^{B_1}\,, \qquad && \text{and} \nonumber\\
    (R_\nu^{B_1})_{A_{p+1}}^{\alpha_1\dots\alpha_p}{}^{M_p}_{\mu_1\dots\mu_{p-1}} &= \delta^{\alpha_1\dots\alpha_p}_{\mu_1\dots\mu_{p-1}\nu} D_{A_{p+1}}{}^{M_p B_1} & & \text{for } p\ge 1\,.
\end{align}
From this set of generators, one can derive the generators which have to be contracted with the various $\rho$'s we encountered. The first of them is
\begin{equation}
    R_{\nu_1\nu_2}^{B_2} := \eta^{B_2}{}_{C_1 D_1} [ R_{\nu_1}^{C_1}, R_{\nu_2}^{D_1} ]\,
\end{equation}
which can be derived from \eqref{eq:Generators}. It comes with only the second sub-diagonal populated by
\begin{align}
    (R_{\nu_1\nu_2}^{B_2})_{A_2}^{\alpha}{}^\mu &= \delta_{\nu_1\nu_2}^{\alpha\mu} \delta_{A_2}^{B_2}\,, \qquad & & \text{and} \nonumber \\
    (R_{\nu_1\nu_2}^{B_2})_{A_{p+2}}^{\alpha_1\dots\alpha_{p+1}}{}^{M_p}_{\mu_1\dots\mu_{p-1}} &= \delta^{\alpha_1 \dots \alpha_{p+1}}_{\mu_1 \dots \mu_{p-1} \nu_1 \nu_2} \eta^{B_2}{}_{C_1 D_1} {D_{A_{p+2}}}^{E_{p+1} C_1} D_{E_{p+1}}{}^{M_p D_1} & & \text{for } p\ge 1\,.
\end{align}
This pattern can be continued recursively to obtain the explicit form of the higher order generators:
\begin{equation}\label{eq:nestedComm}
    R^{B_p}_{\mu_1\dots\mu_p} = \eta^{B_p}{}_{C_{p-1} D_1} [ R^{C_{p-1}}_{[\mu_1\dots\mu_{p-1}}, R^{D_1}_{\mu_p]} ]
\end{equation}
where the normalisation factor is fixed by first row of the resulting matrix as
\begin{equation}
    (R^{B_p}_{\nu_1\dots\nu_p})_{A_p}^{\alpha_1 \dots \alpha_{p-1}}{}^\mu = \delta^{\alpha_1\dots\alpha_{p-1}\mu}_{\nu_1\dots\nu_p} \delta_{A_p}^{B_p} \,.
\end{equation}
As argued in section~\ref{sec:extGG} these generators form the Lie algebra $\widetilde{\mathfrak{R}}_0$ if complemented with the generators of $\mathcal{G}$. At this point, there is no guarantee that $\widetilde{\mathfrak{R}}_0$ is the maximal subalgebra, under the assumptions of the generalised Cartan connection, that gives rise to \eqref{eq:OrthConstraint2}.

Eventually, they find their use in writing
\begin{equation}\label{eq:deltathetafromGens}
    \boxed{\delta\theta = \Omega^\mu_{A_1} R^{A_1}_\mu + \sum_{p\ge 2} \frac{1}{p!} \, \rho^{\mu_1 \dots \mu_p}_{A_p} R^{A_p}_{\mu_1\dots\mu_p}\, . }
\end{equation}
To continue with the computation of curvatures in the next section, the component
\begin{equation}
    \Omega_{L_{p+1}}^{K_1,\mu_1 \dots \mu_p} = - (-1)^p D_{L_{p+1}}{}^{M_p K_1} \rho_{M_p}^{\mu_1 \dots \mu_p}
    \label{eq:ConnectionIdentitySol2}
\end{equation}is needed. It can be either obtained by getting the relevant components of the generators $R^{A_p}_{\mu_1\dots\mu_p}$ and combining them with \eqref{eq:deltathetafromGens} or by employing the conditions \eqref{eq:ConnectionIdentity1}. 

\subsection{A Hierarchy of Curvatures} \label{chap:CartanCurvature}
As the generalised Cartan curvature, we understand the structure functions $\Theta_{\mathcal{A}_1 \mathcal{B}_1}{}^{\mathcal{C}_1}$ in equation \eqref{eq:GeneralisedCartanCurvatureDefinition}. Here, we calculated these components using the brane current algebra. Alternatively, one could obtain the same result using the extended generalised Lie derivative \eqref{eq:hGgeneralisedDiff} -- the generalised Cartan connection $\theta_{\mathcal{A}_1}$ can be interpreted as a $\mathfrak{l}_1$-valued section of the $\bar{\mathfrak{R}}_1$-bundle. But remember that due to \eqref{eq:restrictedV}, the extended Lie bracket only closes for sections $ \theta_{\mathcal{A}_1} \in \mathfrak{h}\oplus\mathcal{R}_1$. However, as we will propose below, this is enough because the independent components of the curvature and torsion are completely contained in 
\begin{itemize}
    \item $\{\theta_\alpha(\sigma),\theta_{B_p}^{\beta_1 \dots \beta_{p-1}}(\sigma^\prime) \}$, which captures the model algebra, and
    \item $\{\theta_{A_1}(\sigma),\theta_{B_p}^{\beta_1 \dots \beta_{p-1}}(\sigma^\prime) \}$, which unifies the generalised torsions and curvatures of the physical space.
\end{itemize}
Let us have a look at the latter. Using the solution \eqref{eq:ConnectionIdentitySol}, its components can be written as
  \begin{align}
     &{} \quad \left\{ \theta_{A_1} (\sigma), \theta_{B_p}^{\beta_1 \dots \beta_{p-1}} (\sigma') \right\} \label{eq:CartanCurvCalculation} \\
     &= (-1)^p \left( \eta^{C_{p+1}}\mathstrut_{A_1 B_p} \theta_{C_{p+1}}^{\beta_1 \dots \beta_{p-1}} (\sigma^\prime) \wedge \mathrm{d}^\prime \delta(\sigma - \sigma^\prime) + f_{\alpha A_1}\mathstrut^{D_1} \eta^{C_{p+1}}\mathstrut_{D_1 B_p} \theta_{C_{p+1}}^{\beta_1 \dots \beta_{p-1} \alpha }(\sigma)  \delta(\sigma - \sigma^\prime) \right) \nonumber \\
     & \begin{aligned}
        &{} \quad + \frac{1}{(p-1)!} \Theta_{A_1 B_p}\mathstrut^{C_p}\mathstrut^{\beta_1 \dots \beta_{p-1}}_{\gamma_1 \dots \gamma_{p-1}}(\sigma)  \theta_{C_p}^{\gamma_1 \dots \gamma_{p-1}}(\sigma)  \delta (\sigma - \sigma') & & \text{($\mathcal{R}_p$-torsion)} \\
        &{} \quad + \frac{1}{(p-2)!} \Theta_{A_1 B_p}\mathstrut^{C_{p-1}}\mathstrut^{\beta_1 \dots \beta_{p-1}}_{\gamma_1 \dots \gamma_{p-2}}(\sigma)  \theta_{C_{p-1}}^{\gamma_1 \dots \gamma_{p-2}}(\sigma)  \delta (\sigma - \sigma') \qquad & & \text{($\mathcal{R}_1$-curvature)} \\
        &{} \qquad \vdots \\
        &{} \quad + \Theta_{A_1 B_p}\mathstrut^{C_1} \mathstrut^{\beta_1 \dots \beta_{p-1}}(\sigma)  \theta_{C_1} (\sigma)  \delta (\sigma - \sigma') & & \text{($\mathcal{R}_{p-1}$-curvature)} \\
        &{} \quad + \Theta_{A_1 B_p}^{\beta_1 \dots \beta_{p-1} \gamma} (\sigma) \theta_\gamma (\sigma) \delta (\sigma - \sigma')  & & \text{($\mathcal{R}_p$-curvature)}
     \end{aligned} \nonumber
\end{align}

From this expression, we extract several fundamental quantities. All of them can be expressed in terms of the independent components of the generalised Cartan connection. We take a generalised coordinate basis with the generalised frame ${E_{A_1}}^{M_1} = \delta_{A_1}^{M_1}$ to obtain
 \begin{itemize}
    \item $\mathcal{R}_p$-torsion:
        \begin{align}
             &{} \quad \Theta_{A_1 B_p}\mathstrut^{C_p}\mathstrut^{\beta_1 \dots \beta_{p-1}}_{\gamma_1 \dots \gamma_{p-1}}  \label{eq:RPTorsion} \\
            &{}= f_{\alpha B_p}\mathstrut^{C_p} \Omega_{A_1}^\alpha \delta^{\beta_1 \dots \beta_{p-1}}_{\gamma_1 \dots \gamma_{p-1}} + {f_{\alpha A_1}}^{D_1} \eta^{L_{p+1}}\mathstrut_{D_1 B_p} D_{L_{p+1}}\mathstrut^{C_p E_1} \Omega_{E_1}^\alpha \delta^{\beta_1 \dots \beta_{p-1}}_{\gamma_1 \dots \gamma_{p-1}} \nonumber \\
             &{} \quad - (p-1) f_{[\gamma_{p-1} A_1}\mathstrut^{D_1} \delta_{B_p}^{C_p} \Omega_{D_1}^\alpha \delta^{\beta_1 \dots \beta_{p-1}}_{\gamma_1 \dots \gamma_{p-2}] \alpha}. \nonumber
        \end{align}
        We appreciate that this result nicely generalises the standard expression for $p = 1$
        \begin{align*}
        	 	\Theta_{A_1 B_1}{}^{C_1} = - 2{f_{\alpha [A_1}}^{C_1} \Omega_{B_1]}^\alpha + {f_{\alpha A_1}}^{D_1} {\eta^{L_{2}}}_{D_1 B_1} {D_{L_{2}}}^{C_1 E_1} \Omega_{E_1}^\alpha
        \end{align*}
of the torsion in exceptional generalised geometry \cite{Cederwall:2013naa}. As in \cite{Hassler:2024hgq}, we naturally identify ${f_{\alpha B_1}}^{C_1} \Omega^\alpha_{A_1}$ with a connection ${\Gamma_{A_1 B_1}}^{C_1}$ on the generalised tangent bundle $\mathcal{R}_1[M]$. Therefore, we call it $\mathcal{R}_p$-torsion, but in the frame formalism one might also understand it as a curvature for the generalised frame.

    \item $\mathcal{R}_p$-curvature:
        \begin{align}
        p=1: &{} \quad \Theta_{A_1 B_1}\mathstrut^\beta = - 2 \partial_{[A_1} \Omega_{B_1]}^\beta + f_{\alpha A_1}\mathstrut^{C_1} \eta^{D_2}\mathstrut_{ C_1  B_1 } \rho^{\beta \alpha}_{D_2} , \label{eq:R1Curvature} \\
        p>1: &{} \quad \Theta_{A_1 B_p}^{\beta_1 \dots \beta_{p}} = - \partial_{A_1} \rho_{B_p}^{\beta_1 \dots \beta_p} + {f_{\alpha A_1}}^{C_1} \eta^{D_{p+1}}\mathstrut_{C_1 B_p} \rho_{D_{p+1}}^{\beta_1 \dots \beta_p \alpha} . \label{eq:RPCurvature}
        \end{align}
\end{itemize}
These formulas reduce to those of \cite{Hassler:2023axp} after linearisation. As suggested in \eqref{eq:CartanCurvCalculation}, we claim that the $\mathcal{R}_q$-curvatures for $q<p$ are also contained in the generalised Cartan curvature. Let us exemplify this by looking at the $\mathcal{R}_1$-curvature. The relevant component takes the form
\begin{equation}
    \Theta_{A_1 B_p}\mathstrut^{C_{p-1}} \mathstrut_{\gamma_1 \dots \gamma_{p-2}}^{\beta_1 \dots \beta_{{p-1}}} = - (-1)^p D_{B_p}\mathstrut^{C_{p-1} D_1} \Theta_{A_1 D_1}\mathstrut^\alpha \delta^{\beta_1 \dots \beta_{p-1}}_{\gamma_1 \dots \gamma_{p-2} \alpha}.
\end{equation}
Hence, under our assumptions many components of the generalised Cartan curvature are described by the $\mathcal{R}_1$-curvature.

Although we do not present a general proof that all components of the generalised Cartan curvature are fixed this way, it seems obvious that at least all dynamical contributions (which contain a derivative of the Cartan connection) are. Derivatives can only carry $\mathcal{R}_1$-indices. Moreover, we cover all independent contributions to the Cartan connection. 

As for the connection, these are only the \textit{linearised} expressions. At the non-linear level, all possible contractions of the spin connections $\Omega_{M_1}$, $\rho_{M_p}$ with $\eta$-symbols and structure constants $f_{\alpha \beta}\mathstrut^\gamma$, ${f_{\alpha M_p}}^{N_p}$ are expected to contribute. Nevertheless, already at the linear level, the key characteristic of these connections -- their hierarchical structure -- and their interplay with the constraints of the generalised Cartan connection become evident.

For the generalised Cartan connection, it was possible to organise all its component into the Lie algebra $\widetilde{\mathfrak{R}}$. One might ask if something along those lines works here, too. Looking at the representations which contribute to \eqref{eq:RPTorsion}, \eqref{eq:R1Curvature}, and \eqref{eq:RPCurvature}, it is suggestive to define the vector space
\begin{equation}
    \mathfrak{R}_{-1} = \mathcal{R}_{-1} \oplus \bigoplus_{q \geq 1} \Big( \mathcal{R}_1 \otimes \mathcal{R}_{q} \otimes \bigwedge\nolimits^{q} \mathfrak{h}^* \Big) \, ,
\end{equation}
with $\mathcal{R}_{-1}$ denoting the representation of the embedding tensor for the duality group $\mathcal{G}$. It hosts the $\mathcal{R}_p$-torsion and all the $\mathcal{R}_p$-curvatures. While the $\mathfrak{h}$ factors are very similar to the other $\mathfrak{R}_p$ spaces we have already encountered, the respective representations of the duality group $\mathcal{G}$ do not fit into the tensor hierarchy any more. A notable exception is the duality group O($d,d$) where one finds
\begin{equation}
    \mathfrak{R}_{-1}^{\mathrm{O}(d,d)} = \mathcal{R}_{-1} \oplus \bigoplus_{q \geq 1} \Big( \mathcal{R}_{q-1} \otimes \bigwedge\nolimits^{q} \mathfrak{h}^* \Big)\,.
\end{equation}

\paragraph{An ambiguity.} The form of the Cartan curvature, as presented in \eqref{eq:RPTorsion}, \eqref{eq:R1Curvature} and \eqref{eq:RPCurvature} is not unique. Given our definition via the current algebra \eqref{eq:CartanCurvCalculation}, additional redundant components of the curvature arise, which do not appear in the desired form. Take for example\footnote{This is defined as the coefficient in $\{\theta_{E_2}^\gamma(\sigma) , \theta_{C_1}(\sigma^\prime) \} = \dots + \Theta_{E_2 C_1}^{\gamma,\alpha} (\sigma) \theta_\alpha(\sigma) \delta(\sigma-\sigma^\prime).$}
\begin{equation}
  \Theta_{E_2 C_1}^{\gamma,\alpha} = - \partial_{C_1} \rho_{E_2}^{\gamma\alpha } - f_{\beta E_2}\mathstrut^{D_2} \eta^{H_3}\mathstrut_{D_2 C_1} \rho_{H_3}^{\alpha \gamma \beta} + \frac{1}{2} f_{\delta \epsilon}\mathstrut^{\gamma} \eta^{H_3}\mathstrut_{E_2 C_1} \rho_{H_3}^{\alpha \delta \epsilon}. \label{eq:RPcurvatureReverse}
\end{equation}
in the conventions of \eqref{eq:CurrentAlgebraExtR1}. It differs from the expected \eqref{eq:RPCurvature} by world-volume boundary terms or 'total'-derivative terms.\footnote{Because up to world-volume boundary terms, $\{\theta_{A_1}(\sigma),\theta_{B_p}^{\beta_1 \dots \beta_{p-1}} (\sigma^\prime) \} = - \{\theta_{B_p}^{\beta_1 \dots \beta_{p-1}} (\sigma^\prime), \theta_{A_1} (\sigma) \} $. } Up to a total sign, one can conveniently put both in the form
\begin{align}
        &{}\text{\small$p=1:  \quad \Theta_{A_1 B_1}\mathstrut^\beta = - 2 \partial_{[A_1} \Omega_{B_1]}^\beta + f_{\alpha A_1}\mathstrut^{C_1} \eta^{D_2}\mathstrut_{ C_1  B_1 } \rho^{\beta \alpha}_{D_2} + \alpha \ \eta^{C_2}{}_{A_1 B_1} \left( f_{\alpha C_2}{}^{D_2} \rho_{D_2}^{\beta\alpha} + \frac{1}{2} f_{\gamma \delta}{}^\beta \rho^{\gamma \delta}_{C_2} \right) ,$} \label{eq:R1CurvatureGen} \\
        &{}\text{\small$p>1: \quad \Theta_{A_1 B_p}^{\quad \beta_1 \dots \beta_{p-1},\beta} = - \partial_{A_1} \rho_{B_p}^{\beta_1 \dots \beta_{p-1}\beta} + {f_{\alpha A_1}}^{C_1} \eta^{D_{p+1}}\mathstrut_{C_1 B_p} \rho_{D_{p+1}}^{\beta_1 \dots \beta_{p-1}\beta\alpha}$} \label{eq:RPCurvatureGen}\\
        &{}\qquad \qquad \qquad \qquad \qquad \text{\small$ + \alpha \ \eta^{C_{p+1}}{}_{A_1 B_p} \left( f_{\alpha C_{p+1}}{}^{D_{p+1}} \rho_{D_{p+1}}^{\beta_1 \dots \beta_{p-1}\beta\alpha} + (-1)^{p+1} \frac{1}{2} f_{\gamma \delta}{}^\beta \rho^{\gamma \delta \beta_1 \dots \beta_{p-1}}_{C_{p+1}} \right) .$}\nonumber
    \end{align}
In general, any real number $\alpha$ should give a reasonable definition of a curvature. The form presented in \eqref{eq:R1Curvature} and \eqref{eq:RPCurvature} (for $\alpha = 0$) is chosen such that it reproduces the expressions from \cite{Hassler:2023axp} and that the curvatures fit nicely into $\mathfrak{R}_{-1}$ defined above. One interpretation of the difference is, that we could use \eqref{eq:RPCurvature} together with \eqref{eq:RPcurvatureReverse} (corresponding to $\alpha = -1$) as the structure 'constants' of a Leibniz algebroid which arises from a \textit{Dorfman} bracket. Alternatively one could take \eqref{eq:RPCurvatureGen} with $\alpha = - \frac{1}{2}$ to obtain a '\textit{Courant} bracket' analogue.

\paragraph{Bianchi identities.} In analogy to the differential and algebraic Bianchi identities for torsion and curvature in Riemannian geometry, we expect to derive consistency conditions relating the different components of $\Theta$. A complete collection of such identities is beyond the scope of this paper, we content ourselves with a sketch of two ways of how to approach Bianchi identities for the curvatures, and what types of relations one can expect. Based on the ambiguity mentioned above,  we have to decide whether we interpret the generalised Cartan curvature as structure coefficients of the Courant bracket, or the Dorfman bracket (resulting in a Leibniz algebra). In both cases, they are derived from the Jacobi identity of the current algebra. Let us demonstrate this with a few crucial examples. The calculations are similar to the ones presented in appendix \ref{App:Jacobi} for the 'flat' current algebra (i.e. the one that is not twisted by the generalised Cartan connection). An additional complication is that the Jacobi identity requires 'generalised Cartan curvature in higher representations', namely the structure constants in 
\begin{equation}
	\{ \theta_{\mathcal{A}_1}(\sigma) , \theta^{(2)}_{\mathcal{B}_2}(\sigma^\prime) \} \sim \ldots + \overset{{(2)}}{\Theta}_{\mathcal{A}_1 \mathcal{B}_2} {}^{\mathcal{C}_2} \theta^{(2)}_{\mathcal{C}_2} (\sigma) \delta(\sigma-\sigma^\prime)\,,
\end{equation}
where $\dots$ refer to the $\mathrm{d}\delta$-terms that are not included explicitly here. We will not present the technical details as they do not offer new insights but only additional complications in comparison to appendix \ref{App:Jacobi}. In principle, the non-trivial\footnote{Other parts give rise to the Jacobi identity of $\mathfrak{h}$ or the identity \eqref{eq:StructureConstantsRepresentation}.} part of the Jacobi identities can be decomposed into two parts: A Jacobi identity of the type 
\begin{equation}
    \{\theta_\alpha(\sigma_1), \{\theta_{A_1}(\sigma_2),\theta_{B_p} (\sigma_3) \} \} + c.p. = 0
\end{equation}
captures the fact that the curvatures are $\mathfrak{h}$-tensors. They transform covariantly as
	\begin{equation}
\nabla_\alpha T_{A_p \dots}^{\beta \dots} = {f_{\alpha A_p}}^{B_p} T_{B_p \dots}^\beta + {f_{\gamma \alpha}}^\beta T_{A_p \dots}^{\gamma \dots}.
	\end{equation} 
Moreover, this Jacobi identity also identifies components of $\overset{(2)}{\Theta}$ with the ones of $\overset{(1)}{\Theta}$
\begin{equation}
     (p-1) \delta_{\alpha}^{[\beta_{p-1}} \overset{(2)}{\Theta}{}_{A_1 B_p, \quad \gamma_1 \dots \gamma_{j-2} }^{\beta_1 \dots \beta_{p-2}], C_j} = \overset{(1)}{\Theta}{}_{A_1 B_p, \quad \gamma_1 \dots \gamma_{j-2} \alpha}^{\beta_1 \dots \beta_{p-1}, C_j},
\end{equation}
for $2 \leq j \leq p$. For example, for $j = p = 2$ we get \begin{equation}
    \delta^{\beta}_{\gamma} \, \overset{\hspace{-1.6em} (2)}{\Theta_{A_1 B_2}}\mathstrut^{C_2} = \overset{ (1)}{\Theta}{}_{A_1 B_2, \gamma}^{\quad \beta, C_2}.
\end{equation}
For the differential Bianchi identities, we obtain different versions, or interpretations.  Depending on $\alpha$ in \eqref{eq:R1CurvatureGen} and \eqref{eq:RPCurvatureGen}, we are dealing with
\begin{itemize}
    \item \textit{'Courant bracket'-type Bianchi identity}: Looking at the $\mathfrak{h}$-component of the Jacobi identity of type 
    \begin{equation}
        \{\theta_{A_1}(\sigma_1), \{\theta_{B_1}(\sigma_2),\theta_{C_p} (\sigma_3) \} \} + c.p. = 0
    \end{equation}
    for $p \geq 2$, one finds at the linearised level in connections the Bianchi identity
\begin{small}
\begin{align}
    0 =& \nabla_{[A_1} \Theta_{B_1 ] C_p}^{\gamma_1 \dots \gamma_{p-1}, \, \delta} + f_{\alpha [A_1| }\mathstrut^{F_1} 
    \left( 
    \Theta_{E_2 C_p}^{\ \alpha \ \gamma_1 \dots \gamma_{p-1}, \, \delta} \eta^{E_2}\mathstrut_{F_1 | B_1]} + \Theta_{| B_1 ] E_{p+1}}^{\quad \alpha \gamma_1 \dots \gamma_{p-1}, \, \delta} \eta^{E_{p+1}}\mathstrut_{F_1 C_p} 
    \right) \nonumber \\
    &+ f_{\alpha [A_1| }\mathstrut^{F_1} 
    \left(
    f_{\beta C_p}\mathstrut^{G_p} \eta^{E_{p+1}}\mathstrut_{F_1 G_p} \eta^{D_{p+2}}\mathstrut_{| B_1 ] E_{p+1}} + f_{\beta | B_1 ]}\mathstrut^{G_1} \eta^{E_{p+1}}\mathstrut_{F_1 C_p} \eta^{D_{p+2}}\mathstrut_{G_1 E_{p+1}} 
    \right)
    \rho^{\alpha \beta \gamma_1 \dots \gamma_{p-1} \delta}_{D_{p+2}} \label{eq:bigBianchi} \\
    &+ \left( \frac{p-1}{2} \right) f_{\alpha \beta}\mathstrut^{[\gamma_1 | } 
    \left( 
    f_{\epsilon [ A_1 |}\mathstrut^{F_1} \eta^{E_{p+1}}\mathstrut_{F_1 C_p} \eta^{D_{p+2}}\mathstrut_{| B_1 ] E_{p+1}} + \frac12 f_{\epsilon F_{p+2}}\mathstrut^{D_{p+2}} \eta^{F_{p+2}}\mathstrut_{C_p E_2} \eta^{E_2}\mathstrut_{A_1 B_1}
    \right) \rho^{\alpha \beta | \gamma_2 \dots \gamma_{p-1} ]  \delta \epsilon}_{D_{p+2}}. \nonumber
\end{align}
\end{small}
where $\nabla_{A_1} = \partial_{A_1} + \Omega_{A_1}\mathstrut^{\mu} \nabla_{\mu}$.\footnote{Of course, in the linearised framework the connection part does not contribute when acting on physical fields: $\nabla_{A_1} \Theta = \partial_{A_1} \Theta$.} Note that this relation contains naked connection components which are not part of the covariant derivative. They originate from derivatives of $\delta$-functions in the brane current algebra. For $p = 1$, \eqref{eq:bigBianchi} becomes
\begin{equation}
    \nabla_{[A_1} \Theta_{B_1 C_1]}\mathstrut^\gamma - 2 f_{\alpha [A_1 |}\mathstrut^{F_1} 
    \left( 
    \eta^{E_2}\mathstrut_{F_1 | B_1} \Theta_{C_1] E_2}^{\quad \alpha \gamma} + f_{\beta B_1 |}\mathstrut^{G_1} \eta^{E_2}\mathstrut_{F_1 G_1} \eta^{D_3}\mathstrut_{| C_1] E_2} \rho^{\alpha \beta \gamma}_{D_3}
    \right) = 0 ,
\end{equation}
and as expected, it can be checked that this identity is satisfied by the $\mathcal{R}_1$-curvature \eqref{eq:R1Curvature}.

Usually, naked connections are not expected in the consistency condition for the curvature. Therefore, our result here gives rise to a new interpretation of the tensor hierarchy of connections: the $\mathcal{R}_1$ connection is introduced to correct the non-covariance of the partial derivative. It has itself a curvature, which is non-covariant unless one introduces a higher $\mathcal{R}_2$ connection. Moreover, as shown in \cite{Hassler:2024hgq}, the $\mathcal{R}_2$ connection takes the role of a higher torsion. Now, the Bianchi identity of the $\mathcal{R}_1$ curvature is also corrected by the introduction of an $\mathcal{R}_3$ connection (which also corrects the curvature of the $\mathcal{R}_2$ connection). As a consistency check, let us note that the last term does not exists for $\mathcal{G} = $ O$(d,d)$ because its tensor hierarchy ends with $\mathcal{R}_2$.  Also in this case, there is no ambiguity in the definition of the generalised curvature and torsion because Courant and Dorfman brackets give the same results.

From this Jacobi identity we also have the following identity relating higher curvatures to lower ones (similar to the above case) for $1 \leq j \leq p $:
\begin{equation}
    \overset{(2)}{\Theta}{}_{A_1 E_{p+1}, \delta_{j-1}}^{\gamma_1 \dots \gamma_{p-1}, D_{j+1}} \eta^{E_{p+1}}\mathstrut_{B_1 C_p} = (-1)^{j} \, {\Theta}_{A_1 C_p, \delta_{j-1}}^{\gamma_1 \dots \gamma_{p-1}, E_j} \eta^{D_{j+1}}\mathstrut_{B_1 E_j}.
\end{equation}
There is a plethora of such Bianchi identities for the other components of the Jacobi identity. We will not consider them here explicitly for the sake of brevity -- typically, they will correspond to algebraic identities, relating different components of the generalised Cartan curvature.

    \item \textit{'Dorfman bracket'-type Bianchi identity.} In contrast to the above construction that only refers to curvatures of the form $\Theta_{A_1 B_p}^{\gamma_1 \dots \gamma_p}$, one can derive an identity that resembles more the Leibniz identity of a Dorfman bracket
\begin{equation}
    2 \partial_{[A_1} \Theta_{B_1 ] C_1}\mathstrut^\gamma + \partial_{C_1} \Theta_{A_1 B_1}\mathstrut^\gamma + 2 f_{\alpha [A_1}\mathstrut^{F_1} \Theta_{B_1] E_2}^{\quad \alpha \gamma} \eta^{E_2}\mathstrut_{F_1 C_1} - f_{\alpha A_1}\mathstrut^{F_1} \Theta_{E_2 C_1}^{\gamma,\alpha} \eta^{E_2}\mathstrut_{F_1 B_1} = 0,
\end{equation}
where $\Theta_{E_2 C_1}^{\gamma,\alpha}$ is defined in \eqref{eq:RPcurvatureReverse}. It can be easily verified by substituting the known curvatures and using the relations \eqref{eq:IdentityEtaEta} and \eqref{eq:StructureConstantsRepresentation}. An advantage of this formulation in comparison to the 'Courant bracket'-type is that no naked connections appear in the Bianchi identity. On the hand, one needs to introduce a curvature \eqref{eq:RPcurvatureReverse} in addition to \eqref{eq:RPCurvature}. Similarly, one can get the Bianchi identity for the $\mathcal{R}_2$-curvature:
\begin{equation}
    2 \partial_{[A_1} \Theta_{B_1 ] C_2}^{\gamma \delta} + 2 f_{\alpha [A_1}\mathstrut^{F_1} \Theta_{B_1] E_3}^{\gamma \delta \alpha} \eta^{E_3}\mathstrut_{F_1 C_2} - f_{\alpha A_1}\mathstrut^{F_1} \Theta_{E_2 C_2}^{ \gamma ,\delta , \alpha} \eta^{E_2}\mathstrut_{F_1 B_1} = 0 \, ,
\end{equation}
with the algebraic curvature\footnote{Meaning that it does not contain derivatives at the linearised level.} $\Theta_{E_2 C_2}^{ \gamma , \delta , \alpha} = - f_{\beta E_2}\mathstrut^{D_2} \eta^{H_4}\mathstrut_{D_2 C_2} \rho_{H_4}^{\alpha \gamma \delta \beta} - \frac{1}{2}
\eta^{H_4}\mathstrut_{E_2 C_2} f_{\rho \sigma}\mathstrut^{\gamma}  \rho_{H_4}^{ \delta \rho \sigma \alpha}$.

The Bianchi identity for the general $\mathcal{R}_p$-curvature for $p \geq 2$ is
\begin{equation}
    2 \partial_{[A_1} \Theta_{B_1 ] C_p}^{\gamma_1 \dots \gamma_p} + 2 f_{\alpha [A_1}\mathstrut^{F_1} \Theta_{B_1] E_{p+1}}^{\: \gamma_1 \dots \gamma_p \alpha} \eta^{E_{p+1}}\mathstrut_{F_1 C_p} - f_{\alpha A_1}\mathstrut^{F_1} \Theta_{E_2 C_p}^{ \quad \gamma_1 \dots \gamma_p , \alpha} \eta^{E_2}\mathstrut_{F_1 B_1} = 0 
\end{equation}
with the curvature
\begin{equation}
    \Theta_{E_2 C_p}^{ \epsilon, \gamma_1 \dots \gamma_{p-1} , \alpha} = - f_{\beta E_2}\mathstrut^{D_2} \eta^{H_{p+2}}\mathstrut_{D_2 C_p} \rho_{H_{p+2}}^{\epsilon \gamma_1 \dots \gamma_{p-1} \alpha \beta} - (-1)^p \frac12 f_{\rho \sigma}\mathstrut^\epsilon \eta^{H_{p+2}}\mathstrut_{E_2 C_p} \rho_{H_{p+2}}^{\rho \sigma \gamma_1 \dots \gamma_{p-1} \alpha},
\end{equation}
which can be obtained from the current algebra.
\end{itemize}

\section{Outlook}\label{chap:Outlook}
In this article, we constructed a \textit{minimal setting} for gauged exceptional geometry, i.e. an extended geometry that simultaneously geometrises the action of generalised diffeomorphisms (and their associated tensor hierarchy) and an underlying compatible gauge symmetry. Building on this and generalising previous work on extensions of Cartan geometry \cite{Polacek:2013nla,Hassler:2022egz,Hassler:2023axp,Hassler:2024hgq}, we proposed a setting for computing curvatures that are covariant under both symmetries. As a proof of concept, we derived in section \ref{chap:CartanCurvature} expressions for a hierarchy of torsion and curvatures at the linearised level. In particular, we derived the $\mathcal{R}_p$-curvatures \eqref{eq:R1Curvature}, \eqref{eq:RPCurvature}, and the $\mathcal{R}_p$-torsion \eqref{eq:RPTorsion}, which are the only independent components which contribute to the generalised Cartan curvature. While the linearised analysis is sufficient in order to obtain insights into the hierarchical structure, with remarkable new features like Dorfman or Courant-like curvatures and Bianchi identities, it would be desirable to eventually find the full non-linear expressions for the curvatures.

Extending our analysis further will eventually require a complete collection of explicit expressions of the tensor hierarchy -- in the language of this article: all $\eta$- and $D$-symbols for all representations $\mathcal{R}_p$. Moreover, the present construction prominently uses the language of differential graded Lie algebras and thereby suggests an index-free version of generalised Cartan geometry which is more in line with the mathematical treatment of Cartan geometry in the language of differential forms. Since differential graded Lie algebras are strict $L_{\infty}$ algebras, a natural extension of our framework would be to allow for generic $L_{\infty}$ algebras with non-trivial higher brackets \cite{Lada:1992wc,Hohm:2017pnh,Borsten:2021ljb}. From supergravity it is known that in "realistic" tensor hierarchies of $E_{d(d)}$, the graded symmetry structure breaks down at high degree. This constitutes one of the major challenges of the \textit{maximal} approach proposed in \cite{Hassler:2023axp}, where the symmetries $H\times\mathcal{G}$ are embedded into an exceptional group of suitable dimension. In practice, this puts severe restrictions on the dimension of the gauge group $H$. Our construction circumvents this problem, albeit at the price of losing the ability to capture generalised U-dualities as explored in \cite{YuhoFalkNew}. On the other hand, as we are now able to deal with gauge groups $H$ of arbitrary dimension, it is possible to address two further questions that are fundamental from a geometric point of view.

Firstly, we focussed solely on the construction of covariant curvatures. But in a physical setting requiring metric-compatibility of the connections (here in particular, what was called the $\mathcal{R}_1$-connection $\Omega_{A_1}^\mu$) is central in order to identify the propagating degrees of freedom of the theory. As is well-known already in O$(d,d)$, vanishing torsion and compatibility with a generalised metric (a symmetric $\bar{\mathcal{R}}_1 \times \bar{\mathcal{R}}_1$-tensor) does not uniquely fix the generalised affine connection \cite{Hohm:2012mf}. The same problem still exists for the generalised Cartan connection. One potential solution is to introduce a new gauge symmetry which shifts the non-fixed components as was suggested in \cite{Butter:2021dtu}. But on its own, this new symmetry does not close; it must be extended, which -- at least for O$(d,d)$ --- leads to a graded, infinite-dimensional symmetry group $H$. Performing a similar construction in M-theory will only be possible with the tools developed here. Moreover, infinite-dimensional gauge groups $H$ are central in the construction of higher-derivative corrections though the generalised Bergshoeff-de Roo mechanism \cite{Baron:2018lve,Baron:2020xel} as has been shown recently in \cite{Gitsis:2024gfb}. Secondly, our hierarchy of connections and curvatures strongly resembles the structures of higher gauge theory \cite{Baez:2010ya,Grutzmann:2014hkn,Ritter:2015zur,Borsten:2021ljb,Borsten:2024gox}. Phrasing our setting in such a language seems feasible and desirable. In particular, our construction might hint at a gravitational analogue of higher gauge theory, consistent with the success of Cartan geometry in formulating gravity as a gauge theory.

Finally, the present construction made essential use of brane current algebras \cite{Alekseev:2004np,Bonelli:2005ti,Hatsuda:2012uk,Hatsuda:2012vm,Hatsuda:2013dya,Hatsuda:2020buq,Osten:2021fil,Arvanitakis:2021wkt,Hatsuda:2023dwx,Osten:2024mjt}, primarily as a computational tool. As suggested in \cite{Osten:2024mjt}, the gauged version of exceptional geometry developed here may find applications in the description of exotic branes (such as the Kaluza-Klein monopole) where one or more transverse directions carry gauge symmetries. First results in that direction were already proposed in \cite{Sakatani:2017vbd}. Potentially, the world-volume theory of such exotic branes could be constructed from first principles using gauged exceptional geometry and the construction outlined in \cite{Osten:2024mjt}. This would not only provide a systematic approach to classify and analyse their dynamics, but also offer a setting in which dualities (U-dualities and their generalisations \cite{Sakatani:2019zrs,Malek:2019xrf}) can be studied explicitly.

\section*{Acknowledgements}
We thank Martin Cederwall, Axel Kleinschmidt and Yuho Sakatani for discussions, and in particular Yuho Sakatani for sharing unpublished results. F.~H.~and A.~S.~are supported by the SONATA BIS grant 2021/42/E/ST2/00304 from the National Science Centre (NCN), Poland. The research of D.~O.~was part of the POLONEZ BIS project No.\linebreak 2022/45/P/ST2/03995 co-funded by NCN and the European Union’s Horizon 2020 research and innovation programme under the Marie Sk\l odowska-Curie grant agreement no. 945339, and the SONATA grant No. 2024/55/D/ST2/01205 funded by NCN.

\vspace{10pt}
\includegraphics[width = 0.09 \textwidth]{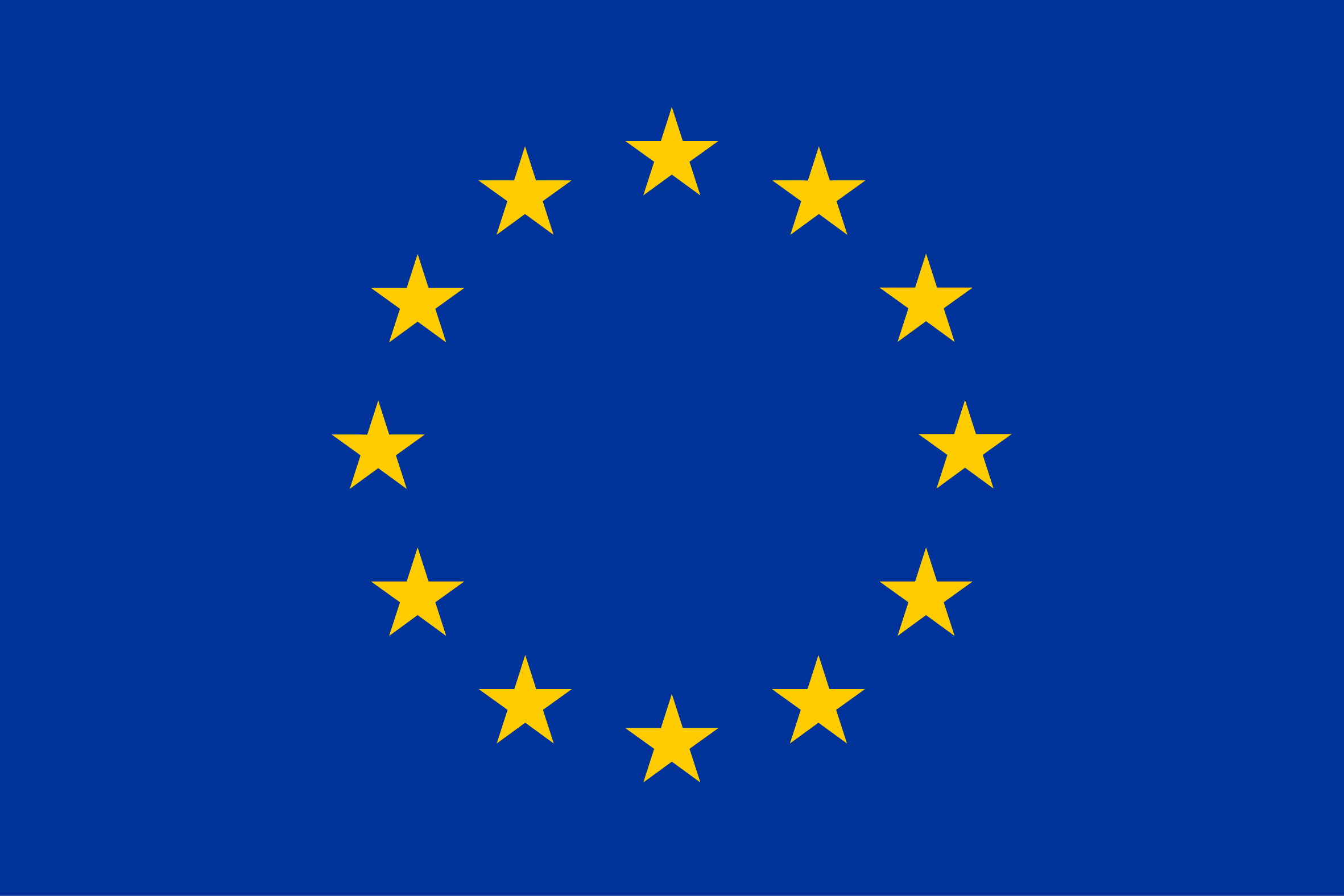} $\quad$
\includegraphics[width = 0.7 \textwidth]{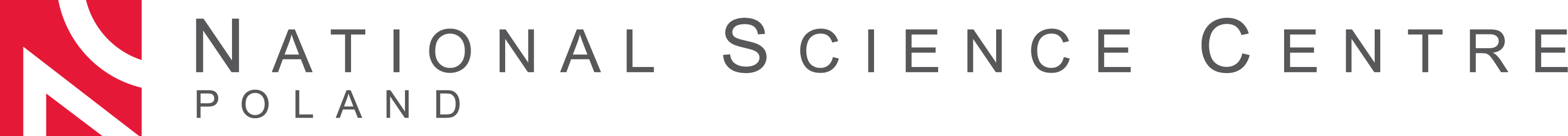}

\appendix

\section{Closure of Extended Generalised Lie derivative}
\label{sec:ClosureExtGenLd}
Assuming closure of the ordinary generalised Lie derivative $\mathcal{L}_{U_1}$, we only need to check the extra terms involving the $D$-symbols and the structure constants $f_{\alpha \beta}\mathstrut^{\gamma}$ in \eqref{eq:hGgeneralisedDiff}. Looking only at the $M_1$ component of the LHS of \eqref{eq:ClosureGenDiffeos}, we get
{\small \begin{align*}
    \left( \left[ \mathcal{L}_{\mathcal{U}} , \mathcal{L}_{\mathcal{V}} \right] \mathcal{W} \right)^{M_1} &= \mathcal{L}_{U_1} \left( D_{L_2}\mathstrut^{M_1 N_1} \partial_{N_1} v^\nu (W_2)^{L_2}_\nu
    \right) - \mathcal{L}_{V_1} \left( D_{L_2}\mathstrut^{M_1 N_1} \partial_{N_1} u^\nu (W_2)^{L_2}_\nu
    \right) 
    \\ &+ D_{L_2}\mathstrut^{M_1 N_1} \partial_{N_1} u^\mu \left(
    \mathcal{L}_{V_1} (W_2)^{L_2}_\mu + 2 D_{K_3}\mathstrut^{L_2 P_1} \partial_{P_1} v^\nu (W_3)_{\mu \nu}^{K_3} + f_{\mu \nu}\mathstrut^{\sigma} v^\nu (W_2)_{\sigma}^{L_2} 
    \right) \\
    &- D_{L_2}\mathstrut^{M_1 N_1} \partial_{N_1} v^\mu \left(
    \mathcal{L}_{U_1} (W_2)^{L_2}_\mu + 2 D_{K_3}\mathstrut^{L_2 P_1} \partial_{P_1} u^\nu (W_3)_{\mu \nu}^{K_3} + f_{\mu \nu}\mathstrut^{\sigma} u^\nu (W_2)_{\sigma}^{L_2} 
    \right) \\
    &= D_{L_2}\mathstrut^{M_1 N_1} (W_2)^{L_2}_\mu \left( U_1^{P_1} \partial_{P_1} \partial_{N_1} v^\mu  -  \partial_{P_1} \partial_{N_1} u^\mu V_1^{P_1} + f_{\nu \sigma}\mathstrut^\mu \partial_{N_1} u^\nu v^\sigma - f_{\nu \sigma}\mathstrut^\mu u^\sigma \partial_{N_1} v^\nu
    \right) \\
    &+ \left( Y^{M_1 P_1}\mathstrut_{R_1 S_1} D_{L_2}\mathstrut^{S_1 N_1} - Y^{K_2 P_1}\mathstrut_{L_2 R_1} D_{K_2}\mathstrut^{M_1 N_1}
    \right) \left( \partial_{P_1} U_1^{R_1} \partial_{N_1} v^\mu - \partial_{N_1} u^\mu \partial_{P_1} V_1^{R_1}
    \right) (W_2)^{L_2}_\mu \\
    &+ 2 D_{K_3}\mathstrut^{L_2 P_1} D_{L_2}\mathstrut^{M_1 N_1} \left( \partial_{N_1} u^\mu \partial_{P_1} v^\nu (W_3)^{K_3}_{\mu \nu} - \partial_{P_1} u^\nu \partial_{N_1} v^\mu (W_3)^{K_3}_{\mu \nu}
    \right).
\end{align*}}
Going from the first to the second equality, we expanded the generalised Lie derivatives and grouped similar terms.
Using the skew-symmetry of $(W_3)^{K_3}_{\mu \nu}$, we observe that the last line combines to give 
\begin{equation*}
    4 D_{K_3}\mathstrut^{L_2 P_1} D_{L_2}\mathstrut^{M_1 N_1} \partial_{(N_1} u^\mu \partial_{P_1)} v^\nu (W_3)^{K_3}_{\mu \nu} = 0,
\end{equation*}
which vanishes thanks to the property \eqref{eq:IdentityDD}. The terms involving the $Y$ tensor simplify using \eqref{eq:IdentityEtaD} and \eqref{eq:IdentityYD} to
\begin{align*}
    & \left( Y^{M_1 P_1}\mathstrut_{R_1 S_1} D_{L_2}\mathstrut^{S_1 N_1} - Y^{K_2 P_1}\mathstrut_{L_2 R_1} D_{K_2}\mathstrut^{M_1 N_1}
    \right) \left( \partial_{P_1} U_1^{R_1} \partial_{N_1} v^\mu - \partial_{N_1} u^\mu \partial_{P_1} V_1^{R_1}
    \right) (W_2)^{L_2}_\mu \\
    \mathrel{\overset{\makebox[0pt]{\mbox{\normalfont\tiny\sffamily \eqref{eq:IdentityEtaD}}}}{=}}& \left( Y^{M_1 P_1}\mathstrut_{R_1 S_1} D_{L_2}\mathstrut^{S_1 N_1} + Y^{M_1 N_1}\mathstrut_{R_1 S_1} D_{L_2}\mathstrut^{S_1 P_1}  
    \right) \left( \partial_{P_1} U_1^{R_1} \partial_{N_1} v^\mu - \partial_{N_1} u^\mu \partial_{P_1} V_1^{R_1}
    \right) (W_2)^{L_2}_\mu \\
    &- D_{L_2}\mathstrut^{M_1 N_1} \left( \partial_{R_1} U_1^{R_1} \partial_{N_1} v^\mu - \partial_{N_1} u^\mu \partial_{R_1} V_1^{R_1} \right) (W_2)^{L_2}_\mu \\
    \mathrel{\overset{\makebox[0pt]{\mbox{\normalfont\tiny\sffamily \eqref{eq:IdentityYD}}}}{=}}& \; D_{L_2}\mathstrut^{M_1 N_1} \left( 2 \partial_{( R_1} U_1^{R_1} \partial_{N_1 )} v^\mu - 2 \partial_{( N_1} u^\mu \partial_{R_1 )} V_1^{R_1} - \partial_{R_1} U_1^{R_1} \partial_{N_1} v^\mu + \partial_{N_1} u^\mu \partial_{R_1} V_1^{R_1} 
    \right) (W_2)^{L_2}_\mu \\
    =& \; D_{L_2}\mathstrut^{M_1 N_1} \left( \partial_{N_1} U_1^{R_1} \partial_{R_1} v^\mu - \partial_{R_1} u^\mu \partial_{N_1} V_1^{R_1}
    \right) (W_2)^{L_2}_\mu .
\end{align*}
This leaves us with the following terms,
{\small\begin{align*}
    \left( \left[ \mathcal{L}_{\mathcal{U}} , \mathcal{L}_{\mathcal{V}} \right] \mathcal{W} \right)^{M_1} &= D_{L_2}\mathstrut^{M_1 N_1} (W_2)^{L_2}_\mu \left( U_1^{P_1} \partial_{P_1} \partial_{N_1} v^\mu  -  \partial_{P_1} \partial_{N_1} u^\mu V_1^{P_1} + f_{\nu \sigma}\mathstrut^\mu \partial_{N_1} u^\nu v^\sigma + f_{\nu \sigma}\mathstrut^\mu u^\nu \partial_{N_1} v^\sigma
    \right. \\
    & \left. + \partial_{N_1} U_1^{P_1} \partial_{P_1} v^\mu - \partial_{P_1} u^\mu \partial_{N_1} V_1^{P_1}
    \right),
\end{align*}}
which is exactly the $M_1$ component of the RHS of \eqref{eq:ClosureGenDiffeos}, namely
{\small\begin{align*}
    \left( \mathcal{L}_{[u + U_1, v + V_1]_C} \mathcal{W} \right)^{M_1} &= D_{L_2}\mathstrut^{M_1 N_1} \partial_{N_1} \left( [u,v]_{\mathfrak{h}}^\mu + \mathcal{L}_{U_1} v^\mu - \mathcal{L}_{V_1} u^\mu \right) (W_2)^{L_2}_\mu \\
    &= D_{L_2}\mathstrut^{M_1 N_1} (W_2)^{L_2}_\mu \left( f_{\nu \sigma}\mathstrut^\mu \partial_{N_1} u^\nu v^\sigma + f_{\nu \sigma}\mathstrut^\mu u^\nu \partial_{N_1} v^\sigma + \partial_{N_1} U_1^{P_1} \partial_{P_1} v^\mu + U_1^{P_1} \partial_{P_1} \partial_{N_1} v^\mu
    \right. \\
    & \left.  - \partial_{P_1} u^\mu \partial_{N_1} V_1^{P_1} -  \partial_{P_1} \partial_{N_1} u^\mu V_1^{P_1}
    \right),
\end{align*}}
and matches the terms above from the LHS, establishing closure.

One can with a bit more work also show closure for all components of $\mathfrak{R}_1$. The LHS is something like,
{\small\begin{align*}
    \left( \left[ \mathcal{L}_{\mathcal{U}} , \mathcal{L}_{\mathcal{V}} \right] \mathcal{W} \right)^{M_p}_{\mu_1 \dots \mu_{p-1}} &= \mathcal{L}_{U_1} \left(
    \mathcal{L}_{V_1} ( W_p )^{M_p}_{\mu_1 \dots \mu_{p-1}} + p \, D_{L_{p+1}}\mathstrut^{M_p N_1} \partial_{N_1} v^\nu  ( W_{p+1} )^{L_{p+1}}_{\mu_1 \dots \mu_{p-1} \nu} \right. \\
    &\left. - (p-1) v^\nu f_{\nu [ \mu_{p-1} }\mathstrut^\lambda ( W_p )^{M_p}_{\mu_1 \dots \mu_{p-2} ] \lambda } 
    \right) \\
    &+ p \, D_{L_{p+1}}\mathstrut^{M_p P_1} \partial_{P_1} u^\rho \left( \mathcal{L}_{V_1} ( W_{p+1} )^{L_{p+1}}_{\mu_1 \dots \mu_{p-1} \rho} + (p+1) D_{Q_{p+2}}\mathstrut^{L_{p+1} N_1} \partial_{N_1} v^\nu ( W_{p+2} )^{Q_{p+2}}_{\mu_1 \dots \mu_{p-1} \rho \nu} \right. \\
    &\left. - p v^\nu f_{\nu [ \rho}\mathstrut^\lambda ( W_{p+1} )^{L_{p+1}}_{\mu_1 \dots \mu_{p-1} ] \lambda}   
    \right) \\ 
    &- (p-1) \, u^\rho f_{\rho [ \mu_{p-1}}\mathstrut^\lambda \left( 
    \mathcal{L}_{V_1} (W_p)^{M_p}_{\mu_1 \dots \mu_{p-2}] \lambda} + p \, D_{L_{p+1}}\mathstrut^{M_p N_1} \partial_{N_1} v^\nu ( W_{p+1} )^{L_{p+1}}_{\mu_1 \dots \mu_{p-2} ] \lambda \nu } \right. \\
    &\left. - (p-1) v^\nu f_{| \nu \lambda }\mathstrut^\sigma ( W_p )^{M_p}_{ | \mu_1 \dots \mu_{p-2}] \sigma} 
    \right) - ( (u, U_1) \leftrightarrow (v, V_1) ).
\end{align*}}
One can immediately see that the term involving $W_{p+2}$ cancels out, thanks to the nilpotency of the $D$-symbols \eqref{eq:IdentityDD}.
The very first term above already closes, by assumption. Expanding the remaining terms, this becomes
\begin{align*}
    \left( \left[ \mathcal{L}_{\mathcal{U}} , \mathcal{L}_{\mathcal{V}} \right] \mathcal{W} \right)^{M_p}_{\mu_1 \dots \mu_{p-1}} &= p \, D_{L_{p+1}}\mathstrut^{M_p N_1} \left[ \left( U_1 \partial_{P_1} \partial_{N_1} v^\nu - \partial_{P_1} \partial_{N_1} u^\nu V_1^{P_1} 
    \right) (W_{p+1})^{L_{p+1}}_{\mu_1 \dots \mu_{p-1} \nu} \right. \\
    &\left. + \left( U_1^{P_1} \partial_{N_1} v^\nu - \partial_{N_1} u^\nu V_1^{P_1} \right) \partial_{P_1} (W_{p+1})^{L_{p+1}}_{\mu_1 \dots \mu_{p-1} \nu} 
    \right] \\
    &+ p \, D_{L_{p+1}}\mathstrut^{P_p N_1} Y^{M_p L_1}\mathstrut_{P_p Q_1} \left( \partial_{L_1} U_1^{Q_1} \partial_{N_1} v^\nu - \partial_{N_1} u^\nu \partial_{L_1} V_1^{Q_1} \right) ( W_{p+1} )^{L_{p+1}}_{\mu_1 \dots \mu_{p-1} \nu}  \\
    &- (p-1)  f_{\nu[ \mu_{p-1} }\mathstrut^\lambda (W_p)^{M_p}_{\mu_1 \dots \mu_{p-2}] \lambda} \left( U_1^{P_1} \partial_{P_1} v^\nu - \partial_{P_1} u^\nu V_1^{P_1} \right) \\
    &- (p-1) f_{\nu[ \mu_{p-1} }\mathstrut^\lambda \partial_{P_1} (W_p)^{M_p}_{\mu_1 \dots \mu_{p-2}] \lambda} \left( U_1^{P_1} v^\nu - u^\nu V_1^{P_1} \right) \\
    &- (p-1) Y^{M_p N_1}\mathstrut_{P_p Q_1} f_{\nu[ \mu_{p-1} }\mathstrut^\lambda (W_p)^{M_p}_{\mu_1 \dots \mu_{p-2}] \lambda} \left( \partial_{N_1} U_1^{Q_1} v^\nu - u^\nu \partial_{N_1} V_1^{Q_1} \right) \\
    &+ p \, D_{L_{p+1}}\mathstrut^{M_p P_1} \left[ \left( \partial_{N_1} u^\nu V_1^{P_1} - U_1^{P_1} \partial_{N_1} v^\nu \right) \partial_{P_1} (W_{p+1})^{L_{p+1}}_{\mu_1 \dots \mu_{p-1} \nu} \right. \\
    &\left. + Y^{L_{p+1} P_1}\mathstrut_{R_{p+1} Q_1} \left( \partial_{N_1} u^\nu \partial_{P_1} V_1^{Q_1} - \partial_{P_1} U_1^{Q_1} \partial_{N_1} v^\nu \right) (W_{p+1})^{R_{p+1}}_{\mu_1 \dots \mu_{p-1} \nu} \right. \\
    &\left. - p f_{\nu[ \rho }\mathstrut^\lambda (W_{p+1})^{L_{p+1}}_{\mu_1 \dots \mu_{p-1}] \lambda} \left( \partial_{N_1} u^\rho v^\nu - u^\nu \partial_{N_1} v^\rho \right)
    \right] \\
    &- (p-1) f_{\nu [ \mu_{p-1} }\mathstrut^\lambda \partial_{N_1} (W_p)^{M_p}_{\mu_1 \dots \mu_{p-2} ] \lambda} \left( u^\nu V_1^{N_1} - U_1^{N_1} v^\nu \right) \\
    &- (p-1) Y^{M_p N_1}\mathstrut_{L_p P_1} \left( u^\rho \partial_{N_1} V_1^{P_1} - \partial_{N_1} U_1^{P_1} v^\rho \right) (W_p)^{L_p}_{\mu_1 \dots \mu_{p-2}] \lambda} \\
    &- p(p-1) D_{L_{p+1}}\mathstrut^{M_p N_1} f_{\rho [ \mu_{p-1} }\mathstrut^\lambda (W_{p+1})^{L_{p+1}}_{\mu_1 \dots \mu_{p-2} ] \lambda \nu} \left( u^\rho \partial_{N_1} V_1^{P_1} - \partial_{N_1} U_1^{P_1} v^\rho \right) \\
    &+ (p-1)^2 ( u^\rho v^\nu - u^\nu v^\rho ) f_{\rho [ \mu_{p-1} | }\mathstrut^\gamma f_{\nu \gamma}\mathstrut^\lambda (W_p)^{M_p}_{| \mu_1 \dots \mu_{p-2} ] \lambda}.
\end{align*}
This should be equal to the RHS, which evaluates to
{\small\begin{align*}
    \big( \mathcal{L}_{[u + U_1, v + V_1]_C} \mathcal{W} \big)^{M_p}_{\mu_1 \dots \mu_{p-1}} &= \mathcal{L}_{[U_1, V_1]_C} (W_p)^{M_p}_{\mu_1 \dots \mu_{p-1}} \\ &+ p \, D_{L_{p+1}}\mathstrut^{M_p N_1} \partial_{N_1} \left( 
    [u,v]_{\mathfrak{h}}^\nu + \mathcal{L}_{U_1} v^\nu - \mathcal{L}_{V_1} u^\nu
    \right) ( W_{p+1} )^{L_{p+1}}_{\mu_1 \dots \mu_{p-1} \nu} \\
    &- (p-1) \left( 
    [u,v]_{\mathfrak{h}}^\nu + \mathcal{L}_{U_1} v^\nu - \mathcal{L}_{V_1} u^\nu
    \right) f_{\nu [ \mu_{p-1}}\mathstrut^\lambda ( W_p )^{M_p}_{\mu_1 \dots \mu_{p-2}] \lambda} \\
    &= p \, D_{L_{p+1}}\mathstrut^{M_p N_1} ( W_{p+1} )^{L_{p+1}}_{\mu_1 \dots \mu_{p-1} \nu} \left( 
    f_{\alpha \beta}\mathstrut^{\nu} ( \partial_{N_1} u^\alpha v^\beta + u^\alpha \partial_{N_1} v^\beta ) + U_1^{P_1} \partial_{N_1} \partial_{P_1} v^\nu \right. \\
    &\left. - \partial_{N_1} \partial_{P_1} u^\nu V_1^{P_1} + \partial_{N_1} U_1^{P_1} \partial_{P_1} v^\nu - \partial_{P_1} u^\nu \partial_{N_1} V_1^{P_1} \right) \\
    &- (p-1) \left( U_1^{P_1} \partial_{P_1} v^\nu - \partial_{P_1} u^\nu V_1^{P_1} \right) f_{\nu [ \mu_{p-1}}\mathstrut^\lambda (W_p)^{M_p}_{\mu_1 \dots \mu_{p-2} ] \lambda} \\
    & - (p-1) f_{\alpha \beta}\mathstrut^\nu f_{\nu [ \mu_{p-1}}\mathstrut^\lambda u^\alpha v^\beta (W_p)^{M_p}_{\mu_1 \dots \mu_{p-2} ] \lambda}.
\end{align*}}
When comparing this to the LHS, we observe that all the terms simplify thanks to the identities \eqref{eq:IdentityDD}, \eqref{eq:IdentityEtaD}, \eqref{eq:Identityfeta}, \eqref{eq:IdentityfD} and the Jacobi identity for $\mathfrak{h}$, establishing closure in much the same way as for the $\mathcal{R}_1$ component.

\section{Properties of the Extended Current Algebra}\label{sec:PropCurrentAlgebra}
\subsection{Derivation from underlying Poisson algebra} \label{App:Derivation}
The fact that the above current algebra is indeed a Poisson algebra can be checked explicitly (below). But it can be understood from a relation to an underlying Poisson algebra:
\begin{align}
    \{ s_\mu(\sigma), y^\nu(\sigma ') \} &= - \delta_\mu^\nu \delta(\sigma - \sigma ') \\
    \{ \tilde{t}_{M_p} (\sigma),  \tilde{t}_{M_q} (\sigma') \} &= - {\eta^{L_{p+q}}}_{M_p N_q} \tilde{t}_{L_{p+q}}(\sigma') \wedge \mathrm{d}' \delta(\sigma-\sigma') \\
    \{s_\mu(\sigma) , \tilde{t}_{M_p} (\sigma') \}&= 0 =  \{y^\mu(\sigma) , \tilde{t}_{M_p} (\sigma') \} = \{s_\mu(\sigma),s_\nu(\sigma') \} = \{y^\mu(\sigma),y^\nu(\sigma')\}.
\end{align}
The coordinates $y^\mu$ are canonically conjugate to the gauge generators $s_\alpha$. This is obviously a Poisson algebra, assuming that the $\tilde{t}^{(p)}$ form a tensor hierarchy algebra (i.e. assuming that the $\eta$-symbols satisfy the graded Jacobi identity \eqref{eq:IdentityEtaEta}). As usual for the brane currents, it is assumed that $\mathrm{d}\tilde{t}^{(p)} = 0$. 

The quantities in \eqref{eq:CurrentAlgebraFlat} are related by
\begin{align*}
    s_\alpha &= {e_\alpha}^\mu(y) s_\mu, \qquad {t^{(p)}_{M_p}} = {e_{M_p}}^{N_p} (y) \tilde{t}^{(p)}_{N_p}, \qquad \Sigma^\alpha = {e^\alpha}_\mu (y) \mathrm{d}y^\mu \, .
\end{align*}
${e_\alpha}^\mu(y)$ is a right-invariant frame on $H$, with dual coframe ${e^\alpha}_\mu (y)$. ${e_{M_p}}^{N_p} (y)$ is an element of the duality group $\mathcal{G}$ in the representation $\mathcal{R}_p \otimes \bar{ \mathcal{R}}_p$.

For this derivation, these 'vielbeine' are chosen such that the structure coefficients are constant and given by
\begin{align}
    {f_{\alpha \beta}}^\gamma &= - 2 {e_{[\alpha|}}^\mu (\partial_{\mu} {e_{| \beta]}}^\nu) {e^\gamma}_{\nu} \\
    {f_{\alpha M_p}}^{N_p} &= - {e_{\alpha}}^\mu (\partial_{\mu} {e_{M_p}}^{L_p}) {e^{-1}_{L_p}}^{N_p}.
\end{align}

\subsection{Jacobi identities} \label{App:Jacobi}
{\small
\begin{align*}
   &{} \quad \{ s_\alpha (\sigma_1) , \{ t_{M_p}^{(p)}(\sigma_2) , t_{N_q}^{(q)}(\sigma_3) \} \} + \{ t_{N_q}^{(q)}(\sigma_3) , \{ s_\alpha(\sigma_1), t_{M_p}^{(p)}(\sigma_2) \} \} - \{ t_{M_p}^{(p)}(\sigma_2) , \{ s_\alpha(\sigma_1), t_{N_q}^{(q)}(\sigma_3) \} \} \\
   &= - \left( - {f_{\alpha K_{p+q}} }^{L_{p+q}} {\eta^{K_{p+q}}}_{M_p N_q} + {f_{\alpha M_{p}} }^{K_{p}} {\eta^{L_{p+q}}}_{K_p N_q} + {f_{\alpha N_{q}} }^{K_{q}} {\eta^{L_{p+q}}}_{M_p K_q} \right) t^{(p+q)}_{L_{p+q}} (\sigma_3) \wedge \mathrm{d} \delta(\sigma_2-\sigma_3) \delta(\sigma_1-\sigma_3) \\
   &{} \quad + t_{P_{p+q}}^{(p+q)} \wedge \Sigma^\beta \delta(\sigma_1 - \sigma_2) \delta(\sigma_2 - \sigma_3) \left( - {f_{\alpha M_p}}^{K_p} {f_{\beta L_{p+q}}}^{P_{p+q}} {\eta^{L_{p+q}}}_{K_p N_q} + {f_{\alpha L_{p+q}}}^{P_{p+q}} {f_{\beta M_{p}}}^{K_{q}} {\eta^{L_{p+q}}}_{K_p N_q} \right. \\
   &{} \qquad \qquad \left. + {f_{\alpha M_p}}^{K_p} {f_{\beta N_{q}}}^{P_{q}} {\eta^{L_{p+q}}}_{K_p P_q} - {f_{\alpha N_q}}^{K_q} {f_{\beta M_{p}}}^{L_{p}} {\eta^{P_{p+q}}}_{L_p N_q} + {f_{\beta \alpha}}^\gamma {f_{\gamma M_p}}^{K_p} {\eta^{P_{p+q}}}_{K_p N_q} \right) \\
   &= 0 
\end{align*}}
Here, manipulations of the $\mathrm{d}\delta$-terms as distributions up to world-volume boundary terms were performed. The first of the $\delta$-terms arises due to \eqref{eq:Identitydt}.

The first term vanishes due to the $\mathfrak{h}$-invariance of the $\eta$-symbols \eqref{eq:Identityfeta}. The second term vanishes due to a combination of $\mathfrak{h}$-invariance of the $\eta$-symbols (first step) and the fact that the ${f_{\alpha M_p}}^{N_p}$ are a representation of $\mathfrak{h}$ (second step), \eqref{eq:StructureConstantsRepresentation}.
\begin{align*}
    &{} \quad - {f_{\alpha M_p}}^{K_p} \left( {f_{\beta L_{p+q}}}^{P_{p+q}} {\eta^{L_{p+q}}}_{K_p N_q} - {f_{\beta N_{q}}}^{P_{q}} {\eta^{L_{p+q}}}_{K_p P_q} \right) \\
    &{} \quad +   {f_{\beta M_{p}}}^{K_{q}} \left(  {f_{\alpha L_{p+q}}}^{P_{p+q}}  {\eta^{L_{p+q}}}_{K_p N_q} - {f_{\alpha N_q}}^{K_q} {\eta^{P_{p+q}}}_{L_p N_q} \right) \\
    &{} \quad + {f_{\beta \alpha}}^\gamma {f_{\gamma M_p}}^{K_p} {\eta^{P_{p+q}}}_{K_p N_q} \\
    &= 2 {f_{[\underline{\beta} M_p}}^{K_p} {f_{[\underline{\alpha} K_p}}^{L_p} {\eta^{P_{p+q}}}_{L_p N_q} + {f_{\beta \alpha}}^\gamma {f_{\gamma M_p}}^{K_p} {\eta^{P_{p+q}}}_{K_p N_q} \\
    &= 0
\end{align*}

Similarly, one can also show the Jacobi identity for $\{t^{(1)}, t^{(1)}, t^{(1)}\}$, which should generalise easily to the generic case $\{t^{(k)}, t^{(l)}, t^{(m)}\}$. This gives,
\begin{align}
    &\{ t_{A_1} (\sigma) , \{ t_{B_1} (\sigma'), t_{C_1} (\sigma'') \} \} + \text{cyc} \left( (A_1, \sigma), (B_1, \sigma'), (C_1, \sigma'')
    \right) \nonumber \\
    &= \eta^{E_3}\mathstrut_{B_1 D_2} \, \eta^{D_2}\mathstrut_{C_1 A_1} t_{E_3} (\sigma) \wedge \mathrm{d} \delta (\sigma - \sigma') \wedge \mathrm{d} \delta (\sigma - \sigma'') \label{eq:JacobiEtaEta}\\
    &+ \left( f_{\alpha A_1}\mathstrut^{F_1} \eta^{E_3}\mathstrut_{C_1 D_2} \, \eta^{D_2}\mathstrut_{F_1 B_1} - f_{\alpha B_1}\mathstrut^{F_1} \eta^{E_3}\mathstrut_{F_1 D_2} \, \eta^{D_2}\mathstrut_{C_1 A_1} \right) t^{\alpha}_{E_3} (\sigma) \wedge \mathrm{d} \delta (\sigma - \sigma'') \delta (\sigma - \sigma') \label{eq:JacobifEtaEta}\\
    &+ f_{\alpha A_1}\mathstrut^{F_1} f_{\beta B_1}\mathstrut^{G_1} \eta^{E_3}\mathstrut_{F_1 D_2} \, \eta^{D_2}\mathstrut_{G_1 C_1} t^{\alpha \beta}_{E_3}  (\sigma) \delta (\sigma - \sigma') \delta (\sigma' - \sigma'') \label{eq:JacobiffEtaEta}\\
    &+ \text{cyc} \left( (A_1, \sigma), (B_1, \sigma'), (C_1, \sigma'')
    \right) \nonumber.
\end{align}
By consistency, this should give zero as a distributional identity, i.e. integrated with a suitable test function $\phi (\sigma, \sigma', \sigma'')$. 

Let us first look at \eqref{eq:JacobiEtaEta}, i.e. the terms proportional to $\mathrm{d} \delta \wedge \mathrm{d} \delta$. Dropping boundary terms after integrating by parts twice to get rid of the delta functions, these terms look like,
{\small \begin{align*}
    &\eta^{E_3}\mathstrut_{B_1 D_2} \, \eta^{D_2}\mathstrut_{C_1 A_1} \int_{\sigma} \int_{\sigma'} \int_{\sigma''} \phi (\sigma, \sigma', \sigma'') t_{E_3} (\sigma) \wedge \mathrm{d} \delta (\sigma - \sigma') \wedge \mathrm{d} \delta (\sigma - \sigma'') + \text{cyc} \left( (A_1, \sigma), (B_1, \sigma'), (C_1, \sigma'')
    \right) \\
    &= \eta^{E_3}\mathstrut_{B_1 D_2} \, \eta^{D_2}\mathstrut_{C_1 A_1} \int \partial_2 \partial_3 \phi (\sigma, \sigma, \sigma) \wedge t_{E_3} (\sigma) + \eta^{E_3}\mathstrut_{C_1 D_2} \, \eta^{D_2}\mathstrut_{A_1 B_1} \int \partial_3 \partial_1 \phi (\sigma, \sigma, \sigma) \wedge t_{E_3} (\sigma) \\
    &+ \eta^{E_3}\mathstrut_{A_1 D_2} \, \eta^{D_2}\mathstrut_{B_1 C_1} \int \partial_1 \partial_2 \phi (\sigma, \sigma, \sigma) \wedge t_{E_3} (\sigma),
\end{align*}}
where $\partial_i \phi(\sigma, \sigma, \sigma)$ for $i = 1, 2, 3$ is short for an exterior derivative with respect to the $i$-th argument of the test function $\phi$, e.g. $\partial_1 \phi (\sigma, \sigma, \sigma) = \int_{'} \int_{''} \mathrm{d} \phi (\sigma, \sigma', \sigma'') \delta (\sigma - \sigma') \delta (\sigma' - \sigma'')$. In particular, after integrating the delta functions, we have $\mathrm{d} = \partial_1 + \partial_2 + \partial_3$. This fact, along with an integration by parts and \eqref{eq:Identitydt} allows us to write these three integrals as
{\small\begin{align*}
    &\left( 
    \eta^{E_3}\mathstrut_{B_1 D_2} \, \eta^{D_2}\mathstrut_{C_1 A_1} + \eta^{E_3}\mathstrut_{C_1 D_2} \, \eta^{D_2}\mathstrut_{A_1 B_1} + \eta^{E_3}\mathstrut_{A_1 D_2} \, \eta^{D_2}\mathstrut_{B_1 C_1}
    \right) \int \partial_1 \partial_2 \phi (\sigma, \sigma, \sigma) \wedge t_{E_3} (\sigma) \\
    &+ f_{\alpha F_3}\mathstrut^{E_3} \eta^{F_3}\mathstrut_{C_1 D_2} \, \eta^{D_2}\mathstrut_{A_1 B_1} \int t^\alpha_{E_3} (\sigma) \wedge \partial_1 \phi (\sigma, \sigma, \sigma) - f_{\alpha F_3}\mathstrut^{E_3} \eta^{F_3}\mathstrut_{B_1 D_2} \, \eta^{D_2}\mathstrut_{C_1 A_1} \int t^\alpha_{E_3} (\sigma) \wedge \partial_2 \phi (\sigma, \sigma, \sigma)
\end{align*}}
The first line above vanishes due to \eqref{eq:IdentityEtaEta}, while the remaining terms will combine with the terms \eqref{eq:JacobifEtaEta} proportional to $\delta \mathrm{d}\delta$, which are now of the form
{\small\begin{align*}
    &\left( 
    f_{\alpha B_1}\mathstrut^{F_1} \eta^{E_3}\mathstrut_{A_1 D_2} \, \eta^{D_2}\mathstrut_{F_1 C_1} - f_{\alpha C_1}\mathstrut^{F_1} \eta^{E_3}\mathstrut_{F_1 D_2} \, \eta^{D_2}\mathstrut_{A_1 B_1} + f_{\alpha F_3}\mathstrut^{E_3} \eta^{F_3}\mathstrut_{C_1 D_2} \, \eta^{D_2}\mathstrut_{A_1 B_1}
    \right) \int t^\alpha_{E_3} (\sigma) \wedge \partial_1 \phi (\sigma, \sigma, \sigma) \\
    + &\left( 
    f_{\alpha C_1}\mathstrut^{F_1} \eta^{E_3}\mathstrut_{B_1 D_2} \, \eta^{D_2}\mathstrut_{F_1 A_1} - f_{\alpha A_1}\mathstrut^{F_1} \eta^{E_3}\mathstrut_{F_1 D_2} \, \eta^{D_2}\mathstrut_{B_1 C_1} - f_{\alpha F_3}\mathstrut^{E_3} \eta^{F_3}\mathstrut_{B_1 D_2} \, \eta^{D_2}\mathstrut_{C_1 A_1}
    \right) \int t^\alpha_{E_3} (\sigma) \wedge \partial_2 \phi (\sigma, \sigma, \sigma)
    \\
    + &\left( 
    f_{\alpha A_1}\mathstrut^{F_1} \eta^{E_3}\mathstrut_{C_1 D_2} \, \eta^{D_2}\mathstrut_{F_1 B_1} - f_{\alpha B_1}\mathstrut^{F_1} \eta^{E_3}\mathstrut_{F_1 D_2} \, \eta^{D_2}\mathstrut_{C_1 A_1}
    \right) \int t^\alpha_{E_3} (\sigma) \wedge \partial_3 \phi (\sigma, \sigma, \sigma) . 
    \end{align*}}
The strategy is to use the identities \eqref{eq:IdentityEtaEta} and \eqref{eq:Identityfeta} to combine the partial derivatives $\partial_i \phi$ to an exterior derivative $\mathrm{d} \phi$. This gives us,
{\small\begin{align*}
    &- \left( 
    f_{\alpha A_1}\mathstrut^{F_1} \eta^{D_2}\mathstrut_{B_1 C_1} + f_{\alpha B_1}\mathstrut^{F_1} \eta^{D_2}\mathstrut_{C_1 A_1} + f_{\alpha C_1}\mathstrut^{F_1} \eta^{D_2}\mathstrut_{A_1 B_1}
    \right) \eta^{E_3}\mathstrut_{F_1 D_2} \int t^\alpha_{E_3} (\sigma) \wedge \mathrm{d} \phi (\sigma, \sigma, \sigma) \\
    &+ \left(- f_{\alpha F_2}\mathstrut^{D_2} \eta^{E_3}\mathstrut_{C_1 D_2} \, \eta^{F_2}\mathstrut_{A_1 B_1} - f_{\alpha A_1}\mathstrut^{F_1} \eta^{E_3}\mathstrut_{B_1 D_2} \, \eta^{F_2}\mathstrut_{C_1 F_1} + f_{\alpha F_3}\mathstrut^{E_3} \eta^{F_3}\mathstrut_{C_1 D_2} \, \eta^{D_2}\mathstrut_{A_1 B_1}
    \right) \int t^\alpha_{E_3} (\sigma) \wedge \partial_1 \phi (\sigma, \sigma, \sigma) \\
    &+ \left(- f_{\alpha F_2}\mathstrut^{D_2} \eta^{E_3}\mathstrut_{A_1 D_2} \, \eta^{F_2}\mathstrut_{B_1 C_1} - f_{\alpha B_1}\mathstrut^{F_1} \eta^{E_3}\mathstrut_{C_1 D_2} \, \eta^{F_2}\mathstrut_{A_1 F_1} - f_{\alpha F_3}\mathstrut^{E_3} \eta^{F_3}\mathstrut_{B_1 D_2} \, \eta^{D_2}\mathstrut_{C_1 A_1}
    \right) \int t^\alpha_{E_3} (\sigma) \wedge \partial_2 \phi (\sigma, \sigma, \sigma) \\
    &+ \left(- f_{\alpha F_2}\mathstrut^{D_2} \eta^{E_3}\mathstrut_{B_1 D_2} \, \eta^{F_2}\mathstrut_{C_1 A_1} - f_{\alpha C_1}\mathstrut^{F_1} \eta^{E_3}\mathstrut_{A_1 D_2} \, \eta^{F_2}\mathstrut_{B_1 F_1}
    \right) \int t^\alpha_{E_3} (\sigma) \wedge \partial_3 \phi (\sigma, \sigma, \sigma) .
\end{align*}}
Using the same aforementioned identities, it turns out that the $t \wedge \partial_i \phi$ integrals all share the same prefactor, namely the one in front of the $\partial_3 \phi$ integral. So we can also complete the exterior derivative, and then integrate by parts and use the identity \eqref{eq:Identitydtalpha}, resulting in,
{\small\begin{align*}
    & \left[- \left( 
    f_{\alpha A_1}\mathstrut^{F_1} \eta^{D_2}\mathstrut_{B_1 C_1} + f_{\alpha B_1}\mathstrut^{F_1} \eta^{D_2}\mathstrut_{C_1 A_1}
    \right) \eta^{E_3}\mathstrut_{F_1 D_2} - f_{\alpha A_1}\mathstrut^{F_1} \eta^{E_3}\mathstrut_{B_1 D_2} \eta^{D_2}\mathstrut_{C_1 F_1} \right] \int \big(
    f_{\alpha E_3}\mathstrut^{F_3} t^\alpha_{F_3} (\sigma) + \frac12 f_{\beta \gamma}^\alpha t^{\beta \gamma}_{E_3} (\sigma)
    \big) \phi . 
\end{align*}}
Simplifying these and grouping terms gives us
\begin{align*}
    - f_{\alpha A_1}\mathstrut^{F_1} f_{\beta B_1}\mathstrut^{G_1} \eta^{E_3}\mathstrut_{F_1 D_2} \, \eta^{D_2}\mathstrut_{G_1 C_1} \int t^{\alpha \beta}_{E_3} (\sigma) \phi (\sigma, \sigma, \sigma) + \text{cyc} (A_1, B_1, C_1),
\end{align*}
which precisely cancels out the contribution coming from the $\delta \delta$ terms \eqref{eq:JacobiffEtaEta}. This proves that the Jacobi identity $\{t^{(1)}, t^{(1)}, t^{(1)}\}= 0$ is satisfied.

\bibliographystyle{jhep}
\bibliography{References}

\providecommand{\href}[2]{#2}\begingroup\raggedright\begin{thebibliography}{100}

\bibitem{Duff:1989tf}
M.~J. Duff, {\it {Duality Rotations in String Theory}},  {\em Nucl. Phys.} {\bf
  B335} (1990) 610.

\bibitem{Tseytlin:1990nb}
A.~A. Tseytlin, {\it {Duality Symmetric Formulation of String World Sheet
  Dynamics}},  {\em Phys. Lett.} {\bf B242} (1990) 163--174.

\bibitem{Tseytlin:1990va}
A.~A. Tseytlin, {\it {Duality symmetric closed string theory and interacting
  chiral scalars}},  {\em Nucl. Phys.} {\bf B350} (1991) 395--440.

\bibitem{Siegel:1993xq}
W.~Siegel, {\it {Two vierbein formalism for string inspired axionic gravity}},
  {\em Phys. Rev.} {\bf D47} (1993) 5453--5459,
  [\href{http://arxiv.org/abs/hep-th/9302036}{{\tt hep-th/9302036}}].

\bibitem{Siegel:1993th}
W.~Siegel, {\it {Superspace duality in low-energy superstrings}},  {\em Phys.
  Rev.} {\bf D48} (1993) 2826--2837,
  [\href{http://arxiv.org/abs/hep-th/9305073}{{\tt hep-th/9305073}}].

\bibitem{Siegel:1993bj}
W.~Siegel, {\it {Manifest duality in low-energy superstrings}},  in {\em
  {International Conference on Strings 93 Berkeley, California, May 24-29,
  1993}}, pp.~353--363, 1993.
\newblock \href{http://arxiv.org/abs/hep-th/9308133}{{\tt hep-th/9308133}}.

\bibitem{Hull:2004in}
C.~M. Hull, {\it {A Geometry for non-geometric string backgrounds}},  {\em
  JHEP} {\bf 10} (2005) 065, [\href{http://arxiv.org/abs/hep-th/0406102}{{\tt
  hep-th/0406102}}].

\bibitem{Hull:2006va}
C.~M. Hull, {\it {Doubled Geometry and T-Folds}},  {\em JHEP} {\bf 07} (2007)
  080, [\href{http://arxiv.org/abs/hep-th/0605149}{{\tt hep-th/0605149}}].

\bibitem{Hitchin:2004ut}
N.~Hitchin, {\it {Generalized Calabi-Yau manifolds}},  {\em Quart. J. Math.}
  {\bf 54} (2003) 281--308, [\href{http://arxiv.org/abs/math/0209099}{{\tt
  math/0209099}}].

\bibitem{Gualtieri:2003dx}
M.~Gualtieri, {\em {Generalized complex geometry}}.
\newblock PhD thesis, Oxford U., 2003.
\newblock \href{http://arxiv.org/abs/math/0401221}{{\tt math/0401221}}.

\bibitem{Grana:2008yw}
M.~Gra{\~{n}}a, R.~Minasian, M.~Petrini, and D.~Waldram, {\it {T-duality,
  Generalized Geometry and Non-Geometric Backgrounds}},  {\em JHEP} {\bf 04}
  (2009) 075, [\href{http://arxiv.org/abs/0807.4527}{{\tt arXiv:0807.4527}}].

\bibitem{Hull:2009mi}
C.~Hull and B.~Zwiebach, {\it {Double Field Theory}},  {\em JHEP} {\bf 09}
  (2009) 099, [\href{http://arxiv.org/abs/0904.4664}{{\tt arXiv:0904.4664}}].

\bibitem{Zwiebach:2011rg}
B.~Zwiebach, {\it {Double Field Theory, T-Duality, and Courant Brackets}},
  {\em Lect. Notes Phys.} {\bf 851} (2012) 265--291,
  [\href{http://arxiv.org/abs/1109.1782}{{\tt arXiv:1109.1782}}].

\bibitem{Geissbuhler:2013uka}
D.~Geissb{\"{u}}hler, D.~Marqu{\'{e}}s, C.~Nu{\~{n}}ez, and V.~Penas, {\it
  {Exploring Double Field Theory}},  {\em JHEP} {\bf 06} (2013) 101,
  [\href{http://arxiv.org/abs/1304.1472}{{\tt arXiv:1304.1472}}].

\bibitem{Aldazabal:2013sca}
G.~Aldazabal, D.~Marqu{\'{e}}s, and C.~Nu{\~{n}}ez, {\it {Double Field Theory:
  A Pedagogical Review}},  {\em Class. Quant. Grav.} {\bf 30} (2013) 163001,
  [\href{http://arxiv.org/abs/1305.1907}{{\tt arXiv:1305.1907}}].

\bibitem{Berman:2013eva}
D.~S. Berman and D.~C. Thompson, {\it {Duality Symmetric String and M-Theory}},
   {\em Phys. Rept.} {\bf 566} (2014) 1--60,
  [\href{http://arxiv.org/abs/1306.2643}{{\tt arXiv:1306.2643}}].

\bibitem{Blair:2013noa}
C.~D.~A. Blair, E.~Malek, and A.~J. Routh, {\it {An $O(D, D)$ invariant
  Hamiltonian action for the superstring}},  {\em Class. Quant. Grav.} {\bf 31}
  (2014), no.~20 205011, [\href{http://arxiv.org/abs/1308.4829}{{\tt
  arXiv:1308.4829}}].

\bibitem{Hohm:2013bwa}
O.~Hohm, D.~L{\"{u}}st, and B.~Zwiebach, {\it {The Spacetime of Double Field
  Theory: Review, Remarks, and Outlook}},  {\em Fortsch. Phys.} {\bf 61} (2013)
  926--966, [\href{http://arxiv.org/abs/1309.2977}{{\tt arXiv:1309.2977}}].

\bibitem{Plauschinn:2018wbo}
E.~Plauschinn, {\it {Non-geometric backgrounds in string theory}},  {\em Phys.
  Rept.} {\bf 798} (2019) 1--122, [\href{http://arxiv.org/abs/1811.11203}{{\tt
  arXiv:1811.11203}}].

\bibitem{Sakatani:2016sko}
Y.~Sakatani and S.~Uehara, {\it {Branes in Extended Spacetime: Brane
  Worldvolume Theory Based on Duality Symmetry}},  {\em Phys. Rev. Lett.} {\bf
  117} (2016), no.~19 191601, [\href{http://arxiv.org/abs/1607.04265}{{\tt
  arXiv:1607.04265}}].

\bibitem{Blair:2017hhy}
C.~D.~A. Blair and E.~T. Musaev, {\it {Five-brane actions in double field
  theory}},  {\em JHEP} {\bf 03} (2018) 111,
  [\href{http://arxiv.org/abs/1712.01739}{{\tt arXiv:1712.01739}}].

\bibitem{Sakatani:2017vbd}
Y.~Sakatani and S.~Uehara, {\it {Exceptional M-brane sigma models and
  $\eta$-symbols}},  {\em PTEP} {\bf 2018} (2018), no.~3 033B05,
  [\href{http://arxiv.org/abs/1712.10316}{{\tt arXiv:1712.10316}}].

\bibitem{Arvanitakis:2018hfn}
A.~S. Arvanitakis and C.~D. Blair, {\it {The Exceptional Sigma Model}},  {\em
  JHEP} {\bf 04} (2018) 064, [\href{http://arxiv.org/abs/1802.00442}{{\tt
  arXiv:1802.00442}}].

\bibitem{Blair:2019tww}
C.~D.~A. Blair, {\it {Open exceptional strings and D-branes}},  {\em JHEP} {\bf
  07} (2019) 083, [\href{http://arxiv.org/abs/1904.06714}{{\tt
  arXiv:1904.06714}}].

\bibitem{Sakatani:2020umt}
Y.~Sakatani and S.~Uehara, {\it {Born sigma model for branes in exceptional
  geometry}},  {\em PTEP} {\bf 2020} (2020), no.~7 073B05,
  [\href{http://arxiv.org/abs/2004.09486}{{\tt arXiv:2004.09486}}].

\bibitem{Duff:1990hn}
M.~J. Duff and J.~X. Lu, {\it {Duality Rotations in Membrane Theory}},  {\em
  Nucl. Phys.} {\bf B347} (1990) 394--419. [,210(1990)].

\bibitem{Alekseev:2004np}
A.~Alekseev and T.~Strobl, {\it {Current algebras and differential geometry}},
  {\em JHEP} {\bf 03} (2005) 035,
  [\href{http://arxiv.org/abs/hep-th/0410183}{{\tt hep-th/0410183}}].

\bibitem{Berman:2010is}
D.~S. Berman and M.~J. Perry, {\it {Generalized Geometry and M theory}},  {\em
  JHEP} {\bf 06} (2011) 074, [\href{http://arxiv.org/abs/1008.1763}{{\tt
  arXiv:1008.1763}}].

\bibitem{Hatsuda:2012uk}
M.~Hatsuda and T.~Kimura, {\it {Canonical approach to Courant brackets for
  D-branes}},  {\em JHEP} {\bf 06} (2012) 034,
  [\href{http://arxiv.org/abs/1203.5499}{{\tt arXiv:1203.5499}}].

\bibitem{Hatsuda:2012vm}
M.~Hatsuda and K.~Kamimura, {\it {SL(5) duality from canonical M2-brane}},
  {\em JHEP} {\bf 11} (2012) 001, [\href{http://arxiv.org/abs/1208.1232}{{\tt
  arXiv:1208.1232}}].

\bibitem{Hatsuda:2013dya}
M.~Hatsuda and K.~Kamimura, {\it {M5 algebra and SO(5,5) duality}},  {\em JHEP}
  {\bf 06} (2013) 095, [\href{http://arxiv.org/abs/1305.2258}{{\tt
  arXiv:1305.2258}}].

\bibitem{Hatsuda:2020buq}
M.~Hatsuda, S.~Sasaki, and M.~Yata, {\it {Five-brane Current Algebras in Type
  II String Theories}},  \href{http://arxiv.org/abs/2011.13145}{{\tt
  arXiv:2011.13145}}.

\bibitem{Duff:2015jka}
M.~J. Duff, J.~X. Lu, R.~Percacci, C.~N. Pope, H.~Samtleben, and E.~Sezgin,
  {\it {Membrane Duality Revisited}},  {\em Nucl. Phys.} {\bf B901} (2015)
  1--21, [\href{http://arxiv.org/abs/1509.02915}{{\tt arXiv:1509.02915}}].

\bibitem{Sakatani:2020iad}
Y.~Sakatani and S.~Uehara, {\it {Non-Abelian $U$-duality for membranes}},  {\em
  PTEP} {\bf 2020} (2020), no.~7 073B01,
  [\href{http://arxiv.org/abs/2001.09983}{{\tt arXiv:2001.09983}}].

\bibitem{Sakatani:2020wah}
Y.~Sakatani, {\it {Extended Drinfel\textquoteright{}d algebras and non-Abelian
  duality}},  {\em PTEP} {\bf 2021} (2021), no.~6 063B02,
  [\href{http://arxiv.org/abs/2009.04454}{{\tt arXiv:2009.04454}}].

\bibitem{Strickland-Constable:2021afa}
C.~Strickland-Constable, {\it {Classical worldvolumes as generalised
  geodesics}},  \href{http://arxiv.org/abs/2102.00555}{{\tt arXiv:2102.00555}}.

\bibitem{Osten:2021fil}
D.~Osten, {\it {Currents, charges and algebras in exceptional generalised
  geometry}},  {\em JHEP} {\bf 06} (2021) 070,
  [\href{http://arxiv.org/abs/2103.03267}{{\tt arXiv:2103.03267}}].

\bibitem{Arvanitakis:2021wkt}
A.~S. Arvanitakis, {\it {Brane current algebras and generalised geometry from
  QP manifolds. Or, \textquotedblleft{}when they go high, we go
  low\textquotedblright{}}},  {\em JHEP} {\bf 11} (2021) 114,
  [\href{http://arxiv.org/abs/2103.08608}{{\tt arXiv:2103.08608}}].

\bibitem{Hatsuda:2022zpi}
M.~Hatsuda, H.~Mori, S.~Sasaki, and M.~Yata, {\it {Gauged double field theory,
  current algebras and heterotic sigma models}},  {\em JHEP} {\bf 05} (2023)
  220, [\href{http://arxiv.org/abs/2212.06476}{{\tt arXiv:2212.06476}}].

\bibitem{Osten:2023iwc}
D.~Osten, {\it {On exceptional QP-manifolds}},  {\em JHEP} {\bf 01} (2024) 028,
  [\href{http://arxiv.org/abs/2306.11093}{{\tt arXiv:2306.11093}}].

\bibitem{Osten:2024mjt}
D.~Osten, {\it {On the universal exceptional structure of world-volume theories
  in string and M-theory}},  {\em Phys. Lett. B} {\bf 855} (2024) 138814,
  [\href{http://arxiv.org/abs/2402.10269}{{\tt arXiv:2402.10269}}].

\bibitem{Hull:2007zu}
C.~M. Hull, {\it {Generalised Geometry for M-Theory}},  {\em JHEP} {\bf 07}
  (2007) 079, [\href{http://arxiv.org/abs/hep-th/0701203}{{\tt
  hep-th/0701203}}].

\bibitem{Pacheco:2008ps}
P.~Pires~Pacheco and D.~Waldram, {\it {M-theory, exceptional generalised
  geometry and superpotentials}},  {\em JHEP} {\bf 09} (2008) 123,
  [\href{http://arxiv.org/abs/0804.1362}{{\tt arXiv:0804.1362}}].

\bibitem{Berman:2011jh}
D.~S. Berman, H.~Godazgar, M.~J. Perry, and P.~West, {\it {Duality Invariant
  Actions and Generalised Geometry}},  {\em JHEP} {\bf 02} (2012) 108,
  [\href{http://arxiv.org/abs/1111.0459}{{\tt arXiv:1111.0459}}].

\bibitem{Coimbra:2011ky}
A.~Coimbra, C.~Strickland-Constable, and D.~Waldram, {\it {$E_{d(d)} \times
  \mathbb{R}^+$ generalised geometry, connections and M theory}},  {\em JHEP}
  {\bf 02} (2014) 054, [\href{http://arxiv.org/abs/1112.3989}{{\tt
  arXiv:1112.3989}}].

\bibitem{Berman:2012vc}
D.~S. Berman, M.~Cederwall, A.~Kleinschmidt, and D.~C. Thompson, {\it {The
  gauge structure of generalised diffeomorphisms}},  {\em JHEP} {\bf 01} (2013)
  064, [\href{http://arxiv.org/abs/1208.5884}{{\tt arXiv:1208.5884}}].

\bibitem{Berman:2012uy}
D.~S. Berman, E.~T. Musaev, and D.~C. Thompson, {\it {Duality Invariant
  M-theory: Gauged supergravities and Scherk-Schwarz reductions}},  {\em JHEP}
  {\bf 10} (2012) 174, [\href{http://arxiv.org/abs/1208.0020}{{\tt
  arXiv:1208.0020}}].

\bibitem{Coimbra:2012af}
A.~Coimbra, C.~Strickland-Constable, and D.~Waldram, {\it {Supergravity as
  Generalised Geometry II: $E_{d(d)} \times \mathbb{R}^+$ and M theory}},  {\em
  JHEP} {\bf 03} (2014) 019, [\href{http://arxiv.org/abs/1212.1586}{{\tt
  arXiv:1212.1586}}].

\bibitem{Hohm:2013pua}
O.~Hohm and H.~Samtleben, {\it {Exceptional Form of D=11 Supergravity}},  {\em
  Phys. Rev. Lett.} {\bf 111} (2013) 231601,
  [\href{http://arxiv.org/abs/1308.1673}{{\tt arXiv:1308.1673}}].

\bibitem{Lee:2014mla}
K.~Lee, C.~Strickland-Constable, and D.~Waldram, {\it {Spheres, generalised
  parallelisability and consistent truncations}},  {\em Fortsch. Phys.} {\bf
  65} (2017), no.~10-11 1700048, [\href{http://arxiv.org/abs/1401.3360}{{\tt
  arXiv:1401.3360}}].

\bibitem{Hohm:2014qga}
O.~Hohm and H.~Samtleben, {\it {Consistent Kaluza-Klein Truncations via
  Exceptional Field Theory}},  {\em JHEP} {\bf 01} (2015) 131,
  [\href{http://arxiv.org/abs/1410.8145}{{\tt arXiv:1410.8145}}].

\bibitem{Berman:2020tqn}
D.~S. Berman and C.~D.~A. Blair, {\it {The Geometry, Branes and Applications of
  Exceptional Field Theory}},  {\em Int. J. Mod. Phys. A} {\bf 35} (2020),
  no.~30 2030014, [\href{http://arxiv.org/abs/2006.09777}{{\tt
  arXiv:2006.09777}}].

\bibitem{Sterckx:2024vju}
C.~Sterckx, {\it {Modave lecture notes: Introduction to Exceptional Field
  Theory}},  {\em PoS} {\bf Modave2023} (2025) 004,
  [\href{http://arxiv.org/abs/2410.19600}{{\tt arXiv:2410.19600}}].

\bibitem{Samtleben:2025fta}
H.~Samtleben, {\it {Exceptional field theories}},
  \href{http://arxiv.org/abs/2503.16947}{{\tt arXiv:2503.16947}}.

\bibitem{Marsden:1974dsb}
J.~Marsden and A.~Weinstein, {\it {Reduction of symplectic manifolds with
  symmetry}},  {\em Rept. Math. Phys.} {\bf 5} (1974), no.~1 121--130.

\bibitem{Sharpe:1997}
R.~W. Sharpe, {\em Differential Geometry: Cartan’s Generalization of
  Klein’s Erlangen Program}, vol.~166 of {\em Graduate Texts in Mathematics}.
\newblock Springer, 1997.

\bibitem{cap2009parabolic}
A.~Cap and J.~Slov{\'a}k, {\em Parabolic Geometries: Background and general
  theory}.
\newblock Mathematical surveys and monographs. American Mathematical Society,
  2009.

\bibitem{Mackenzie_2005}
K.~C.~H. Mackenzie, {\em General Theory of Lie Groupoids and Lie Algebroids}.
\newblock London Mathematical Society Lecture Note Series. Cambridge University
  Press, 2005.

\bibitem{Blaom:2006}
A.~D. Blaom, {\it Geometric structures as deformed infinitesimal symmetries},
  {\em Trans. Amer. Math. Soc.} {\bf 358} (2006), no.~8 3651--3671.

\bibitem{crampin2016cartan}
M.~Crampin and D.~Saunders, {\em Cartan Geometries and their Symmetries: A Lie
  Algebroid Approach}.
\newblock Atlantis Studies in Variational Geometry. Atlantis Press, 2016.

\bibitem{Attard:2019pvw}
J.~Attard, J.~Fran{\c{c}}ois, S.~Lazzarini, and T.~Masson, {\it {Cartan
  Connections and Atiyah Lie Algebroids}},  {\em J. Geom. Phys.} {\bf 148}
  (2020) 103541, [\href{http://arxiv.org/abs/1904.04915}{{\tt
  arXiv:1904.04915}}].

\bibitem{Liu:1995lsa}
Z.-J. Liu, A.~Weinstein, and P.~Xu, {\it {Manin Triples for Lie Bialgebroids}},
   {\em J. Diff. Geom.} {\bf 45} (1997), no.~3 547--574,
  [\href{http://arxiv.org/abs/dg-ga/9508013}{{\tt dg-ga/9508013}}].

\bibitem{baraglia2012leibniz}
D.~Baraglia, {\it Leibniz algebroids, twistings and exceptional generalized
  geometry},  {\em Journal of Geometry and Physics} {\bf 62} (2012), no.~5
  903--934.

\bibitem{Bugden:2021wxg}
M.~Bugden, O.~Hulik, F.~Valach, and D.~Waldram, {\it {G-Algebroids: A Unified
  Framework for Exceptional and Generalised Geometry, and
  Poisson{\textendash}Lie Duality}},  {\em Fortsch. Phys.} {\bf 69} (2021),
  no.~4-5 2100028, [\href{http://arxiv.org/abs/2103.01139}{{\tt
  arXiv:2103.01139}}].

\bibitem{Coimbra:2011nw}
A.~Coimbra, C.~Strickland-Constable, and D.~Waldram, {\it {Supergravity as
  Generalised Geometry I: Type II Theories}},  {\em JHEP} {\bf 11} (2011) 091,
  [\href{http://arxiv.org/abs/1107.1733}{{\tt arXiv:1107.1733}}].

\bibitem{Hohm:2012mf}
O.~Hohm and B.~Zwiebach, {\it {Towards an invariant geometry of double field
  theory}},  {\em J. Math. Phys.} {\bf 54} (2013) 032303,
  [\href{http://arxiv.org/abs/1212.1736}{{\tt arXiv:1212.1736}}].

\bibitem{Garcia-Fernandez:2016ofz}
M.~Garcia-Fernandez, {\it {Ricci flow, Killing spinors, and T-duality in
  generalized geometry}},  {\em Adv. Math.} {\bf 350} (2019) 1059--1108,
  [\href{http://arxiv.org/abs/1611.08926}{{\tt arXiv:1611.08926}}].

\bibitem{Cortes:2025lns}
V.~Cort{\'e}s, M.~Mackevicius, T.~Mohaupt, and O.~Schiller, {\it {The canonical
  generalised Levi-Civita connection and its curvature}},
  \href{http://arxiv.org/abs/2507.17604}{{\tt arXiv:2507.17604}}.

\bibitem{Gualtieri:2007bq}
M.~Gualtieri, {\it Branes on poisson varieties},  in {\em The Many Facets of
  Geometry: A Tribute to Nigel Hitchin}.
\newblock Oxford University Press, 07, 2010.
\newblock \href{http://arxiv.org/abs/0710.2719}{{\tt arXiv:0710.2719}}.

\bibitem{Hohm:2011si}
O.~Hohm and B.~Zwiebach, {\it {On the Riemann Tensor in Double Field Theory}},
  {\em JHEP} {\bf 05} (2012) 126, [\href{http://arxiv.org/abs/1112.5296}{{\tt
  arXiv:1112.5296}}].

\bibitem{Jurco:2016emw}
B.~Jurco and J.~Vysoky, {\it {Courant Algebroid Connections and String
  Effective Actions}},  in {\em {Workshop on Strings, Membranes and Topological
  Field Theory}}, pp.~211--265, 2017.
\newblock \href{http://arxiv.org/abs/1612.01540}{{\tt arXiv:1612.01540}}.

\bibitem{Garcia-Fernandez:2013gja}
M.~Garcia-Fernandez, {\it {Torsion-free generalized connections and Heterotic
  Supergravity}},  {\em Commun. Math. Phys.} {\bf 332} (2014), no.~1 89--115,
  [\href{http://arxiv.org/abs/1304.4294}{{\tt arXiv:1304.4294}}].

\bibitem{Cavalcanti:2024uky}
G.~R. Cavalcanti, J.~Pedregal, and R.~Rubio, {\it {On the Equivalence of
  Generalized Ricci Curvatures}},  {\em Proc. Am. Math. Soc.} {\bf 153} (2025)
  2639--2648, [\href{http://arxiv.org/abs/2406.06695}{{\tt arXiv:2406.06695}}].

\bibitem{Cederwall:2013naa}
M.~Cederwall, J.~Edlund, and A.~Karlsson, {\it {Exceptional geometry and tensor
  fields}},  {\em JHEP} {\bf 07} (2013) 028,
  [\href{http://arxiv.org/abs/1302.6736}{{\tt arXiv:1302.6736}}].

\bibitem{Polacek:2013nla}
M.~Pol\'a\v{c}ek and W.~Siegel, {\it {Natural curvature for manifest
  T-duality}},  {\em JHEP} {\bf 01} (2014) 026,
  [\href{http://arxiv.org/abs/1308.6350}{{\tt arXiv:1308.6350}}].

\bibitem{Hassler:2024hgq}
F.~Hassler, O.~Hulik, and D.~Osten, {\it {Current algebra and generalized
  Cartan geometry}},  {\em Phys. Rev. D} {\bf 110} (2024), no.~12 126022,
  [\href{http://arxiv.org/abs/2409.00176}{{\tt arXiv:2409.00176}}].

\bibitem{Hassler:2023axp}
F.~Hassler and Y.~Sakatani, {\it {Hierarchy of curvatures in exceptional
  geometry}},  {\em Phys. Rev. D} {\bf 109} (2024), no.~10 106002,
  [\href{http://arxiv.org/abs/2311.12095}{{\tt arXiv:2311.12095}}].

\bibitem{Aschieri:2019qku}
P.~Aschieri, F.~Bonechi, and A.~Deser, {\it {On Curvature and Torsion in
  Courant Algebroids}},  {\em Annales Henri Poincare} {\bf 22} (2021), no.~7
  2475--2496, [\href{http://arxiv.org/abs/1910.11273}{{\tt arXiv:1910.11273}}].

\bibitem{Cueca_2020}
M.~Cueca and R.~A. Mehta, {\it Courant cohomology, cartan calculus,
  connections, curvature, characteristic classes},  {\em Communications in
  Mathematical Physics} {\bf 381} (Nov., 2020) 1091–1113.

\bibitem{Batakidis:2020ylk}
P.~Batakidis and F.~Petalidou, {\it {Courant-Dorfman algebras of differential
  operators and Dorfman connections of Courant algebroids}},  {\em J. Geom.
  Phys.} {\bf 199} (2024) 105142, [\href{http://arxiv.org/abs/2002.10175}{{\tt
  arXiv:2002.10175}}].

\bibitem{Chatzistavrakidis:2023otk}
A.~Chatzistavrakidis and L.~Jonke, {\it {Basic curvature {\&} the Atiyah
  cocycle in gauge theory}},  {\em J. Phys. A} {\bf 57} (2024), no.~46 465401,
  [\href{http://arxiv.org/abs/2302.04956}{{\tt arXiv:2302.04956}}].

\bibitem{Chatzistavrakidis:2024utp}
A.~Chatzistavrakidis, T.~Kod{\v{z}}oman, and Z.~{\v{S}}koda, {\it {Brane
  mechanics and gapped Lie n-algebroids}},  {\em JHEP} {\bf 08} (2024) 231,
  [\href{http://arxiv.org/abs/2404.14126}{{\tt arXiv:2404.14126}}].

\bibitem{Palmkvist:2013vya}
J.~Palmkvist, {\it {The tensor hierarchy algebra}},  {\em J. Math. Phys.} {\bf
  55} (2014) 011701, [\href{http://arxiv.org/abs/1305.0018}{{\tt
  arXiv:1305.0018}}].

\bibitem{Greitz:2013pua}
J.~Greitz, P.~Howe, and J.~Palmkvist, {\it {The tensor hierarchy simplified}},
  {\em Class. Quant. Grav.} {\bf 31} (2014) 087001,
  [\href{http://arxiv.org/abs/1308.4972}{{\tt arXiv:1308.4972}}].

\bibitem{Cederwall:2018aab}
M.~Cederwall and J.~Palmkvist, {\it {$L_{\infty }$ Algebras for Extended
  Geometry from Borcherds Superalgebras}},  {\em Commun. Math. Phys.} {\bf 369}
  (2019), no.~2 721--760, [\href{http://arxiv.org/abs/1804.04377}{{\tt
  arXiv:1804.04377}}].

\bibitem{Cederwall:2019qnw}
M.~Cederwall and J.~Palmkvist, {\it {Tensor hierarchy algebras and extended
  geometry. Part I. Construction of the algebra}},  {\em JHEP} {\bf 02} (2020)
  144, [\href{http://arxiv.org/abs/1908.08695}{{\tt arXiv:1908.08695}}].

\bibitem{Cederwall:2019bai}
M.~Cederwall and J.~Palmkvist, {\it {Tensor hierarchy algebras and extended
  geometry. Part II. Gauge structure and dynamics}},  {\em JHEP} {\bf 02}
  (2020) 145, [\href{http://arxiv.org/abs/1908.08696}{{\tt arXiv:1908.08696}}].

\bibitem{Bonezzi:2019ygf}
R.~Bonezzi and O.~Hohm, {\it {Leibniz Gauge Theories and Infinity Structures}},
   {\em Commun. Math. Phys.} {\bf 377} (2020), no.~3 2027--2077,
  [\href{http://arxiv.org/abs/1904.11036}{{\tt arXiv:1904.11036}}].

\bibitem{Bonezzi:2019bek}
R.~Bonezzi and O.~Hohm, {\it {Duality Hierarchies and Differential Graded Lie
  Algebras}},  \href{http://arxiv.org/abs/1910.10399}{{\tt arXiv:1910.10399}}.

\bibitem{Lavau:2017tvi}
S.~Lavau, {\it {Tensor hierarchies and Leibniz algebras}},  {\em J. Geom.
  Phys.} {\bf 144} (2019) 147--189,
  [\href{http://arxiv.org/abs/1708.07068}{{\tt arXiv:1708.07068}}].

\bibitem{Lavau:2019oja}
S.~Lavau and J.~Palmkvist, {\it {Infinity-enhancing of Leibniz algebras}},
  {\em Lett. Math. Phys.} {\bf 110} (2020), no.~11 3121--3152,
  [\href{http://arxiv.org/abs/1907.05752}{{\tt arXiv:1907.05752}}].

\bibitem{Lavau:2020pwa}
S.~Lavau and J.~Stasheff, {\it {From Lie algebra crossed modules to tensor
  hierarchies}},  {\em J. Pure Appl. Algebra} {\bf 227} (2023) 107311,
  [\href{http://arxiv.org/abs/2003.07838}{{\tt arXiv:2003.07838}}].

\bibitem{Sakatani:2017xcn}
Y.~Sakatani and S.~Uehara, {\it {$\eta$-symbols in exceptional field theory}},
  {\em PTEP} {\bf 2017} (2017), no.~11 113B01,
  [\href{http://arxiv.org/abs/1708.06342}{{\tt arXiv:1708.06342}}].

\bibitem{Hohm:2015xna}
O.~Hohm and Y.-N. Wang, {\it {Tensor hierarchy and generalized Cartan calculus
  in SL(3) \texttimes{} SL(2) exceptional field theory}},  {\em JHEP} {\bf 04}
  (2015) 050, [\href{http://arxiv.org/abs/1501.01600}{{\tt arXiv:1501.01600}}].

\bibitem{Wang:2015hca}
Y.-N. Wang, {\it {Generalized Cartan Calculus in general dimension}},  {\em
  JHEP} {\bf 07} (2015) 114, [\href{http://arxiv.org/abs/1504.04780}{{\tt
  arXiv:1504.04780}}].

\bibitem{Samtleben:2008pe}
H.~Samtleben, {\it {Lectures on Gauged Supergravity and Flux
  Compactifications}},  {\em Class. Quant. Grav.} {\bf 25} (2008) 214002,
  [\href{http://arxiv.org/abs/0808.4076}{{\tt arXiv:0808.4076}}].

\bibitem{YuhoFalkNew}
F.~Hassler and Y.~Sakatani, {\it {Consistent truncations and generalized
  dualities based on exceptional generalized cosets}}, . to appear.

\bibitem{Bossard:2017aae}
G.~Bossard, M.~Cederwall, A.~Kleinschmidt, J.~Palmkvist, and H.~Samtleben, {\it
  {Generalized diffeomorphisms for $E_9$}},  {\em Phys. Rev. D} {\bf 96}
  (2017), no.~10 106022, [\href{http://arxiv.org/abs/1708.08936}{{\tt
  arXiv:1708.08936}}].

\bibitem{Bossard:2018utw}
G.~Bossard, F.~Ciceri, G.~Inverso, A.~Kleinschmidt, and H.~Samtleben, {\it
  {E$_{9}$ exceptional field theory. Part I. The potential}},  {\em JHEP} {\bf
  03} (2019) 089, [\href{http://arxiv.org/abs/1811.04088}{{\tt
  arXiv:1811.04088}}].

\bibitem{Bossard:2021jix}
G.~Bossard, F.~Ciceri, G.~Inverso, A.~Kleinschmidt, and H.~Samtleben, {\it
  {E$_{9}$ exceptional field theory. Part II. The complete dynamics}},  {\em
  JHEP} {\bf 05} (2021) 107, [\href{http://arxiv.org/abs/2103.12118}{{\tt
  arXiv:2103.12118}}].

\bibitem{Cederwall:2021ymp}
M.~Cederwall and J.~Palmkvist, {\it {Tensor Hierarchy Algebra Extensions of
  Over-Extended Kac{\textendash}Moody Algebras}},  {\em Commun. Math. Phys.}
  {\bf 389} (2022), no.~1 571--620,
  [\href{http://arxiv.org/abs/2103.02476}{{\tt arXiv:2103.02476}}].

\bibitem{Cederwall:2025muh}
M.~Cederwall and J.~Palmkvist, {\it {Gradient structures from extensions of
  over-extended Kac-Moody algebras}},  {\em JHEP} {\bf 08} (2025) 200,
  [\href{http://arxiv.org/abs/2503.17779}{{\tt arXiv:2503.17779}}].

\bibitem{Bossard:2019ksx}
G.~Bossard, A.~Kleinschmidt, and E.~Sezgin, {\it {On supersymmetric E$_{11}$
  exceptional field theory}},  {\em JHEP} {\bf 10} (2019) 165,
  [\href{http://arxiv.org/abs/1907.02080}{{\tt arXiv:1907.02080}}].

\bibitem{Bossard:2021ebg}
G.~Bossard, A.~Kleinschmidt, and E.~Sezgin, {\it {A master exceptional field
  theory}},  {\em JHEP} {\bf 06} (2021) 185,
  [\href{http://arxiv.org/abs/2103.13411}{{\tt arXiv:2103.13411}}].

\bibitem{Polacek:2017hnq}
M.~Polacek, {\em {Aspects of T-dually extended Superspaces}}.
\newblock PhD thesis, SUNY, Stony Brook, 5, 2017.

\bibitem{Butter:2022iza}
D.~Butter, F.~Hassler, C.~N. Pope, and H.~Zhang, {\it {Consistent truncations
  and dualities}},  {\em JHEP} {\bf 04} (2023) 007,
  [\href{http://arxiv.org/abs/2211.13241}{{\tt arXiv:2211.13241}}].

\bibitem{Hassler:2022egz}
F.~Hassler and Y.~Sakatani, {\it {All maximal gauged supergravities with
  uplift}},  \href{http://arxiv.org/abs/2212.14886}{{\tt arXiv:2212.14886}}.

\bibitem{Osten:2019ayq}
D.~Osten, {\it {Current algebras, generalised fluxes and non-geometry}},  {\em
  J. Phys. A} {\bf 53} (2020), no.~26 265402,
  [\href{http://arxiv.org/abs/1910.00029}{{\tt arXiv:1910.00029}}].

\bibitem{Hatsuda:2023dwx}
M.~Hatsuda, O.~Hul\'\i{}k, W.~D. Linch, W.~D. Siegel, D.~Wang, and Y.-P. Wang,
  {\it {$ \mathcal{A} $-theory \textemdash{} A brane world-volume theory with
  manifest U-duality}},  {\em JHEP} {\bf 10} (2023) 087,
  [\href{http://arxiv.org/abs/2307.04934}{{\tt arXiv:2307.04934}}].

\bibitem{Butter:2021dtu}
D.~Butter, {\it {Exploring the geometry of supersymmetric double field
  theory}},  {\em JHEP} {\bf 01} (2022) 152,
  [\href{http://arxiv.org/abs/2101.10328}{{\tt arXiv:2101.10328}}].

\bibitem{Lada:1992wc}
T.~Lada and J.~Stasheff, {\it {Introduction to SH Lie algebras for
  physicists}},  {\em Int. J. Theor. Phys.} {\bf 32} (1993) 1087--1104,
  [\href{http://arxiv.org/abs/hep-th/9209099}{{\tt hep-th/9209099}}].

\bibitem{Hohm:2017pnh}
O.~Hohm and B.~Zwiebach, {\it {$L_{\infty}$ Algebras and Field Theory}},  {\em
  Fortsch. Phys.} {\bf 65} (2017), no.~3-4 1700014,
  [\href{http://arxiv.org/abs/1701.08824}{{\tt arXiv:1701.08824}}].

\bibitem{Borsten:2021ljb}
L.~Borsten, H.~Kim, and C.~Saemann, {\it {$EL_\infty$-algebras, Generalized
  Geometry, and Tensor Hierarchies}},
  \href{http://arxiv.org/abs/2106.00108}{{\tt arXiv:2106.00108}}.

\bibitem{Baron:2018lve}
W.~H. Baron, E.~Lescano, and D.~Marqu{\'e}s, {\it {The generalized
  Bergshoeff-de Roo identification}},  {\em JHEP} {\bf 11} (2018) 160,
  [\href{http://arxiv.org/abs/1810.01427}{{\tt arXiv:1810.01427}}].

\bibitem{Baron:2020xel}
W.~Baron and D.~Marques, {\it {The generalized Bergshoeff-de Roo
  identification. Part II}},  {\em JHEP} {\bf 01} (2021) 171,
  [\href{http://arxiv.org/abs/2009.07291}{{\tt arXiv:2009.07291}}].

\bibitem{Gitsis:2024gfb}
A.~Gitsis and F.~Hassler, {\it {Unraveling the generalized Bergshoeff-de Roo
  identification}},  {\em JHEP} {\bf 06} (2025) 048,
  [\href{http://arxiv.org/abs/2412.17900}{{\tt arXiv:2412.17900}}].

\bibitem{Baez:2010ya}
J.~C. Baez and J.~Huerta, {\it {An Invitation to Higher Gauge Theory}},  {\em
  Gen. Rel. Grav.} {\bf 43} (2011) 2335--2392,
  [\href{http://arxiv.org/abs/1003.4485}{{\tt arXiv:1003.4485}}].

\bibitem{Grutzmann:2014hkn}
M.~Gr{\"u}tzmann and T.~Strobl, {\it {General Yang{\textendash}Mills type gauge
  theories for $p$-form gauge fields: From physics-based ideas to a
  mathematical framework or From Bianchi identities to twisted Courant
  algebroids}},  {\em Int. J. Geom. Meth. Mod. Phys.} {\bf 12} (2014) 1550009,
  [\href{http://arxiv.org/abs/1407.6759}{{\tt arXiv:1407.6759}}].

\bibitem{Ritter:2015zur}
P.~Ritter, C.~S{\"a}mann, and L.~Schmidt, {\it {Generalized Higher Gauge
  Theory}},  {\em JHEP} {\bf 04} (2016) 032,
  [\href{http://arxiv.org/abs/1512.07554}{{\tt arXiv:1512.07554}}].

\bibitem{Borsten:2024gox}
L.~Borsten, M.~Jalali~Farahani, B.~Jur{\v{c}}o, H.~Kim,
  J.~N{\'a}ro{\v{z}}n{\'y}, D.~Rist, C.~Saemann, and M.~Wolf, {\it {Higher
  Gauge Theory}},  \href{http://arxiv.org/abs/2401.05275}{{\tt
  arXiv:2401.05275}}.

\bibitem{Bonelli:2005ti}
G.~Bonelli and M.~Zabzine, {\it {From current algebras for p-branes to
  topological M-theory}},  {\em JHEP} {\bf 09} (2005) 015,
  [\href{http://arxiv.org/abs/hep-th/0507051}{{\tt hep-th/0507051}}].

\bibitem{Sakatani:2019zrs}
Y.~Sakatani, {\it {$U$-duality extension of Drinfel\textquoteright{}d double}},
   {\em PTEP} {\bf 2020} (2020), no.~2 023B08,
  [\href{http://arxiv.org/abs/1911.06320}{{\tt arXiv:1911.06320}}].

\bibitem{Malek:2019xrf}
E.~Malek and D.~C. Thompson, {\it {Poisson-Lie U-duality in Exceptional Field
  Theory}},  {\em JHEP} {\bf 04} (2020) 058,
  [\href{http://arxiv.org/abs/1911.07833}{{\tt arXiv:1911.07833}}].

\end{thebibliography}\endgroup

\end{document}